\documentclass[12pt,a4paper]{article}
\pdfoutput=1

\usepackage{array}
\usepackage{epsfig}
\usepackage{amssymb}
\usepackage{graphics,graphpap}
\usepackage{amsmath}
\usepackage{wasysym}
\usepackage{slashed}
\usepackage{url}
\usepackage{placeins}

\newcommand{\gev}{\,\, \mathrm{GeV}}

\newcommand{\MHp}{M_{H^\pm}}
\newcommand{\matr}[1]{{\mathbf{#1}}}

\newcommand{\be}{\begin{equation}}
\newcommand{\ee}{\end{equation}}
\newcommand{\bea}{\begin{eqnarray}}
\newcommand{\eea}{\end{eqnarray}}

\newcommand{\bi}{\begin{itemize}}
\newcommand{\ei}{\end{itemize}}

\newcommand{\re}{\mathrm{Re}\,}

\newcommand{\UCha}{U}
\newcommand{\VCha}{V}
\newcommand{\ZNeu}{N}

\usepackage{cite}

\newcommand\T{\rule{0pt}{2.5 ex}}

\newcommand\B{\rule[-1.5ex]{0pt}{0pt}}

\begin{document}
\begin{titlepage}
\renewcommand{\thefootnote}{\fnsymbol{footnote}}
\begin{flushright}
{\tt DESY 12-015}
\end{flushright}
\begin{center}
{\large \bf \boldmath
Consistent on shell renormalisation of electroweakinos\\[.4em]
in the complex MSSM: LHC and LC predictions}
\vskip 0.4cm 
{\sc
Aoife Bharucha\footnote{aoife.bharucha@desy.de}$^{,1}$,
Alison Fowler\footnote{former address}$^{,2}$,
Gudrid Moortgat-Pick\footnote{gudrid.moortgat-pick@desy.de}$^{,1,3}$ 
and  Georg~Weiglein\footnote{georg.weiglein@desy.de}$^{,3}$
} \vskip 0.4cm
\begin{small}
        $^1$ {\em
II. Institut f\"{u}r Theoretische Physik, University of Hamburg, Luruper Chaussee 149, D-22761 Hamburg, Germany}\\
\vskip 0.1cm
        $^2$ {\em IPPP, Department of Physics,
University of Durham, Durham DH1 3LE, UK}\\
\vskip 0.1cm
$^3$ {\em DESY, Deutsches Elektronen-Synchrotron, Notkestr. 85, D-22607 Hamburg, Germany}\\
\vskip 0.4cm
\end{small}
{\large\bf Abstract\\[9.pt]}\parbox[t]{\textwidth}{
We extend the formalism developed in Ref.~\cite{Fowler:2009ay} for the 
renormalisation of the chargino--neutralino sector to the most general
case of the MSSM with complex parameters. We show that products of
imaginary parts arising from MSSM parameters and from absorptive parts
of loop integrals can already contribute to predictions for physical 
observables at the one-loop level, and demonstrate that the consistent
treatment of such contributions gives rise to non-trivial
structure, either in the field renormalisation constants or the
corrections associated with the external legs of the considered
diagrams. We furthermore point out that the phases of the parameters in the
chargino--neutralino sector do not need to be renormalised at the one-loop
level, and demonstrate that the appropriate choice for the mass parameters 
used as input for the on-shell conditions depends both on the process
and the region of MSSM parameter space under consideration. As an
application, we compute the complete one-loop results in the MSSM with
complex parameters for the process
$h_a\to\tilde\chi_i^+\tilde\chi_j^-$ 
(Higgs-propagator corrections have been incorporated up to the two-loop
level), which may be of interest for SUSY
Higgs searches at the LHC, and for chargino pair-production at an 
$e^+e^-$ Linear Collider, $e^+e^-\to\tilde\chi_i^+\tilde\chi_j^-$. 
We investigate the dependence of the theoretical predictions on the
phases of the MSSM parameters, 
analysing in particular the numerical relevance of the absorptive
parts of loop integrals.
}
\vfill

\end{center}
\end{titlepage}

\setcounter{footnote}{0}

\section{Introduction}\label{sec:1}

The search for physics beyond the Standard Model (SM) is one of the main
goals of the physics programme of the Large Hadron Collider (LHC).
Supersymmetry (SUSY) continues to be a particularly attractive extension
of the SM. The minimal version, the Minimal Supersymmetric Standard Model 
(MSSM), predicts superpartners for all the fermions and gauge bosons 
of the SM as well as an extended Higgs sector consisting of two Higgs
doublets. 

The recent signal discovered in the Higgs searches at
ATLAS~\cite{ATLASdisc} and CMS~\cite{CMSdisc} with a mass of about 
$126\gev$, which is also compatible with the excess observed at the
Tevatron~\cite{TevHiggs}, is well in keeping with an interpretation in
the MSSM in terms of the
light (see e.g.\
\cite{125-susypapers,Heinemeyer:2011aa,Benbrik:2012uy,Bechtle:2012jw})
or even the heavy CP-even Higgs
boson~\cite{Heinemeyer:2011aa,Benbrik:2012uy,MH126,Bechtle:2012jw}. 
On the other hand, the direct
searches for superpartners at the LHC have so far not revealed any sign
of a signal. The searches up to now have mainly been sensitive
to the production of squarks of the first two generations and the
gluino~\cite{Atlas-sqgl,CMS-sqgl}.
The analyses are just starting to become sensitive to the direct
production of the squarks of the third generation and to the direct 
production of colour-neutral states of the MSSM, see 
e.g.~Refs.~\cite{Atlas-stga,CMS-stga}. In particular, in the
chargino sector the most important limit is still the one from LEP of 
about $100 \gev$~\cite{PDG}. The neutralino sector of the MSSM is even less
constrained. If one drops the assumption of the GUT relation between
the parameters $M_2$, which appears in the chargino sector, and $M_1$,
which appears only in the neutralino sector, the lightest neutralino can 
be arbitrarily light without violating the existing experimental
constraints~\cite{Dreiner:2009ic}.

In general, many of the MSSM parameters may take complex values. If this
is the case, it leads to CP violation beyond that provided by the phase 
of the Cabibbo-Kobayashi-Maskawa (CKM) matrix, i.e.\ the matrix that 
governs the mixing of the quarks in the SM. The CKM picture of CP
violation has been remarkably successful in passing many experimental
tests, which has led to stringent constraints on the possible structure
of new physics contributions. On the other hand, additional sources of 
CP violation are in
fact needed to explain the baryon asymmetry in the universe via a first 
order electro-weak phase transition, as the CKM phase can only partially
account for the observed asymmetry (for a review see e.g.\
Ref.~\cite{Ibrahim:2007fb}).

It is therefore of interest to investigate the collider phenomenology of 
CP-violating effects in the MSSM with complex parameters, see for
instance Refs.~\cite{Kittel:2009fg,MoortgatPick:2010wp,Dreiner:2010wj,Bornhauser:2011ab} and references
therein for studies of CP-violating effects at the LHC 
via rate asymmetries or triple products. While in some cases CP-violating
effects occur already at tree-level, it is important to take into
account also loop-induced CP-violating effects. In order to make
predictions for CP-violating effects at the loop-level it is thus
necessary to perform the renormalisation of the MSSM for the general
case of complex parameters. It should be noted in this context that in
the MSSM with complex parameters there are two sources of imaginary
parts occurring at the loop level, namely complex pararameters and 
the absorptive parts of loop integrals. While absorptive parts of loop
integrals are often neglected in one-loop calculations, a consistent
treatment of these contributions is essential in the MSSM with complex
parameters since imaginary
parts of the loop integrals can combine with imaginary parts of the MSSM
parameters to contribute to the real part of the one-loop contribution.

In this paper we work out the on-shell renormalisation of the
chargino--neutralino sector of the MSSM for the most general case of
complex parameters, which involves in particular a consistent treatment
of the absorptive parts of loop integrals. 
At leading order (LO) the 
chargino--neutralino sector of the complex MSSM
depends on the gaugino masses $M_1$ and $M_2$, the
higgsino mass $\mu$ and $\tan\beta$, the ratio of the vacuum expectation
values of the two neutral Higgs doublet fields, while at higher orders
many more parameters become relevant. 
For the field
renormalisation, we follow the formalism developed in
Ref.~\cite{Fowler:2009ay}, where it was shown that it is convenient to
choose different field renormalisation constants for incoming and
outgoing charginos and neutralinos in order to ensure the correct
on-shell properties of the external particles. While in
Ref.~\cite{Fowler:2009ay} the parameters in the chargino and neutralino
sector were assumed to be real, we perform the parameter renormalisation
for the general case of complex parameters.
We find that at the one-loop level the phases of the parameters in the
chargino--neutralino sector do not need to be renormalised. Furthermore
we demonstrate that an appropriate choice of the mass parameters 
used as input for the on-shell conditions depends both on the process
under consideration and the region of MSSM parameter space. 

Recently, in Refs.~\cite{Fritzsche:2011nr,Heinemeyer:2011gk,Heinemeyer:2012wp} 
an on-shell scheme for the renormalisation of the chargino and neutralino 
sector in the complex MSSM has also been 
employed. The differences in the approach
of treating the CP-violating phases as compared to the present paper
have been discussed in Ref.~\cite{Bharucha:2012re} 
(see also Refs.~\cite{Bharucha:2012qr,Bharucha:2012wu}).

The on-shell renormalisation of the chargino-neutralino sector has been
investigated for the MSSM with real parameters in
Refs.~\cite{Lahanas:1993ib,Pierce:1993gj,Pierce:1994ew,Eberl:2001eu,Fritzsche:2002bi,Oller:2003ge,Drees:2006um,Fowler:2009ay,AlisonsThesis}.
More recently, in Ref.~\cite{Chatterjee:2011wc} a method to
systematically choose on-shell conditions for the parameter
renormalisation in any chosen scenario was presented, including the case of the
parameters being strongly mixed. These results are in accordance with
the earlier investigations of Refs.~\cite{AlisonsThesis,Fowler:2009ay}.

As an application of the framework for the renormalisation developed in
this work, we compute the complete one-loop results in the MSSM with complex 
parameters for the processes $h_a\to\tilde\chi_i^+\tilde\chi_j^-$ (for $a$=2,3) and 
$e^+e^-\to\tilde\chi_i^+\tilde\chi_j^-$. Note that in the complex MSSM 
the neutral Higgs bosons $(h,H,A)$ mix to give $(h_1,h_2,h_3)$, as described later.
In our numerical analysis we 
study in particular the dependence of the results on the phases of
the complex parameters and we investigate the numerical 
impact of products of imaginary parts arising from 
complex pararameters and from absorptive parts of loop integrals.

The decay of heavy Higgs bosons to charginos and neutralinos, 
$h_a\to\tilde\chi_i^+\tilde\chi_j^-$, is important in the context of 
SUSY Higgs searches at the LHC. The experimental signature of this
process comprises four leptons and missing transverse 
energy~\cite{Baer:1992kd,Baer:1994fx,Moortgat:2001pp,Bisset:2000ud,
Bisset:2007mi}.
In the search for the heavy neutral Higgs bosons of the MSSM this
channel may provide sensitivity also in the ``LHC wedge
region'' (see e.g.\ Refs.~\cite{CMSTDR,arXiv:0704.0619}), where the 
standard searches for heavy MSSM Higgs bosons in $\tau^+\tau^-$ and 
$b \bar b$ final states are expected to be not sufficiently significant 
for a discovery.
Electroweak one-loop corrections to this class of processes have been
evaluated in Refs.~\cite{Zhang:2002fu,Eberl:2004ic,Frisch:2010gw} for
the case of real parameters. As explained above, we have obtained the
complete one-loop result for the decay of a heavy MSSM Higgs boson
in the general
case of complex parameters (which gives rise to a mixing of 
all three neutral Higgs bosons 
to form the mass eigenstates). We have incorporated into our
result Higgs propagator corrections up to the two-loop level.

Since charginos and neutralinos are expected to be among the lightest 
supersymmetric particles and, as mentioned above, their mass range is
only weakly constrained from SUSY searches at the LHC so far, the
direct production of these particles via $e^+e^-$ annihilation is of particular interest for
physics at a future $e^+e^-$ Linear Collider (LC). We focus here on 
chargino pair-production, $e^+e^-\to\tilde\chi_i^+\tilde\chi_j^-$.
High-precision measurements of this process in the clean experimental
environment of an $e^+e^-$ LC could be crucial for
uncovering the fundamental parameters of this sector and for determining
the nature of the underlying physics. 
Based on a leading-order treatment, the determination of the parameters 
$M_1$, $M_2$, $\mu$ and $\tan\beta$ is expected to be possible at the 
percent level, providing also sensitivity to non-zero phases of complex
parameters~\cite{Desch:2003vw}. At this level of accuracy, higher-order
corrections need to be incorporated. For the case of real parameters, 
one-loop corrections to chargino pair production at a future LC have
been investigated in Refs.~\cite{Oller:2005xg,Fritzsche:2005}.
Loop-induced CP-violating effects have been studied in 
Refs.~\cite{Osland:2007xw,Rolbiecki:2007se}, in particular the effect of
complex parameters on certain asymmetries, but the considered
quantities were UV finite and therefore did not require
renormalisation. We extend the previous results to the general case of
complex parameters.

In the following section we will introduce the MSSM with complex
parameters and briefly summarise our renormalisation procedure for the 
Higgs, (s)fermion and gauge boson sectors. 
In Sec.~\ref{sec:3} we present a framework for the on-shell renormalisation 
of the chargino and neutralino sector of the MSSM 
for the general case of complex parameters, where in particular the
imaginary parts arising from complex parameters and from absorptive
parts of loop integrals are consistently treated. As an application of
this framework, in Sec.~\ref{sec:4} we derive new results for two
phenomenologically interesting processes, Higgs decays into charginos
and chargino pair production at a future LC, and we study the dependence
of the results on the complex parameters and the numerical impact of
products of imaginary parts. 
We conclude in Sec.~\ref{sec:5}.

\section{The MSSM with complex parameters and its renormalisation in the on-shell scheme}\label{sec:2}

As mentioned above, complex parameters arise naturally in SUSY, 
inducing CP violation.
In the most general MSSM (for the case of massless neutrinos) 
there are 40 possible phases.
Under the assumption, however, of Minimal Flavour Violation, 
which is motivated by the strong
constraints on supersymmetric contributions to flavour-changing neutral
current processes such as $b\to s\gamma$, the number of MSSM parameters
that may be complex reduces to 14:
the phases of the sfermion trilinear couplings
$\phi_f$, where $f=u,c,t,d,s,b,e,\mu,\tau$; the phases of the gaugino
mass parameters $\phi_{M_i}$, where $i=1,2,3$; the phase of the Higgsino mass
parameter $\phi_\mu$, and the phase of the Higgsino mass mixing
parameter $\phi_{m_{12}}$~\cite{Dimopoulos:1995ju}.
Out of these, the freedom to redefine fields means that any two phases 
may be rotated away,
and as in Refs.~\cite{Barger:2001nu,Fowler:2009ay} we choose those to be 
$\phi_{M_2}$ and $\phi_{m_{12}}$.
We therefore consider 12 non-vanishing phases in the following. 
The most restrictive experimental constraints on those phases
arise from bounds on the electric dipole moments (EDMs) of the
neutron $d_n$, mercury $d_{\rm Hg}$ and Thallium, $d_{\rm
Tl}$~\cite{Baker:2006ts,Regan:2002ta,Griffith:2009zz}\footnote{In
addition, EDMs of the heavy quarks~\cite{Hollik:1997vb}, the
electron~\cite{Demir:2003js,Pilaftsis:1999td} and the
deuteron~\cite{Lebedev:2004va} can have a significant impact.},
\begin{align}
\nonumber  |d_n|&<2.9\,10^{-26} e\,\mathrm{cm}\,(90\% \mathrm{C.L.}),\\
\nonumber  |d_{\rm Tl}|&<9.0\,10^{-25} e\,\mathrm{cm}\,(90\% \mathrm{C.L.}),\\
 |d_{\rm Hg}|&<3.1\,10^{-29} e\,\mathrm{cm}\,(95\% \mathrm{C.L.}).
\end{align}
These EDMs are functions of CP-odd operators, e.g.\ the electron EDM
$d_e$, the quark EDM $d_q$, the chromo EDM $\tilde{d}_q$ (for a recent
review see Ref.~\cite{Li:2010ax}), via atomic or hadronic matrix
elements which can have large theoretical uncertainties, ranging from
$\sim 10\%$ to $50\%$~\cite{Ellis:2008hq,Demir:2003js}. 
Dominant contributions to $d_n$, $d_{\rm Hg}$ and $d_{\rm Tl}$ come from
$d_e$, $d_{u,d}$, $\tilde{d}_{u,d}$. The MSSM contributions to these
EDMs have been studied in detail in the literature, see
Refs.~\cite{Ellis:2008hq,Li:2010ax} and references therein. The dominant
contributions to those EDMs involve the first two
generations of squarks and sleptons, thereby imposing severe constraints
on $A_{q,l}$ for $q=u,d,s,c$ and $l=e,\mu$
\footnote{Possible circumstances under which those bounds are evaded 
are discussed in
Refs.~\cite{Ibrahim:1997nc,Brhlik:1998zn}}. In our numerical evaluation
below we set the severely constrained phases of the trilinear couplings 
to zero.
In contrast, the third generation trilinear couplings of the squarks and
sleptons are much less constrained by the EDM's, and can possibly result
in large effects on observables.
The phase $\phi_\mu$ of the higgsino mass parameter is also severely
constrained in the convention where $\phi_{M_2}$ is rotated away.
However, since it is the only phase present at tree level in the
chargino sector, we will nevertheless investigate below 
the numerical impact of varying this phase.
The bino phase $\phi_{M_1}$, on the other hand, is less constrained by the 
EDMs, so that variations of this phase can potentially have interesting 
consequences in the neutralino sector.
Cosmological effects of this phase include favouring
bino-driven electroweak baryogenesis~\cite{Li:2008ez}, and modifying the
relic density as well as both direct and indirect detection rates of
neutralino dark matter.
Concerning collider phenomenology, it can affect neutralino production rates
and CP-violating observables at the LHC, see e.g.\
Ref.~\cite{Ellis:2008hq,MoortgatPick:2009jy}, and at a future LC, see
e.g.\ Refs.~\cite{Barger:2001nu,Bartl:2004ut,Kittel:2009fg,Deppisch:2010nc,
Kittel:2011rk}. Note that, unfortunately, complex 
phases are often neglected in publically available tools for calculating
MSSM spectra, meaning that these areas of SUSY parameter space are not 
sufficiently explored.

Below we will describe a systematic approach to the on-shell
renormalisation of the chargino and neutralino sector of the MSSM with
complex parameters. In order to illustrate this approach, we will derive
new one-loop results for two processes involving external charginos, 
namely $h_a\to\tilde\chi_i^+\tilde\chi_j^-$ and
$e^+e^-\to\tilde\chi_i^+\tilde\chi_j^-$, where $i,j=1,2$ and $a=2,3$.
Besides the renormalisation of the 
chargino and neutralino sectors, these processes also require the renormalisation of the MSSM
Higgs sector, the sfermion sector, as well as the gauge boson and
SM-fermion sector. 
We therefore first describe the renormalisation of these sectors before
turning to the renormalisation of the chargino and neutralino sectors in
the following section.

\subsection{Renormalisation in the Higgs sector}
 
The renormalisation of the Higgs sector plays an important role in the
following, firstly because we study the process
$h_a\to\tilde\chi^+_i\tilde\chi^-_j$, where the external Higgs
must be renormalised, and secondly because it enters the
chargino-neutralino sector through the renormalisation of the parameter
$\tan\beta$.
Expressing the Higgs potential in terms of the soft masses $m_1$, $m_2$,
and the mass mixing parameter $m_{12}$ leads to
\begin{eqnarray}
 V_H&=&m_1^2 H^{\ast}_{1i} H_{1i}+m_2^2 H^{\ast}_{2i} H_{2i}-\epsilon^{ij}(m_{12}^2 H_{1i}H_{2j}+m_{12}^{2\;\ast} 
H_{1i}^{\ast}H_{2j}^{\ast})\nonumber \\
&&+\frac{1}{8}(g^2+g^{\prime\,2})(H_{1i}^{\ast}H_{1i}-H_{2i}^{\ast}H_{2i})^2+\frac{1}{2}g^{\prime\,2}|H_{1i}^{\ast}H_{2i}|^2,
\end{eqnarray}
where $g$ and $g^\prime$ are the $U(1)$ and $SU(2)_L$ couplings, and the two Higgs doublets can be expressed through
\begin{eqnarray}
\mathcal{H}_1&=\left(\begin{array}{c} H_{11}\\H_{12}\end{array}\right)
             &=\left( \begin{array}{cc} v_1+\frac{1}{\sqrt{2}}(\phi_1-i \chi_1)\\
                            -\phi_1^{-}
                            \end{array} \right),\\
\mathcal{H}_2&=\left(\begin{array}{c} H_{21}\\H_{22}\end{array}\right)
             &=e^{i\xi}\left( \begin{array}{cc}\phi_2^{+}\\
                            v_2+\frac{1}{\sqrt{2}}(\phi_2+i \chi_2)
                            \end{array} \right),
\end{eqnarray}
with the vacuum expectation values of the two Higgs doublets, $v_1$ and
$v_2$, and $\tan\beta \equiv v_2/v_1$.
The MSSM Higgs sector is CP-conserving at tree-level, i.e.\ the phase of
the parameter $m_{12}$ and the relative phase 
$\xi$ between the two Higgs doublets can be rotated away and vanish upon
the minimisation of the Higgs potential, respectively.
Expanding $V_H$ in terms of the neutral fields $\phi_{1/2}$ and $\chi_{1/2}$ 
as well as the charged fields $\phi^\pm_{1/2}$ results in tadpole terms and mass mixing terms,
\begin{align}
 V_H=\,& ... -T_{\phi_1} \phi_1-T_{\phi_2} \phi_2-T_{\chi_1} \chi_1-T_{\chi_2} \chi_2\nonumber\\
&+\frac{1}{2}(\phi_1\; \phi_2 \; \chi_1\;\chi_2)\,\mathbf{M}_{\phi\phi\chi\chi}\left( \begin{array}{c}
\phi_1 \\
\phi_2 \\
\chi_1 \\
\chi_2 \end{array} \right)  
+\frac{1}{2}(\phi_1^-\; \phi_2^-)\,\mathbf{M}_{\phi^{\pm}\phi^{\pm}}\left( \begin{array}{c}
\phi_1^+ \\
\phi_2^+ \end{array} \right)+...\,\,.
\end{align}
The two mass mixing matrices $\mathbf{M}_{\phi\phi\chi\chi}$ and
$\mathbf{M}_{\phi^{\pm}\phi^{\pm}}$ can be diagonalised by rotation
matrices parametrised by mixing angles $\alpha$ and $\beta_n$ and
$\beta_c$, resulting in the
neutral Higgs bosons, $h$, $H$ and $A$, and the neutral Goldstone boson $G$, 
as well as the charged Higgs bosons $H^\pm$ and the charged Goldstone bosons 
$G^\pm$.
By renormalising the Higgs doublet fields,
\begin{equation}
\mathcal{H}_{1,2} \rightarrow (1+\frac{1}{2}\delta Z_{\mathcal{H}_{1,2}})\mathcal{H}_{1,2},
\end{equation}
one can obtain all the required Higgs field renormalisation constants, 
which can be written in terms of $\delta Z_{\mathcal{H}_{1,2}}$, as
discussed in Ref.~\cite{Frank:2006yh}.
Similarly choosing to renormalise the tadpole coefficients,
\begin{equation}
T_{h,H,A} \rightarrow T_{h,H,A}+ \delta T_{h,H,A},
\end{equation}
the charged Higgs mass
\begin{equation}
M_{H^\pm}^2\rightarrow  M_{H^\pm}^2+\delta M_{H^\pm}^2, 
\end{equation}
and $\tan\beta$ via
\begin{equation}
\tan{\beta} \rightarrow \tan{\beta}(1+\delta \tan{\beta}),
\end{equation}
all parameter renormalisation constants can be obtained using relations
connecting them to $\delta T_{h,H,A}$, $\delta M_{H^\pm}^2$ and $\delta
\tan{\beta}$, as also given in Ref.~\cite{Frank:2006yh}. It is
convenient to renormalise $\tan\beta$ in the $\overline{\rm DR}$ scheme, 
see the discussion in
Refs.~\cite{Brignole:1992uf,Frank:2002qa,Freitas:2002pe,Frank:2006yh}.
With the Higgs field renormalisation constants in the 
$\overline{\rm DR}$ scheme,
\begin{equation}\delta
Z_{\mathcal{H}_1}^{\overline{\mathrm{DR}}}=-\re\;\Sigma_{HH,\alpha=0}^{\prime\;\mathrm{div}}
\label{eqn:dZH1}\end{equation}
\begin{equation}\delta
Z_{\mathcal{H}_2}^{\overline{\mathrm{DR}}}=-\re\;\Sigma_{hh,\alpha=0}^{\prime\;\mathrm{div}
} , \label{eqn:dZH2}\end{equation}
this yields
\begin{equation}
 \delta \tan{\beta}^{\overline{\mathrm{DR}}}=
\frac{1}{2}(\delta Z^{\overline{\mathrm{DR}}}_{\mathcal{H}_2}
-\delta Z^{\overline{\mathrm{DR}}}_{\mathcal{H}_1}).
\end{equation}
The mass of the charged Higgs is renormalised accoring to the usual
on-shell condition, yielding
\begin{equation}
 \delta M_{H^\pm}^2 = \re\;\Sigma_{H^+H^-}(M_{H^\pm}^2),
\end{equation}
and the renormalisation constants of the tadpole coefficients are fixed via the condition  that the renormalised coefficient should vanish, leading to 
\begin{equation} \delta T_{h,H,A}=-T_{h,H,A}.\end{equation}

The numerically important Higgs propagator-type corrections in the MSSM
Higgs sector not only affect the predictions for the Higgs boson masses,
but also give rise to a loop-induced mixing between the neutral Higgs
bosons. In order to ensure the correct on-shell properties of the
external particles in the S-matrix elements, the mixing between
different states has to vanish on-shell, and the residues of the
propagators have to be normalised to one. We achieve this by applying 
finite wave function normalisation factors $\hat{\bf Z}_{ij}$, which
contain the complete one-loop contributions of the Higgs boson
self-energies as well as the dominant two-loop corrections, as
implemented in the program
\texttt{FeynHiggs}~\cite{Heinemeyer:1998np,Heinemeyer:1998yj,Degrassi:2002fi,Fra
nk:2006yh,Hahn:2009zz}. 

The wave function normalisation factors $\hat{\bf Z}_{ij}$, for which we
use the definition given in Refs.~\cite{Williams:2007dc,Williams:2011bu}, 
can be written as a non-unitary matrix $\hat{\bf Z}$. In this way
a one-particle
irreducible n-point vertex-function $\hat G_{h_a}$ involving a single external
Higgs $h_a$ can be expressed as
\begin{align}
\begin{pmatrix} \hat G_{h_a} \\ \hat G_{h_b} \\ \hat G_{h_c}
\end{pmatrix} = \matr{\hat Z} \cdot
\begin{pmatrix} \hat G_h \\ \hat G_H \\\hat  G_A \end{pmatrix},
\end{align}
in terms of the vertex functions $\hat G_h$, $\hat G_H$, $\hat G_A$ in 
the $(h, H, A)$ basis.
Here $(h_a,h_b,h_c)$ denotes some combination of $(h_1,h_2,h_3)$. For
definiteness we choose $h_a=h_1$, $h_b=h_2$ and $h_c=h_3$.

In addition to the mixing between the physical Higgs fields, a complete
one-loop prediction for a process in the MSSM involving a neutral Higgs
boson as external particle will in general also involve mixing
contributions with the neutral Goldstone boson and with the Z~boson.
These contributions must explicitly be included in the calculation at
the one-loop level, as discussed in detail in
Refs.~\cite{Williams:2007dc,Williams:2011bu}.

\subsection{Renormalisation in the sfermion sector}

As stated earlier, the calculation of
$e^+e^-\to\tilde\chi^+_i\tilde\chi^-_j$ at one-loop order requires the
renormalisation of the sfermion sector, as $\tilde\nu_e$ enters the
tree-level t-channel diagram. The renormalisation in the sfermion sector 
is furthermore needed for the evaluation of higher-order corrections in
the Higgs sector, see above.
At lowest order, the squarks and charged sleptons are mixed via
\begin{equation}
M_{\tilde{f}}=
\left( \begin{array}{cc}
M_L^2+m_f^2+\widetilde{M}_{Z}^2 (I^f_3-Q_f s_W^2) & m_f X^{\ast}_f  \\[.5em]
m_f X_f  & M_{\tilde{f}_R}^2+m_f^2+ \widetilde{M}_{Z}^2\,Q_f s_W^2
\end{array} \right),
\label{sfermion}
\end{equation}
for $s_w$ as defined in the following subsection, making use of the abbreviation $\widetilde{M}_{Z}^2\equiv M_Z^2\cos{2\beta}$, where $M_Z$ is the mass of the $Z$ boson, and defining $X_f$ by
\begin{equation}
 X_f=A_f-\mu^{\ast} \left\{\cot\beta,\tan\beta\right\},
\end{equation}
where $\cot\beta$ applies for the up-type squarks, $f=u,c,t$, and
$\tan\beta$ applies for the down-type sfermions, $f=d,s,b,e,\mu,\tau$
(we treat the neutrinos as being massless). Note that $m_f$, $Q_f$ and $I_3^f$ are the mass, charge and isospin projection of the fermion $f$ respectively.
Here $M_L^2$, $M_{\tilde{f}_R}^2$ and the trilinear coupling $A_f$ are soft SUSY breaking parameters,  of which only the latter may be complex,
\begin{equation}
  A_f=|A_f|e^{i \phi_{A_f}}.
\end{equation}
As we do not consider right-handed neutrinos, 
the sneutrino masses can be expressed by
\begin{equation}
 m_{\tilde{\nu}}^2=M_L^2+\frac{1}{2}M_Z^2\cos{2\beta},
\end{equation}
where $\tilde{\nu}=\tilde{\nu_e},\,\tilde{\nu_\mu},\,\tilde{\nu_\tau}$.

In order to renormalise the sneutrino sector, we define the field and mass renormalisation constants by,
\begin{equation}
 \tilde{\nu}\rightarrow (1+\frac{1}{2}\delta Z_{\tilde{\nu}})\tilde{\nu}\qquad 
 m_{\tilde{\nu}}^2\rightarrow  m_{\tilde{\nu}}^2+\delta m_{\tilde{\nu}}^2.
\end{equation}
Imposing on-shell conditions in the sneutrino sector yields
\begin{equation}
\delta Z_{\tilde{\nu}}=-\Sigma_{\tilde{\nu}}^{\prime}(m_{\tilde{\nu}}^2) \quad\mbox{and}\quad\delta m_{\tilde{\nu}}^2=\re\,\Sigma_{\tilde{\nu}}^{\prime}(m_{\tilde{\nu}}^2).
\end{equation}
It should be noted that choosing the sneutrino mass as an independent
input parameter in this way implies that the renormalisation constant
for the left-handed selectron mass is a derived quantity (following from
SU(2) invariance). 

\subsection{Renormalisation in the gauge boson and fermion sector}

For the gauge-boson masses, $M_W$ and $M_Z$, we choose on-shell
conditions. The weak mixing angle $\theta_W$ is a derived quantity,
following from
\begin{equation}
\sin^2\theta_W \equiv s_W^2=1-\frac{M_W^2}{M_Z^2}.
\end{equation}
With the renormalisation transformations 
\begin{align}
M_Z^2 &\rightarrow  M_Z^2+\delta M_Z^2,&\quad
M_W^2 &\rightarrow  M_W^2+\delta M_W^2,\label{eq:renormtransform1}\\
s_W &\rightarrow  s_W+\delta s_W,&\quad
c_W &\rightarrow  c_W+\delta c_W,
\label{eq:renormtransform2}
\end{align}
where $c_W^2 \equiv \cos^2\theta_W$, this yields
\begin{equation}
 \delta M_{W}^2 = \re\Sigma_T^{WW}(M_{W}^2) \;\;\;\mathrm{and}\;\;\; \delta M_{Z}^2 = \re\Sigma_T^{ZZ}(M_{Z}^2),
\end{equation}
where the transverse part $\Sigma_T(p^2)$ of a self-energy
$\Sigma_{\mu\nu}(p)$ is defined according to
\begin{equation}
 \Sigma_{\mu\nu}(p)=\bigg(- g_{\mu\nu}
+\frac{p_{\mu}p_{\nu}}{p^2}\bigg)\Sigma_T(p^2)
-\frac{p_{\mu}p_{\nu}}{p^2}\Sigma_L(p^2).
\end{equation}
For the renormalisation of the weak mixing angle this results in
\begin{equation}
\delta s_W = \frac{c_W^2}{2 s_W} \left(\frac{\delta
M_Z^2}{M_Z^2}-\frac{\delta M_W^2}{M_W^2}\right) \;\;\;\;\;\mathrm{and} \;\;\;\;\; \delta c_W = - \frac{s_W}{c_W} \delta s_W .\label{eq:RCswcw}
\end{equation}

The renormalisation of the electric charge
\begin{equation}
e\rightarrow e(1+\delta Z_e)
\end{equation}
in the on-shell scheme yields 
\begin{equation}            
\delta Z_e = \frac{1}{2} \Pi^{\gamma\gamma}(0) +
\frac{s_W}{c_W}\frac{\Sigma^{\gamma Z}_T(0)}{M_Z^2} ,
\label{eq:RCZe}
\end{equation} 
where $\Pi^{\gamma\gamma}(0) \equiv \frac{d\, \Sigma^{\gamma\gamma}_T(p^2)}{d\,
p^2}\Big|_{p^2 = 0}$. In order to avoid sensitivity to the light quark
masses this is usually re-expressed in terms of the shift in the fine
structure constant, $\Delta \alpha$, where
\begin{equation}
\Delta \alpha = \Pi^{\gamma\gamma}_{\rm light\; fermions}(0) -
\frac{1}{M_Z^2}\re
\Sigma^{\gamma\gamma}_{T, {\rm light\; fermions}}(M_Z^2) .
\end{equation}
Here ``light fermions'' refers to the contributions of all quarks and
leptons except the top quark. While the leptonic contribution to 
$\Delta \alpha$ can directly be calculated, the hadronic contribution is
obtained from experimental data via a dispersion relation. In our
numerical analysis below we will express our lowest-order results in
terms of the fine structure constant at the scale $M_Z$, 
$\alpha(M_Z^2) = \alpha/(1 - \Delta \alpha)$, so that the contribution of 
$\Delta \alpha$ is absorbed into the lowest-order coupling.

Since we consider processes with external electrons, fermion field
renormalisation constants are needed. We define them according to 
the transformation
\begin{equation}
f_{L,R}\rightarrow (1+\frac{1}{2}\delta Z^f_{L,R})f_{L,R}.
\end{equation}
In the case where the mass of the fermion can be neglected, the 
on-shell condition leads to the simple expression
\begin{equation}
\delta Z^f_{L,R} = - \Sigma^f_{L,R}(0) ,
\end{equation}
where $\Sigma^f_{L}$, $\Sigma^f_{R}$ are the left- and right-handed
components of the fermion self-energy, respectively.

While we do not consider processes with external gauge bosons, so that
the field renormalisation constants of the gauge bosons drop out in our 
physical results, these renormalisation constants do appear in
expressions for individual vertices given below. For completeness, we
therefore also list the expressions for the field renormalisation
constants of the gauge bosons. 
We define the field renormalisation constants for the
charged $W$ bosons via 
\begin{equation}
W^{\pm}\rightarrow (1+\frac{1}{2}\delta Z_{WW})W^{\pm},
\end{equation}
and for the neutral $Z$ boson and photon by
\begin{equation}
 \left(\begin{array}{c} Z \\ \gamma \end{array}\right)\rightarrow
\left(\begin{array}{cc} 1+\frac{1}{2}\delta Z_{ZZ} & \frac{1}{2}\delta Z_{Z\gamma} \\
\frac{1}{2}\delta Z_{\gamma Z} & 1+\frac{1}{2}\delta Z_{\gamma\gamma} \end{array}\right)
\left(\begin{array}{c} Z \\ \gamma \end{array}\right).
\end{equation}
The on-shell conditions for the field renormalisation constants ensure
that on-shell external particles have diagonal propagators with unity
residues (see e.g.\ Ref.~\cite{Denner:1991kt} and references therein). 
This leads to
\begin{equation}
 \delta Z_{VV}=-(\Sigma^{VV}_{T})^{\prime}(M_V^2),
\end{equation}
where $V=W,\,Z,\,\gamma$ and 
$\Sigma^{\prime}(k^2)\equiv\frac{\partial \Sigma(p^2)}{\partial p^2}|_{p^2=k^2}$, and 
\begin{equation}
 \delta Z_{\gamma Z}=-\frac{2\Sigma^{\gamma Z}_T(M_Z^2)}{M_Z^2},\quad
\delta Z_{Z \gamma}=\frac{2\Sigma^{Z \gamma}_T(0)}{M_Z^2}.
\end{equation}

\section{Renormalisation of the chargino and neutralino sector}\label{sec:3}

We now turn to the renormalisation of the chargino and neutralino sector
of the MSSM for the general case of arbitrary complex parameters. 
For the field
renormalisation, we follow the formalism developed in the earlier work
of Ref.~\cite{Fowler:2009ay}. We list here the relevant expressions for
completeness. For the parameter renormalisation, we extend the results
of Ref.~\cite{Fowler:2009ay}, which were restricted to the case of real
parameters in the chargino and neutralino sector, to the general case of
complex parameters.

The charginos and neutralinos are the mass eigenstates of the gauginos and higgsinos, as seen from the relevant part of the MSSM Lagrangian,
\begin{align*}
\mathcal{L}_{\tilde{\chi}} = &\overline{\tilde{\chi}^-_i}( \displaystyle{\not}p \,\delta_{ij}-\omega_L (U^{*}X V^{\dagger})_{ij}-\omega_R (V X^\dagger U^{T})_{ij})\tilde{\chi}^-_j\\
&+\frac{1}{2} \overline{\tilde{\chi}^0_i}( \displaystyle{\not}p\, 
\delta_{ij}-\omega_L (N^{*}Y N^{\dagger})_{ij}-\omega_R (N Y^\dagger N^{T})_{ij})\tilde{\chi}^0_j,
\end{align*}
where $\omega_{L/R}=1/2(1\mp\gamma_5)$. The mass matrix 

for the charginos is given by 
\begin{equation}\label{eq:X}
X=
\left( \begin{array}{cc}
M_2 & \sqrt{2} M_W s_\beta  \\
\sqrt{2} M_W c_\beta  & \mu
\end{array} \right),
\end{equation}
where $s_\beta/c_\beta=\sin\beta/\cos\beta$. This matrix is diagonalised via the bi-unitary transform $\displaystyle\mathbf{M}_{\tilde{\chi^+}}=U^* X V^\dag$.
The mass matrix for the neutralinos 

in the $(\tilde{B},\tilde{W},\tilde{H}_1,\tilde{H}_2)$ basis is given by
\begin{equation}\label{eq:Y}
Y =\left( \begin{array}{cccc}
M_1 & 0 & -M_Z c_\beta s_W & M_Z s_\beta s_W \\
0   & M_2 & M_Z c_\beta c_W & -M_Z s_\beta c_W \\
-M_Z c_\beta s_W & M_Z c_\beta c_W & 0 & -\mu \\
M_Z s_\beta s_W & -M_Z s_\beta c_W & -\mu & 0 \end{array} \right),
\end{equation}
where $M_1$ is the bino mass. Since $Y$ is complex symmetric, its diagonalisation requires only one unitary matrix $N$, via $\mathbf{M}_{\tilde{\chi^0}}=N^*Y N^\dag$.
The additional parameters that enter this sector are $M_1$, $M_2$ and $\mu$.

In the MSSM with complex parameters, absorptive parts arising from loop
integrals of unstable particles in general contribute to squared matrix
elements already at the one-loop level, since they can be multiplied by
imaginary coefficients involving complex parameters.
It has been shown in Ref.~\cite{Fowler:2009ay} that a proper treatment
of the absorptive parts from loop integrals of unstable particles
implies that full on-shell conditions giving rise to vanishing mixing
contributions on-shell can only be satisfied by the field
renormalisation constants in the chargino and neutralino sector 
if they are allowed to differ for incoming
and outgoing fields (see Ref.~\cite{Espriu:2002xv} for an earlier
discussion of this issue in the context of the SM). Accordingly, we 
define the renormalisation of the chargino and neutralino fields in 
the most general way, i.e.\ we introduce separate renormalisation constants 
$\delta Z^{L/R}_{\pm,ij}$ and $\bar{Z}^{L/R}_{\pm,ij}$ for incoming and 
outgoing (left- and right-handed) charginos, respectively. The
renormalisation constants for incoming and outgoing (left- and
right-handed) neutralinos are denoted as $\delta Z^{L/R}_{0,ij}$ and
$\bar{Z}^{L/R}_{0,ij}$, respectively.
Therefore, the renormalisation transformations for the chargino and
neutralino fields read
\begin{align}
  \omega_L \tilde{\chi}^-_i &\rightarrow (1+\tfrac{1}{2} \delta Z_\pm^L)_{ij} \omega_L  \tilde{\chi}^-_j,&
 \overline{\tilde{\chi}^{-}_i} \omega_R &\rightarrow \overline{\tilde{\chi}^-_i} (1+\tfrac{1}{2} \delta \bar{Z}_\pm^L)_{ij} \omega_R,  \nonumber\\
\omega_R \tilde{\chi}^-_i &\rightarrow (1+\tfrac{1}{2} \delta Z_\pm^{R})_{ij} \omega_R  \tilde{\chi}^-_j    
,& 
\overline{\tilde{\chi}^-_i} \omega_L &\rightarrow \overline{\tilde{\chi}^-_i} (1+\tfrac{1}{2} \delta \bar{Z}_\pm^R)_{ij}\omega_L,\nonumber\\
\omega_L \tilde{\chi}^0_i &\rightarrow (1+\tfrac{1}{2} \delta Z_0^L)_{ij} \omega_L  \tilde{\chi}^0_j,&
\overline{\tilde{\chi}^0_i} \omega_R &\rightarrow \overline{\tilde{\chi}^0_i} (1+\tfrac{1}{2} \delta \bar{Z}_0^L)_{ij} \omega_R,\nonumber\\
 \omega_R \tilde{\chi}^0_i &\rightarrow (1+\tfrac{1}{2} \delta Z_0^{R})_{ij} \omega_R  \tilde{\chi}^0_j  ,&
 \overline{\tilde{\chi}^0_i} \omega_L&\rightarrow
\overline{\tilde{\chi}^0_i} (1+\tfrac{1}{2} \delta \bar{Z}_0^R)_{ij}
\omega_L ,
\end{align}
where the indices $i,j$ can take values up to 2 for charginos and 4 for 
neutralinos, respectively.

Concerning the parameter renormalisation, we treat $M_1$, $M_2$ and
$\mu$ as independent free parameters that are determined by imposing 
on-shell renormalisation
conditions in the chargino and neutralino sector. On the other hand 
the parameter $\tan\beta$ has been renormalised in the Higgs sector, 
and the parameters $e$, $M_W$ and $M_Z$ (as well as the dependent parameter
$\sin\theta_W$) have been renormalised in the gauge sector, as described
above, see Sec.~\ref{sec:2}. The renormalisation transformations for 
$M_1$, $M_2$ and $\mu$ read
\begin{eqnarray}
\label{eqn:paramb}
 |M_1|&\rightarrow|M_1|+\delta|M_1|,\;\;&\;\;\phi_{M_1}\rightarrow \phi_{M_1}+\delta\phi_{M_1}, \nonumber\\
 |M_2|&\rightarrow|M_2|+\delta|M_2|,\;\;&\nonumber\\
 |\mu|&\rightarrow|\mu|+\delta|\mu|,\;\;&\;\;\phi_{\mu}\rightarrow \phi_{\mu}+\delta\phi_{\mu}\label{deltamu},
\end{eqnarray} 
where we treat the general case of complex parameters
in the chargino and neutralino sector. 
As mentioned above, we have adopted the convention where the phase
$\phi_{M_2}$ is rotated away. The parameter renormalisation in the
chargino and neutralino sectors therefore amounts to the renormalisation
of the five real parameters $|M_1|$, $|M_2|$, $|\mu|$, $\phi_{M_1}$ and
$\phi_\mu$. These parameter renormalisations induce a renormalisation of
the mass matrices $X$ and $Y$ via
\begin{eqnarray}
 X &\rightarrow X+ \delta X,\;\;\;\;\; Y &\rightarrow Y+ \delta Y.
\end{eqnarray}

Our renormalisation scheme builds on and extends the work of 
Refs.~\cite{Fritzsche:2002bi,Fritzsche:2005,Fowler:2009ay,AlisonsThesis}.
Our scheme, besides addressing the general case of
complex parameters, differs from the methods followed in
Refs.~\cite{Guasch:2002ez,Eberl:2004ic}, where the renormalisation was
carried out for the case of real parameters. The approach of
Ref.~\cite{Guasch:2002ez} differs from ours since in
Ref.~\cite{Guasch:2002ez} the mixing matrices are left unrenormalised. 
In Ref.~\cite{Eberl:2004ic} the mixing matrices are renormalised using
the proposal of Ref.~\cite{Kniehl:1996bd}, where the renormalisation of 
$\tan\beta$, $M_W$, $M_Z$ (and also $\sin\theta_W$) differs from our
prescription. While the divergent parts of the prescription in
Ref.~\cite{Eberl:2004ic} agree with the ones of the corresponding
quantities in our approach, the finite parts differ, see the discussion
in Ref.~\cite{Fritzsche:2002bi}. In an explicit comparison carried out for
the case of real parameters in Ref.~\cite{Fritzsche:2002bi}
it was found that the resulting differences
in the predictions for the physical chargino and neutralino masses based
on the different methods were numerically small.

As mentioned above, we determine the field renormalisation constants
from full on-shell conditions that ensure vanishing mixing
contributions on-shell for all chargino and neutralino fields. The
propagators are required to have unity residues. The corresponding
renormalisation conditions read
 \begin{align}
\hat{\Gamma}^{(2)}_{ij} \tilde{\chi}_{i}(p)|_{p^2=m_{\tilde{\chi}_j}^2} &=0,\quad&\quad
\overline{\tilde{\chi}}_{i}(p)\hat{\Gamma}^{(2)}_{ij}|_{p^2=m_{\tilde{\chi}_i}^2} &=0,\label{eqn:nomixing}\\
\lim_{p^2 \rightarrow m_{\tilde{\chi}_i}^2} \frac{1}{\displaystyle{\not}p-m_{\tilde{\chi}_i}} 
\hat{\Gamma}^{(2)}_{ii} \tilde{\chi}_{i}(p) &= i\tilde{\chi}_{i},\quad&\quad
\displaystyle{\lim_{p^2 \rightarrow m_{\tilde{\chi}_i}^2}}
 \overline{\tilde{\chi}}_{i}(p)\hat{\Gamma}^{(2)}_{ii}
\frac{1}{\displaystyle{\not}p-m_{\tilde{\chi}_i}} &=
i\overline{\tilde{\chi}}_{i},
\label{eqn:residues}
\end{align}
where $\tilde{\chi}_{i}=\tilde{\chi}^-_{i}\;(i,j=1,2)$ or
$\tilde{\chi}^0_{i}\;(i,j=1,2,3,4)$, and $i\neq j$.
Here the renormalised propagator $\hat{S}^{(2)}_{ij}(p^2)$ can be
obtained from the 1PI vertex function $\hat{\Gamma}^{(2)}_{ij}(p^2)$, 
which can be expressed in terms of the renormalised self-energy 
$\hat{\Sigma}_{ij}(p^2)$,
\begin{equation}
 \hat{S}^{(2)}_{ij}(p^2)= - (\hat{\Gamma}^{(2)}_{ij}(p^2))^{-1},\quad \hat{\Gamma}^{(2)}_{ij}(p^2)=i(\displaystyle{\not}p-m_i)\delta_{ij}+i\hat{\Sigma}_{ij}(p^2).
\label{eqn:fermion2pt}
\end{equation}
For convenience, we decompose the self-energy in terms of the coefficients $\Sigma^{L/R}_{ij}(p^2)$ and $\Sigma^{SL/SR}_{ij}(p^2)$ via
\begin{equation}
 \Sigma_{ij}(p^2)=\displaystyle{\not}p\, \omega_L  \Sigma^L_{ij}(p^2)+\displaystyle{\not}p\, \omega_R  \Sigma^R_{ij}(p^2)
+\omega_L  \Sigma^{SL}_{ij}(p^2)+ \omega_R \Sigma^{SR}_{ij}(p^2),
\label{eqn:Lorentzse}
\end{equation}
and define the left and right handed vector and scalar coefficients, $\displaystyle\hat{\Sigma}^{L/R}_{ij}(p^2)$ and $\displaystyle\hat{\Sigma}^{SL/SR}_{ij}(p^2)$ respectively, for the renormalised self-energy analogously.
Note that the conditions in Eqs.~(\ref{eqn:nomixing}) and
(\ref{eqn:residues}) do not specify the wavefunction renormalisation
constants completely, and so in addition we impose the conditions that
the renormalised propagators retain the same Lorentz structure as the
tree level propagators in the on-shell limit, i.e. 
\begin{eqnarray}
 \hat{\Sigma}^{L}_{ii}(m_{\tilde{\chi}_i}^2)&=&
\hat{\Sigma}^{R}_{ii}(m_{\tilde{\chi}_i}^2) , \label{eqn:nochiralLR}\\
 \hat{\Sigma}^{SL}_{ii}(m_{\tilde{\chi}_i}^2)&=&
\hat{\Sigma}^{SR}_{ii}(m_{\tilde{\chi}_i}^2) \label{eqn:nochiralSLSR} .
\end{eqnarray}

Together Eqs.~(\ref{eqn:nomixing}), (\ref{eqn:residues}),
(\ref{eqn:nochiralLR}) and (\ref{eqn:nochiralSLSR}) result in the
following expressions for the diagonal chargino wave function
renormalisation constants,
\begin{align}
\delta Z^{L/R}_{\pm,ii}=\,&-\Sigma_{\pm,ii}^{L/R}(m_{\tilde{\chi}^{\pm}_i}^2)\!-\!m_{\tilde{\chi}^{\pm}_i}^2
 \big[\Sigma_{\pm,ii}^{L^{\prime}}(m_{\tilde{\chi}^{\pm}_i}^2)\!+\!\Sigma_{\pm,ii}^{R^{\prime}}(m_{\tilde{\chi}^{\pm}_i}^2)\big]\! \nonumber\\
 &-\!
m_{\tilde{\chi}^{\pm}_i}
\big[\Sigma_{\pm,ii}^{SL^{\prime}}(m_{\tilde{\chi}^{\pm}_i}^2)\!+\!\Sigma_{\pm,ii}^{SR^{\prime}}(m_{\tilde{\chi}^{\pm}_i}^2)\big] \pm\frac{1}{2
m_{\tilde{\chi}^{\pm}_i}}\big[\Sigma_{\pm,ii}^{SL}(m_{\tilde{\chi}^{\pm}_i}^2)\!\nonumber\\
 &-\!\Sigma_{\pm,ii}^{SR}(m_{\tilde{\chi}^{\pm}_i}^2)+(V
\delta X^{\dagger} U^{T})_{ii}\!-\!(U^{\ast}\delta X
V^{\dagger})_{ii}\big] , \\
\delta\bar{Z}^{L/R}_{\pm,ii}=\,&-\Sigma_{\pm,ii}^{L/R}(m_{\tilde{\chi}^{\pm}_i}^2)\!-\!m_{\tilde{\chi}^{\pm}_i}^2
 \big[\Sigma_{\pm,ii}^{L^{\prime}}(m_{\tilde{\chi}^{\pm}_i}^2)\!+\!\Sigma_{\pm,ii}^{R^{\prime}}(m_{\tilde{\chi}^{\pm}_i}^2)\big]\!- \nonumber\\
 &\! m_{\tilde{\chi}^{\pm}_i} \big[\Sigma_{\pm,ii}^{SL^{\prime}}(m_{\tilde{\chi}^{\pm}_i}^2)\!+\!\Sigma_{\pm,ii}^{SR^{\prime}}(m_{\tilde{\chi}^{\pm}_i}^2)\big]\mp\frac{1}{2
m_{\tilde{\chi}^{\pm}_i}}\big[\Sigma_{\pm,ii}^{SL}(m_{\tilde{\chi}^{\pm}_i}^2)\!\nonumber\\
 &-\!\Sigma_{\pm,ii}^{SR}(m_{\tilde{\chi}^{\pm}_i}^2)+(V
\delta X^{\dagger} U^{T})_{ii}\!-\!(U^{\ast}\delta X
V^{\dagger})_{ii}\big],\label{eqn:chargdiagfieldren}
\end{align}
and the following off-diagonal chargino wave function renormalisation constants,
\begin{eqnarray}
\delta Z^{L/R}_{\pm,ij}&=&\frac{2}{m_{\tilde{\chi}^{\pm}_i}^2-m_{\tilde{\chi}^{\pm}_j}^2}\big[m_{\tilde{\chi}^{\pm}_j}^2 \Sigma_{\pm,ij}^{L/R}
(m_{\tilde{\chi}^{\pm}_j}^2)+m_{\tilde{\chi}^{\pm}_i}m_{\tilde{\chi}^{\pm}_j} \Sigma_{\pm,ij}^{R/L}
(m_{\tilde{\chi}^{\pm}_j}^2)\nonumber\\&&+m_{\tilde{\chi}^{\pm}_i}\Sigma_{\pm,ij}^{SL/SR}(m_{\tilde{\chi}^{\pm}_j}^2)
+m_{\tilde{\chi}^{\pm}_j}\Sigma_{\pm,ij}^{SR/SL}(m_{\tilde{\chi}^{\pm}_j}^2)\nonumber\\&&
- m_{\tilde{\chi}^{\pm}_{i/j}}\big(U^\ast \delta X V^\dagger\big)_{ij}-
  m_{\tilde{\chi}^{\pm}_{j/i}}\big(V \delta X^\dagger U^T\big)_{ij}\big]\\
\delta \bar{Z}^{L/R}_{\pm,ij}&=&\frac{2}{m_{\tilde{\chi}^{\pm}_j}^2-m_{\tilde{\chi}^{\pm}_i}^2}\big[m_{\tilde{\chi}^{\pm}_i}^2 
\Sigma_{\pm,ij}^{L/R}
(m_{\tilde{\chi}^{\pm}_i}^2)+m_{\tilde{\chi}^{\pm}_i}m_{\tilde{\chi}^{\pm}_j} \Sigma_{\pm,ij}^{R/L}
(m_{\tilde{\chi}^{\pm}_i}^2)\nonumber\\&&+m_{\tilde{\chi}^{\pm}_i}\Sigma_{\pm,ij}^{SL/SR}(m_{\tilde{\chi}^{\pm}_i}^2)
+m_{\tilde{\chi}^{\pm}_j}\Sigma_{\pm,ij}^{SR/SL}(m_{\tilde{\chi}^{\pm}_i}^2)\nonumber\\&&
- m_{\tilde{\chi}^{\pm}_{i/j}}\big(U^\ast \delta X V^\dagger\big)_{ij}-
  m_{\tilde{\chi}^{\pm}_{j/i}}\big(V \delta X^\dagger U^T\big)_{ij}\big].\label{eqn:chargoffdiagfieldren}
\end{eqnarray}

As a consequence of their Majorana nature, the renormalisation constants for the neutralinos satisfy the relations
\begin{equation}
 \delta Z^{L/R}_{0,ij}=\delta\bar{Z}^{R/L}_{0,ji}.
\label{eqn:neutrenconst}
\end{equation}
The diagonal and off-diagonal wave function renormalisation constants
are given by
\begin{eqnarray}
\delta Z^{L/R}_{0,ii}&=&\!\!\!\!-\Sigma_{0,ii}^{L/R}(m_{\tilde{\chi}^0_i}^2)\!-\!m_{\tilde{\chi}^0_i}^2
 \big[\Sigma_{0,ii}^{L^{\prime}}(m_{\tilde{\chi}^0_i}^2)\!+\!\Sigma_{0,ii}^{R^{\prime}}(m_{\tilde{\chi}^0_i}^2)\big]\! \nonumber\\
 &&-\! m_{\tilde{\chi}^0_i} \big[\Sigma_{0,ii}^{SL^{\prime}}(m_{\tilde{\chi}^0_i}^2)\!+\!\Sigma_{0,ii}^{SR^{\prime}}(m_{\tilde{\chi}^0_i}^2)\big]\pm\frac{1}{2
m_{\tilde{\chi}^0_i}}\big[\Sigma_{0,ii}^{SL}(m_{\tilde{\chi}^0_i}^2)\!-\nonumber\\
 &&\!\Sigma_{0,ii}^{SR}(m_{\tilde{\chi}^0_i}^2)+(N
\delta Y^{\dagger} N^{T})_{ii}\!-\!(N^{\ast}\delta Y
N^{\dagger})_{ii}\big] ,
\label{eqn:neutdiagfieldrena} \\[.3em]
\delta Z^{L/R}_{0,ij}&=&\!\!\!\!\frac{2}{m_{\tilde{\chi}^0_i}^2-m_{\tilde{\chi}^0_j}^2} \big[m_{\tilde{\chi}^0_j}^2 \Sigma_{0,ij}^{L/R}(m_{\tilde{\chi}^0_j}^2)+m_{\tilde{\chi}^0_i}m_{\tilde{\chi}^0_j} \Sigma_{0,ij}^{R/L}(m_{\tilde{\chi}^0_j}^2)\nonumber\\
 &&+m_{\tilde{\chi}^0_i}\Sigma_{0,ij}^{SL/SR}(m_{\tilde{\chi}^0_j}^2)+m_{\tilde{\chi}^0_j}\Sigma_{0,ij}^{SR/SL}(m_{\tilde{\chi}^0_j}^2)\nonumber\\
 &&- m_{\tilde{\chi}^0_{i/j}}\big(N^\ast \delta Y N^\dagger\big)_{ij}-
  m_{\tilde{\chi}^0_{j/i}}\big(N \delta Y^\dagger N^T\big)_{ij}\big].
\label{eqn:neutdiagfieldrenb}
\end{eqnarray}

It should be noted that the barred constants,
$\delta\bar{Z}^{L/R}_{ij}$, are not related to 
$(\delta Z^{L/R}_{ij})^{\dagger}$ via hermiticity relations,
\begin{equation}
\delta\bar{Z}^{L/R}_{ij} \neq (\delta Z^{L/R}_{ij})^{\dagger} .
\label{eqn:hermcond}
\end{equation}
For both the charginos and neutralinos, $\delta\bar{Z}^{L/R}_{ij}$
differ from
 $(\delta Z^{L/R}_{ij})^{\dagger}$ in their absorptive parts only
(arising from loop integrals of unstable particles), while this
difference vanishes in the CP-conserving MSSM, see the discussion in 
Ref.~\cite{Fowler:2009ay}.

We briefly comment in this context on the similar issue arising
in the fermion sector of the SM in connection with the renormalisation of
the CKM matrix, see
Refs.~\cite{Denner:1990yz,Denner:1991kt,Gambino:1998ec,Yamada:2001px,
Espriu:2002xv,Zhou:2005mc,Kniehl:2009kk}.
It was realised that the hermiticity constraint for the incoming and
outgoing fermion field renormalisation is incompatible with the
demand of fulfilling the standard on-shell conditions. This is a consequence
of absorptive parts of loop integrals, which are gauge-parameter
dependent. Attempts to phrase the renormalisation prescription such that those
absorptive parts do not enter the field renormalisation constants, restoring
in this way the hermiticity relation, turned out to be problematic.
The field renormalisation constants were found to be
related via a Ward identity to the renormalisation constant for the 
CKM matrix, and the above prescription would lead to a 
gauge-dependent result~\cite{Gambino:1998ec,Yamada:2001px}.
Alternative methods to renormalise the CKM matrix have been 
proposed~\cite{Gambino:1998ec,Kniehl:2009kk}, however in order to ensure 
that on-shell external propagators are flavour diagonal it was advocated
to relax the hermiticity condition and to allow independent renormalisation 
constants for incoming and outgoing fields~\cite{Espriu:2002xv}.
This approach was found to be consistent with the gauge invariance of 
the SM~\cite{Zhou:2005mc} and the CPT theorem~\cite{Espriu:2002xv}.
Due to the interference with the CKM phase, inclusion of the imaginary parts 
was found to give rise to numerically relatively small shifts in the
predictions for the relevant observables of $\sim0.5\%$~\cite{Espriu:2002xv}. 
In Sec.~\ref{sec:4.1} we investigate the size of the effect of a proper
treatment of the absorptive parts for the chargino and neutralino case and 
we analyse the numerical impact on
predictions for physical observables.


We now turn to the renormalisation of the parameters in the chargino and
neutralino sector, which as discussed above comprises the
renormalisation of the five parameters $|M_1|$, $|M_2|$, $|\mu|$,
$\phi_{M_1}$ and $\phi_\mu$. On-shell conditions for the parameters
$|M_1|$, $|M_2|$ and $|\mu|$ can be obtained from the requirement that 
three of the chargino and neutralino masses are renormalised on-shell, 
in analogy to the case where $M_1$, $M_2$ and $\mu$ are real, see
Ref.~\cite{Fowler:2009ay}. For the two phases $\phi_{M_1}$ and
$\phi_\mu$ there exists no obvious on-shell condition (the same is true
for several other MSSM parameters, for instance the parameter
$\tan\beta$). A possible
choice would be to employ the $\overline{\mathrm{DR}}$ scheme, as 
advocated in the ``SPA conventions''~\cite{AguilarSaavedra:2005pw}.
However, we have verified explicitly that no renormalisation of
$\phi_{M_1}$ and $\phi_\mu$ is required at all in order to render the
relevant Green's functions finite. This result can be understood as
follows: starting from the Lagrangian expressed in the 
gaugino--higgsino basis, the diagonalisation of the mass matrices upon
making the transition to the mass eigenstate basis leads to expressions
in terms of the real combinations 
$U^{*}X V^{\dagger}$ and $N^{*}Y N^{\dagger}$. In those expressions the
phases of $M_1$ and $\mu$ that are present in the mass matrices $X$ and $Y$
have been compensated by the corresponding elements of the
transformation matrices $U$, $V$ and $N$. Thus, the phases of $M_1$ and
$\mu$ appearing in the couplings of neutralinos and charginos to other
particles can be related to elements of the transformation matrices $U$,
$V$ and $N$. The elements of those transformation matrices, however, do
not need to be renormalised. This is in analogy,
for instance, to the transformations of fields in the Higgs sector of
the MSSM, where it is well known that the mixing angles $\alpha$,
$\beta_{\rm n}$ and $\beta_{\rm c}$ (using the notation of
Ref.~\cite{Frank:2006yh}) do not require renormalisation. 
We therefore adopt a renormalisation scheme where the phases
$\phi_{M_1}$ and $\phi_\mu$ of the parameters in the chargino and
neutralino sector are left unrenormalised. This is convenient both
from a technical and a conceptual point of view.

We define the physical masses of the charginos and neutralinos
${\tilde{\chi}_i}$ according to the real part of the complex pole
\begin{equation}
{\cal M}^2_{\tilde{\chi}_i} = M^2_{\tilde{\chi}_i} - i
M_{\tilde{\chi}_i} \Gamma_{\tilde{\chi}_i} ,
\end{equation}
where $\Gamma_{\tilde{\chi}_i}$ is the width of particle
$\tilde{\chi}_i$. The physical mass at the one-loop level, 
$M_{\tilde{\chi}_i}$, in general differs from the tree-level mass,
$m_{\tilde{\chi}_i}$, by a finite amount, $\Delta m_{\tilde{\chi}_i}$,
which is obtained from the relation
\begin{align}
\nonumber M_{\tilde{\chi}_i}&=m_{\tilde{\chi}_i} (1-\frac{1}{2}\re[\hat{\Sigma}^L_{ii}(m_{\tilde{\chi}_i}^2)+\hat{\Sigma}^{R}_{ii}(m_{\tilde{\chi}_i}^2)])-\frac{1}{2}\re[\hat{\Sigma}^{SL}_{ii}(m_{\tilde{\chi}_i}^2)+\hat{\Sigma}^{SR}_{ii}(m_{\tilde{\chi}_i}^2)]&\qquad\qquad\\
&\equiv m_{\tilde{\chi}_i}+\Delta m_{\tilde{\chi}_i}.&
\end{align}
The renormalisation conditions for the three independent parameters 
$|M_1|$, $|M_2|$ and $|\mu|$ can be chosen such that three of the
chargino and neutralino masses are renormalised on-shell, i.e.\ for
those three particles the physical mass at the one-loop level is equal
to the mass value at tree level, $m_{\tilde{\chi}_i}$. Accordingly,
these conditions can be written as
\begin{align}
\nonumber \Delta m_{\tilde{\chi}_i}&\equiv-\frac{m_{\tilde{\chi_i}}}{2}\re[\hat{\Sigma}^L_{ii}(m_{\tilde{\chi}_i}^2)+\hat{\Sigma}^{R}_{ii}(m_{\tilde{\chi}_i}^2)]-\frac{1}{2}\re[\hat{\Sigma}^{SL}_{ii}(m_{\tilde{\chi}_i}^2)+\hat{\Sigma}^{SR}_{ii}(m_{\tilde{\chi}_i}^2)]&\\
\label{eqn:deltami}&=0.&
\end{align}
The resulting expressions 
for $\delta|M_1|$, $\delta|M_2|$, $\delta|\mu|$ depend on the choice
that has been made for the three
masses that are renormalised on-shell. There are obviously three
generic possibilities, namely selecting three neutralinos (NNN), 
two neutralinos and one chargino (NNC), or one neutralino and two
charginos (NCC).
Using the shorthands
\begin{align}
C_{(i)}\equiv\,&\re \big[m_{\tilde{\chi}^+_i}[\Sigma^L_{-,ii}(m_{\tilde{\chi}^+_i}^2)+ \Sigma^R_{-,ii}(m_{\tilde{\chi}^+_i}^2)]+
      \Sigma^{SL}_{-,ii}(m_{\tilde{\chi}^+_i}^2)+\Sigma^{SR}_{-,ii}(m_{\tilde{\chi}^+_i}^2)\big]\nonumber\\&- 
     \sum_{\tiny\substack{j=1,2\\k=1,2}}2\delta X_{jk}
\re(\UCha_{ij}\VCha_{ik}), \\
%
N_{(i)}\equiv\,&\re \big[m_{\tilde{\chi}^0_i}[\Sigma^L_{0,ii}(m_{\tilde{\chi}^0_i}^2)+\Sigma^R_{0,ii}(m_{\tilde{\chi}^0_i}^2)] +\Sigma^{SL}_{0,ii}(m_{\tilde{\chi}^0_i}^2)+ \Sigma^{SR}_{0,ii}(m_{\tilde{\chi}^0_i}^2)\big]\nonumber
\\& - \sum_{\tiny\substack{j=1,2\\k=3,4}}
       4\delta Y_{jk}\re(\ZNeu_{ij}\ZNeu_{ik}),
\label{eqn:CiNishorthand}
\end{align}
the condition that the $i^{\prime}$th neutralino mass is on-shell reads
\begin{align}
\label{eqn:nmassonshellph} N_{i^{\prime}}=\,&2\delta|M_2|(\re(e^{-i\phi_{M_1}}\,N_{i^{\prime}1}^2)
+\re N_{i^{\prime}2}^2)-4\delta|\mu|\re(e^{-i\phi_{\mu}}N_{i^{\prime}3}
N_{i^{\prime}4}) , & 
\end{align}
while the condition that the $i^{\prime\prime}$th chargino mass is
on-shell reads
\begin{align}
\label{eqn:cmassonshellph} C_{i^{\prime\prime}}=\,&2\delta|M_2|\re(U_{i^{\prime\prime}1} V_{i^{\prime\prime}1})+2\delta|\mu|\re(e^{-i\phi_{\mu}}U_{i^{\prime\prime}2} V_{i^{\prime\prime}2}).
&\end{align}

In the NNN case, where neutralinos $\tilde{\chi}^0_{i}$, $\tilde{\chi}^0_{j}$ and $\tilde{\chi}^0_{k}$ are chosen on-shell, we obtain $\delta|M_1|$, $\delta|M_2|$, $\delta|\mu|$ by solving Eq.~(\ref{eqn:nmassonshellph}) with $i^{\prime}=i,j,k$ simultaneously,
\begin{align}
 \nonumber \delta |M_1|=&\frac{1}{Q}\Big((\re(e^{-i\phi_{\mu}}N_{i3} N_{i4}) \re N_{j2}^2- \re(e^{-i\phi_{\mu}}N_{j3} N_{j4})\,\re N_{i2}^2)N_{k}\\
\nonumber& +(\re(e^{-i\phi_{\mu}}N_{j3}N_{j4})\,\re N_{k2}^2-\re(e^{-i\phi_{\mu}}N_{k3} N_{k4})\,\re N_{j2}^2)N_{i}\qquad\qquad\\
&+(\re(e^{-i\phi_{\mu}}N_{k3} N_{k4})\,\re N_{i2}^2-\re(e^{-i\phi_{\mu}}N_{i3} N_{i4})\,\re N_{k2}^2)N_{j}\Big),\\
\nonumber \delta |M_2|=\,&\frac{1}{Q}\Big((\re(e^{-i\phi_{\mu}}N_{j3} N_{j4})\,\re(e^{-i\phi_{M_1}}N_{i1}^2)\\
\nonumber&- \re(e^{-i\phi_{\mu}}N_{i3} N_{i4})\,\re(e^{-i\phi_{M_1}}N_{j1}^2))N_{k}\\
\nonumber& +(\re(e^{-i\phi_{\mu}}N_{k3} N_{k4})\,\re(e^{-i\phi_{M_1}}N_{j1}^2)\\
\nonumber&-\re(e^{-i\phi_{\mu}}N_{j3} N_{j4})\,\re(e^{-i\phi_{M_1}}N_{k1}^2))N_{i}\\&
\nonumber+(\re(e^{-i\phi_{\mu}}N_{i3} N_{i4})\,\re(e^{-i\phi_{M_1}}N_{k1}^2)\\
&-\re(e^{-i\phi_{\mu}}N_{k3} N_{k4})\,\re(e^{-i\phi_{M_1}}N_{i1}^2))N_{j}\Big),\\
 \delta |\mu|=\,&-\frac{1}{2\,Q}\Big((\re N_{i2}^2\,\re(e^{-i\phi_{M_1}}\,N_{j1}^2)- \re(e^{-i\phi_{M_1}}\,N_{i1}^2)\,\re N_{j2}^2)N_{k}\nonumber\\
&+(\re N_{j2}^2\,\re(e^{-i\phi_{M_1}}\,N_{k1}^2)-\re(e^{-i\phi_{M_1}}\,N_{k1}^2)\,\re N_{k2}^2)N_{i}\nonumber\\
&+(N_{i1}^2\,\re N_{k2}^2- \re N_{i2}^2\,\re(e^{-i\phi_{M_1}}\,N_{k1}^2))N_{j}\Big)
\label{eq:NNN}\\
\nonumber &\mbox{where}\\
\nonumber  Q =\,& 2 \left(\re(e^{-i\phi_{\mu}}\,N_{i3}\,N_{i4})\,\re N_{j2}^2\,\re(e^{-i\phi_{M_1}}\,N_{k1}^2)\right.\\
\nonumber&-\re N_{i2}^2\,\re(e^{-i\phi_{\mu}}\,N_{j3} N_{j4})\,\re(e^{-i\phi_{M_1}}\,N_{k1}^2)\\
\nonumber& - \re(e^{-i\phi_{\mu}}\,N_{i3} N_{i4})\,\re(e^{-i\phi_{M_1}}\,N_{j1}^2)\,\re N_{k2}^2 \\
\nonumber&+\re(e^{-i\phi_{M_1}}\,N_{i1}^2)\,\re(e^{-i\phi_{\mu}}\!N_{j3} N_{j4})\,\re N_{k2}^2\\
\nonumber &+ \re N_{i2}^2\,\re(e^{-i\phi_{M_1}}\!N_{j1}^2)\,\re(e^{-i\phi_{\mu}}\!N_{k3} N_{k4})\\
&\left.- \re(e^{-i\phi_{M_1}}\!N_{i1}^2)\,\re N_{j2}^2\,\re(e^{-i\phi_{\mu}}\!N_{k3} N_{k4})\right).
\end{align}
For the NNC case, when neutralinos $\tilde{\chi}^0_{i}$, $\tilde{\chi}^0_{j}$ and chargino $\tilde{\chi}^{\pm}_{k}$ are on-shell, we solve Eq.~(\ref{eqn:nmassonshellph}) with $i^{\prime}=i,j$ and Eq.~(\ref{eqn:cmassonshellph}) with $i^{\prime\prime}=k$ simultaneously, finding
\begin{align}
\nonumber \delta |M_1|=\,&\frac{1}{R}\Big(2 \big(\re(e^{-i\phi_{\mu}}N_{i3} N_{i4})\,\re N_{j2}^2\\
\nonumber& - \re(e^{-i\phi_{M_2}} N_{i2}^2)\,\re(e^{-i\phi_{\mu}}N_{j3} N_{j4})\big)C_{k}\qquad\\
\nonumber&+\big(2 \re(U_{k1} V_{k1})\,\re(e^{-i\phi_{\mu}}N_{i3} N_{i4})\qquad\\
\nonumber& +\re(e^{-i\phi_{\mu}}U_{k2} V_{k2}) \re N_{j2}^2\big)N_{i}-\big(\re(e^{-i\phi_{\mu}}U_{k2} V_{k2})\,\re N_{i2}^2\\
&+2 \re(U_{k1} V_{k1})\,\re(e^{-i\phi_{\mu}}N_{i3} N_{i4})\big)N_{j}\Big),
\end{align}

\begin{align}
\nonumber\delta |M_2|=\,&\frac{1}{R}\Big(-2[\re(e^{-i\phi_{\mu}}N_{i3} N_{i4})\,\re(e^{-i\phi_{M_1}}\,N_{j1}^2)\\
\nonumber& - \re(e^{-i\phi_{\mu}}N_{j3} N_{j4})\,\re(e^{-i\phi_{M_1}}\,N_{i1}^2)]C_{k}\\
\nonumber&-\re(e^{-i\phi_{\mu}}U_{k2} V_{k2})\,\re(e^{-i\phi_{M_1}}\,N_{j1}^2) N_{i}\\
&+\!\re(e^{-i\phi_{\mu}}\!U_{k2} V_{k2}) \re(e^{-i\phi_{M_1}}N_{i1}^2)N_{j}\Big),
\\
\nonumber \delta |\mu|=\,&\frac{1}{R}\Big(-\big(\re N_{i2}^2\re(e^{-i\phi_{M_1}}\,N_{j1}^2)- \re N_{j2}^2\re(e^{-i\phi_{M_1}}\,N_{i1}^2)\big)\,C_{k}\\ 
\nonumber&+\re(U_{k1} V_{k1})\,\re(e^{-i\phi_{M_1}}\,N_{j1}^2)N_{i}\\
\label{eq:NNC}&-\re(U_{k1} V_{k1})\,\re(e^{-i\phi_{M_1}}\,N_{i1}^2) N_{j}\Big)\\
\nonumber &\mbox{where}\\
 \nonumber R=\,& 2\re(e^{-i\phi_{\mu}}U_{k2} V_{k2})\big(-\re N_{i2}^2\re(e^{-i\phi_{M_1}}\,N_{j1}^2)+N_{i1}^2\re N_{j2}^2\big)\\
\nonumber&+ 4\re(U_{k1} V_{k1})\big(-\re(e^{-i\phi_{\mu}}N_{i3} N_{i4})\,\re(e^{-i\phi_{M_1}}N_{j1}^2) \!\\
&+\! \re(e^{-i\phi_{M_1}}N_{i1}^2)\,\re(e^{-i\phi_{\mu}}N_{j3} N_{j4})\big).
\end{align}
Finally, if masses of one neutralino $\tilde{\chi}^0_{i}$, and two
charginos, $\tilde{\chi}^{\pm}_{j}$, $\tilde{\chi}^{\pm}_{k}$, are
on-shell, corresponding to the NCC case, we solve
Eq.~(\ref{eqn:nmassonshellph}) with $i^{\prime}=i$ and
Eq.~(\ref{eqn:cmassonshellph}) with $i^{\prime\prime}=j,k$
simultaneously, resulting in 
\begin{align}
\nonumber\delta |M_1|=\,&-\frac{1}{\re(e^{-i\phi_{M_1}}\,N_{i1}^2) S}\Big((2\re(e^{-i\phi_{\mu}}N_{i3} N_{i4}) \re(U_{j1} V_{j1})\\
\nonumber&+ \re N_{i2}^2 \re(e^{-i\phi_{\mu}}U_{j2} V_{j2}))C_{k}+(\re(U_{j1} V_{j1})\,\re(e^{-i\phi_{\mu}}U_{k2} V_{k2})\\
\nonumber&-\re(e^{-i\phi_{\mu}}U_{j2} V_{j2})\,\re(U_{k1} V_{k1}))N_{i}\qquad\\
\nonumber&-(\re N_{i2}^2\re(e^{-i\phi_{\mu}}U_{k2} V_{k2})\\
\label{eq:NCCa} & +2\re(e^{-i\phi_{\mu}}N_{i3} N_{i4})\,\re(U_{k1} V_{k1}))C_{j}\Big),\\
\label{eq:NCCb} \delta |M_2|=\,&\frac{1}{S}\Big(\re(e^{-i\phi_{\mu}}U_{j2} V_{j2})C_{k}-\re(e^{-i\phi_{\mu}}U_{k2} V_{k2}) C_{j}],\\
 \delta |\mu|=\,&-\frac{1}{S}\Big(\re(U_{j1} V_{j1})C_{k}-\re(U_{k1} V_{k1})C_{j}\Big), \label{eq:NCCc}
\end{align}
where
\begin{align}
  S=2\big(\re(U_{k1} V_{k1}) \re(e^{-i\phi_{\mu}}U_{j2} V_{j2})-\re(U_{j1} V_{j1})\,\re(e^{-i\phi_{\mu}}U_{k2} V_{k2})\big).
\end{align}

In order to apply the above renormalisation prescription to a certain process,
it is necesary to investigate which of the possible choices of the
three masses that are renormalised on-shell is in fact appropriate and
results in a well-behaved renormalisation scheme. 
It is usually convenient to impose on-shell conditions for the 
external particles of the process under consideration. However, some
more care is necessary in order to ensure that the imposed conditions
are indeed suitable for determining the parameters 
$|M_1|$, $|M_2|$ and $|\mu|$. This issue was investigated in detail 
in Ref.~\cite{AlisonsThesis} for the case of the CPX benchmark 
scenario and a higgsino-like variant of the CPX scenario.

We define here the CPX scenario such that the 
parameters take the values $M_1=(5/3) (s_W^2/c_W^2) M_2$, 
$M_{\rm SUSY}= 500 \gev$, $A_{q,l}=900\gev$, $\phi_{M_1}=0$,
$\phi_{\mu}=0$, $\phi_{M_3}=\pi/2$, $\phi_{A_{f3}}=\pi/2$,  
$\phi_{A_{f1,2}}=\pi$.
%
For the gaugino-like case we use $M_2=200 \gev$ and $\mu=2000 \gev$,
whereas for the higgsino-like case we choose $\mu=200\gev$ and 
$M_2=1000\gev$.
In our numerical example below we furthermore use $\tan\beta=5.5$ and
$M_{H^{\pm}}=132.1\gev$.

\begin{table}[htb]
\begin{tabular}{c||c|c|c|c|c||c|c}
\hline
\T
\B
 & NNN & NNC & NCC & NCCb& NCCc&NCCb* & NCCc*\\
\hline
\hline
\T $\delta |M_1|$ & -1.468 & -1.465 &  -1.468 & 2517 & -3685&-365.4& -4.671\\
 $\delta |M_2|$ & -9.265 & -9.265 &  -9.410 & -9.410 & -9.410 &13.23 &13.23\\
 $\delta |\mu|$ & -18.48 & -18.98 &  -18.98 & -18.98 & -18.98 & -5.333 & -5.333\\
 $\Delta m_{\tilde{\chi}^0_1}$ & 0 & 0 & 0 & 2517. & -3681&-5.809&-0.522\\
 $\Delta m_{\tilde{\chi}^0_2}$ & 0& 0& -0.1446 & 0& 0.3560 &0& -0.4806\\
 $\Delta m_{\tilde{\chi}^0_3}$& 0 & -0.5018 &  -0.5016 & -0.8447 & 0& -354.9& 0\\
 $\Delta m_{\tilde{\chi}^0_4}$ & 0.3238 & -0.1775 & -0.1775 & 0.6851& -1.439& -0.1734&-0.1548\\
 $\Delta m_{\tilde{\chi}^{\pm}_1}$ & 0.1446 & 0.1445 &  0 & 0& 0&0&0\\
\B $\Delta m_{\tilde{\chi}^{\pm}_2}$ & 0.5012 & 0 &  0 & 0& 0 &0&0\\
\hline
\end{tabular} 
\caption{Finite parts of parameter renormalisation constants and mass
corrections in $\mathrm{GeV}$ for the CPX scenario (gaugino-like case). The last two columns, denoted with an asterisk, show the results for a higgsino-like scenario. The parameters for both scenarios are given in the text.}
\label{tab:RenormScheme}\
\end{table}

In Tab.~\ref{tab:RenormScheme} we show the finite parts of the
renormalisation constants $\delta|M_1|$, $\delta|M_2|$ and
$\delta|\mu|$ for the gaugino-like case of
the CPX scenario, using five different choices of 
parameter renormalisation: 
\begin{itemize}
\item
NNN with $\tilde{\chi}^0_{1}$, $\tilde{\chi}^0_{2}$ and 
$\tilde{\chi}^0_{3}$ on-shell

\item
NNC with $\tilde{\chi}^0_{1}$, $\tilde{\chi}^0_{2}$ and
$\tilde{\chi}^{\pm}_{2}$ on-shell


\item
NCC with $\tilde{\chi}^0_{1}$, $\tilde{\chi}^{\pm}_{1}$ and
$\tilde{\chi}^{\pm}_{2}$ on-shell

\item
NCCb with $\tilde{\chi}^0_{2}$, $\tilde{\chi}^{\pm}_{1}$ and
$\tilde{\chi}^{\pm}_{2}$ on-shell

\item
NCCc with $\tilde{\chi}^0_{3}$, $\tilde{\chi}^{\pm}_{1}$ and
$\tilde{\chi}^{\pm}_{2}$ on-shell
\end{itemize}
Also shown are the resulting one-loop corrections to those masses that
are not renormalised on-shell. For two scenarios, NCCb and NCCc, also
the results for the higgsino-like case of the CPX scenario are displayed
(denoted as NCCb$^*$ and NCCc$^*$ in Tab.~\ref{tab:RenormScheme}).

For the gaugino-like case of the CPX scenario one can see from
Tab.~\ref{tab:RenormScheme} that NNN, NNC and NCC are all suitable 
schemes, giving similar (and relatively small) values for 
the finite parts of the three renormalisation constants and modest
corrections (in this example at the sub-GeV level) to the masses.
On the other hand, the NCCb and NCCc prescriptions yield a huge
value for the finite part of $\delta|M_1|$ and correspondingly an
unphysically large correction to the mass $m_{\tilde{\chi}^0_{1}}$. 
This is due to the fact that this (gaugino-like) scenario has the
hierarchy $|M_1|<|M_2|\ll|\mu|$, which implies that 
the parameters $|M_1|$, $|M_2|$ and $|\mu|$ 
broadly determine the values of the masses of $\tilde\chi^0_1$,
$\tilde\chi^0_2$/$\tilde\chi^\pm_1$ and 
$\tilde\chi^0_{3/4}$/$\tilde\chi^\pm_2$, respectively.
Consequently, since the prescriptions NCCb and NCCc do not use the mass 
$m_{\tilde\chi^0_1}$ as input, the parameter $|M_1|$ is only weakly
constrained, yielding an unphysically large value for its counterterm 
and correspondingly an unphysically large correction to
$m_{\tilde\chi^0_1}$. Thus, in order to avoid unphysically 
large contributions to $\delta|M_1|$, the only bino-like particle in
this scenario, $\tilde\chi^0_1$, should be chosen on-shell. 
More generally, the renormalisation conditions must be chosen such that 
they provide sufficient sensitivity to each of the three underlying
parameters that are renormalised, $|M_1|$, $|M_2|$ and $|\mu|$.

On the other hand in the higgsino-like scenario, 
where $|\mu|\ll |M_1|<|M_2|$,
the parameters $|M_1|$, $|M_2|$ and
$|\mu|$ form the dominant component of the masses of $\tilde\chi^0_3$,
$\tilde\chi^0_4$/$\chi^\pm_2$ and $\tilde\chi^0_{1/2}$/$\chi^\pm_1$,
respectively.
Tab.~\ref{tab:RenormScheme} shows that in this case the results in
scheme NCCc are well-behaved, i.e.\ it yields moderate contributions to the
counterterms and the masses. This is a consequence of the fact that the 
bino-like $\tilde\chi^0_3$ has been chosen to be renormalised on-shell.
Scheme NCCb, on the other hand, where $\tilde\chi^0_2$ instead of
$\tilde\chi^0_3$ is renormalised on-shell, shows unphysical bavaviour
since the parameter $|M_1|$ is only weakly constrained.

The comparison of the gaugino-like and higgsino-like scenarios in 
Tab.~\ref{tab:RenormScheme} illustrates that an appropriate choice of
renormalisation prescription depends not only on the process in
question but also on the considered scenario of parameter values. It is
therefore in general not possible to make a choice of the three masses
that are renormalised on-shell in such a way 
that this prescription can safely be
applied to all possible parameter configurations.%
\footnote{One might wonder whether the problems related to selecting
three out of six masses for an on-shell renormalisation could be avoided
by using the $\overline{\mathrm{DR}}$ scheme where the predictions for
all the physical masses receive loop corrections. However, such a scheme
will in general lead to a situation where the mass value inserted for a
particle at an external line will be different from the mass value of
the same particle if it appears as an internal line of a Feynman
diagram. Such a mismatch is problematic in view of a consistent
treatment of infrared divergent contributions associated with external
particles that carry electric charge or colour.}
Instead, it is necessary to adjust the renormalisation prescription such
that at least one of the three masses that are chosen on-shell has a 
sizable bino component, at least one has a sizable wino component and at
least one has a sizable higgsino component. Failing to fulfill this
requirement will result in renormalisation constants being essentially
unconstrained, and therefore taking large unphysical values.
This issue was recently discussed for the case of real MSSM parameters 
in Ref.~\cite{Chatterjee:2011wc}, where it was similarly argued that 
one bino-, wino- and higgsino-like mass should be set to be on-shell.%
\footnote{
In addition, in Ref.~\cite{Chatterjee:2011wc}
the case of large mixing was discussed in detail, and it 
was found that there the most stable results were obtained in an 
NNC scheme where the mass of a wino-like
chargino is chosen on-shell.}

\section{NLO predictions at the LHC and the LC}\label{sec:4}

We now wish to utilise the above renormalisation framework to make NLO predictions for the LHC and LC.
This serves the purpose of illustrating the possible sensitivity of
collider observables to the details of the renormalisation procedure, in 
particular the treatment of the imaginary parts,
as well as highlighting the dependence of such observables on the CP violating phases introduced in the complex MSSM, which has so far not been studied in the on-shell scheme at one-loop.
 
Specifically we calculate $\Gamma(h_{a}\to\tilde{\chi}^+_i\tilde{\chi}^-_j)$ and $\sigma(e^+e^-\to\tilde{\chi}^+_i\tilde{\chi}^-_j)$ at NLO, including full MSSM corrections and allowing $A_{f_i}$, $\mu$, $M_1$ and $M_3$ to be complex.
The diagrams are generated and the amplitudes calculated using
\texttt{FeynArts}~\cite{Kublbeck:1990xc,Denner:1992vza,FAorig,Hahn:2000kx,Hahn:2001rv}, which however requires the counterterms for the relevant couplings as input.
We calculated these counterterms by renormalising the fields and parameters as described in detail in Sec.~\ref{sec:3}. 
Explicit expressions for the necessary MSSM counterterms are given in the following subsections.
\texttt{FormCalc}~\cite{Hahn:1998yk,FormCalc2,FormCalc3}
then was used to calculate the matrix elements and 
\texttt{LoopTools}~\cite{Hahn:1998yk}
to perform the necessary loop integrals.
The loop integrals are regularised via dimensional
reduction~\cite{DRED,DRED2,0503129}, which ensures that SUSY is 
preserved, via the implementation described in 
Refs.~\cite{Hahn:1998yk,delAguila:1998nd}.
We assume a unit CKM matrix.
\begin{table}[tbh!]
\begin{center}
 \begin{tabular}{c|c||c|c}
\hline
\T \B Parameter & Value & Parameter & Value\\
\hline
\hline
\T $|M_1|$ & 100 GeV& $M_2$ & 200 GeV\\
$|\mu|$ & 420 GeV & $M_{H^+}$ & 800 GeV\\
$|M_3|$ & 1000 GeV & $\tan\beta$ & 20\\
$M_{\tilde{q}_{1,2}}$ & 1000 GeV & $M_{\tilde q_3}$ & 500-800 GeV\\ 
$M_{\tilde{l}_{1,2}}$ & 400 GeV & $M_{\tilde l_3}$ & 500 GeV\\ 
$|A_q|$ & 1300 GeV & $|A_l|$ & 1000 GeV\\
\B $\alpha_s(M_Z)$ & 0.118 & $m_t$ & 173.2\\
\hline
\end{tabular}
\caption{Table of parameters used in our numerical analysis, where $A_q$
and $A_l$ denote the common trilinear couplings for the quarks and
leptons, respectively. $\alpha_s(M_Z)$ and $m_t$ are taken from Refs.~\cite{Nakamura:2010zzi} and \cite{Lancaster:2011wr} respectively.\label{tab:Params}}
\end{center}
\end{table}

For both processes, we present our results for the scenario given in
Tab.~\ref{tab:Params}.
In light of the current LHC results~\cite{Atlas-sqgl,CMS-sqgl}, 
we take the masses of the first two generations of squarks and the gluino to be at 1 TeV.
As the bounds on the third generation squark masses are much less 
constraining, we consider $M_{\tilde q_3}$ between 500 and 800 GeV. 
Here $M_{\tilde{q}_{1}} = (M_{L})_{\tilde{q}_{1}} =
(M_{f_{R}})_{\tilde{q}_{1}}$ denotes the soft-SUSY breaking parameters
as defined in Eq.~(\ref{sfermion}) for the first generation squarks,
etc.
%
In view of the fact that the LHC up to now places hardly any constraint 
on the charginos and neutralinos, we choose relatively low values for
the mass parameters, $|M_1|=100 \gev$, $M_2=200\gev$ and $|\mu|=420 \gev$, 
adopting a CMSSM-like scenario in the chargino and neutralino sector.
As we will be considering $h_a\to\tilde\chi^+_1\tilde\chi^-_2$, 
where $a=2,3$, we choose $M_{H^\pm}$ such as to ensure that the masses of 
$h_{2,3}$ are above the threshold for these decay channels to be open.
In view of the prospects for observing this decay via a signature
comprising four leptons and missing transverse energy we choose 
relatively low slepton masses, i.e.\ $M_{\tilde{l}_{1,2}}= 400 \gev$ and 
$M_{\tilde l_3} = 500 \gev$ (where the value of the 
ratio $M_{\tilde{l}_{1,2}}/M_{\tilde l_3}$ has an impact on the relative
amount of electrons and muons in the final state as compared to tau
leptons),
as well as relatively high $\tan\beta$, i.e.\ $\tan\beta = 20$.
Although further reducing the slepton masses and increasing $\tan\beta$ 
would enhance the signal, the EDM bounds would be tighter, as discussed below. 
Using the program 
\texttt{FeynHiggs}~\cite{Heinemeyer:1998np,Heinemeyer:1998yj,Degrassi:2002fi,Frank:2006yh,Hahn:2009zz}
and taking the current theoretical uncertainties from unknown
higher-order corrections into account, we have checked the
predictions for the MSSM Higgs masses arising from 
the parameters in Tab.~\ref{tab:Params}.
Although the chosen parameters give predictions for the light Higgs above 
114 GeV, respecting the LEP limits~\cite{LEP-I,LEP-II}, 
they are not in keeping with the recent discovery of a scalar resonance at the 
LHC~\cite{ATLASdisc,CMSdisc}. As the purpose of this paper is not to study
the detailed phenomenological consequences of the presented renormalization scheme,
we do not discuss this issue further, but note that a compatible light Higgs mass 
could be achieved, for example, by decreasing $A_t$ to 1050 GeV, which would have a small impact on
the loop corrections.

We study the effect of varying the phases $\phi_{A_t}$, $\phi_{A_b}$, 
$\phi_{A_\tau}$, $\phi_{M_1}$, $\phi_{M_3}$ and $\phi_\mu$ 
(using the convention that $M_2$ is real).
As discussed in Sec.~\ref{sec:2}, the EDM bounds on these phases can be
quite restrictive, and we therefore evaluate the predictions for the EDMs 
explicitly 
using \texttt{CPSuperH2.2}~\cite{Lee:2003nta}, incorporating further 
two-loop contributions using \texttt{2LEDM}~\cite{Li:2010ax}.
We find that for $M_{\tilde q_3}=600 \gev$  the approximate bounds
on the phases are $\phi_{A_t}\lesssim\pi/6$, $\phi_{M_1}\lesssim\pi/50$ and $\phi_{\mu}\lesssim\pi/1000$. For
$M_{\tilde q_3}=800 \gev$ the phase $\phi_{A_t}$ of the trilinear
coupling in  the stop sector is essentially unconstrained. The phases of 
$M_3$, $A_b$, $A_\tau$ are also found to be unconstrained. 
In obtaining these values, we took into account that while the
prediction for $|d^{\rm Tl}|$ is robust, the prediction for $|d^{\rm Hg}|$ 
involves atomic matrix elements which are only known up to a factor 2 to 3.
While the EDMs are mainly sensitive to the phases of the trilinear
couplings of the first two generations, the relatively large value of
$\tan\beta=20$ results in a non-trivial bound on the phase 
$\phi_{A_t}$ from the mercury EDM if $M_{\tilde q_3}$ is sufficiently
light.
For $\tan\beta=10$, on the other hand, any value of $\phi_{A_t}$ would
be allowed by the EDM constraints.
Similarly, due to the choice of $\tan\beta$ in combination with the
relatively low values of the slepton masses, the bound on $\phi_{M_1}$
is rather tight. For $\tan\beta=10$, on the other hand, the bound 
on $\phi_{M_1}$ would be $\sim\phi_{M_1}<\pi/10$ and, for example, 
upon additionally increasing $M_{\tilde l_i}$, for $i=1,2,3$, to a common 
value of 600 GeV, $\phi_{M_1}$ would be unrestricted.
We will discuss below the numerical impact of varying the different
phases in our results within the context of the bounds on the phases
arising from the EDM constraints.

\subsection{Heavy Higgs decays to charginos}
\label{sec:4.1}

The recent discovery of a light Higgs-like state is not 
sufficient to distinguish the hypothesis of a SM Higgs boson from the
hypothesis that the new state belongs to an extended Higgs sector. 
For instance, in the decoupling region of the MSSM the new state could
be interpreted as the lightest neutral MSSM Higgs boson. This state
behaves SM-like in the decoupling region, while the heavy MSSM Higgs
bosons decouple from the gauge bosons.
In this region the class of processes involving heavy Higgs bosons 
decaying to pairs of neutralinos and charginos are of particular
interest, as they could provide experimental evidence for an extended
Higgs sector. Detection of these processes at the LHC could be possible
in the final state with four leptons and missing transverse 
energy~\cite{Baer:1992kd,Baer:1994fx,Moortgat:2001pp,Bisset:2000ud}. 
A study in the MSSM with real parameters came to the conclusion 
that with $300\, \rm{fb}^{-1}$ it may be possible at the LHC to detect 
heavy Higgs bosons $H$ or $A$ with masses up to $\sim 800$ GeV at the 
$5 \sigma$ level~\cite{Bisset:2007mi}. 
This could cover part of the ``LHC wedge region'' (see e.g.\ 
Refs.~\cite{CMSTDR,arXiv:0704.0619}), where the standard searches for
heavy MSSM Higgs bosons in $\tau^+\tau^-$ (or $b \bar b$) final states 
will not be sufficiently sensitive to establish a signal.
 
In the general case of the MSSM with complex parameters, where the three 
neutral Higgs bosons mix to form the mass eigenstates, we calculate the 
decay widths $\mathit\Gamma(h_{a}\to\tilde{\chi}^+_i\tilde{\chi}^-_j)$
for the two heavy MSSM Higgs bosons, i.e.\ $a=2,3$. Since in the parameter 
region of sufficiently high values of $\MHp$ where these decays are open
kinematically, the two states $h_2$ and $h_3$ are nearly
mass-degenerate, it will experimentally be very difficult to distinguish
between these two states\footnote{ 
Complementarily, in Higgs decays to $\tau$ leptons it may be possible to
use certain asymmetries involving the $\tau$ polarisation to analyse the 
CP properties of the decaying scalar 
particle~\cite{Berge:2008dr,Berge:2008wi,Berge:2011ij}.}
in the signature with four leptons and missing $E_T$.
This fact is also apparent from the analysis of
Ref.~\cite{Bisset:2007mi}, which was restricted to the case of real
parameters, as the distribution of events arising from $H$ and $A$ 
in this analysis did not show considerable
differences.

\begin{figure}[ht]
\begin{center}
\includegraphics[scale=0.8]{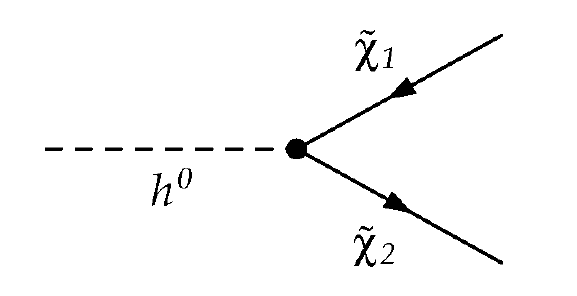}\includegraphics[scale=0.8]{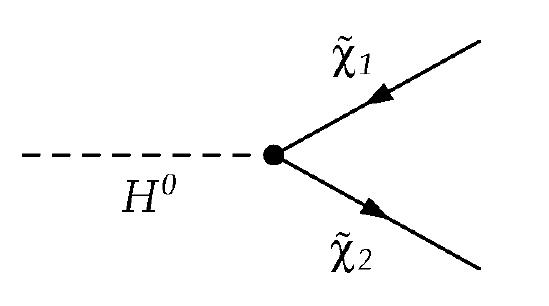}\includegraphics[scale=0.16]{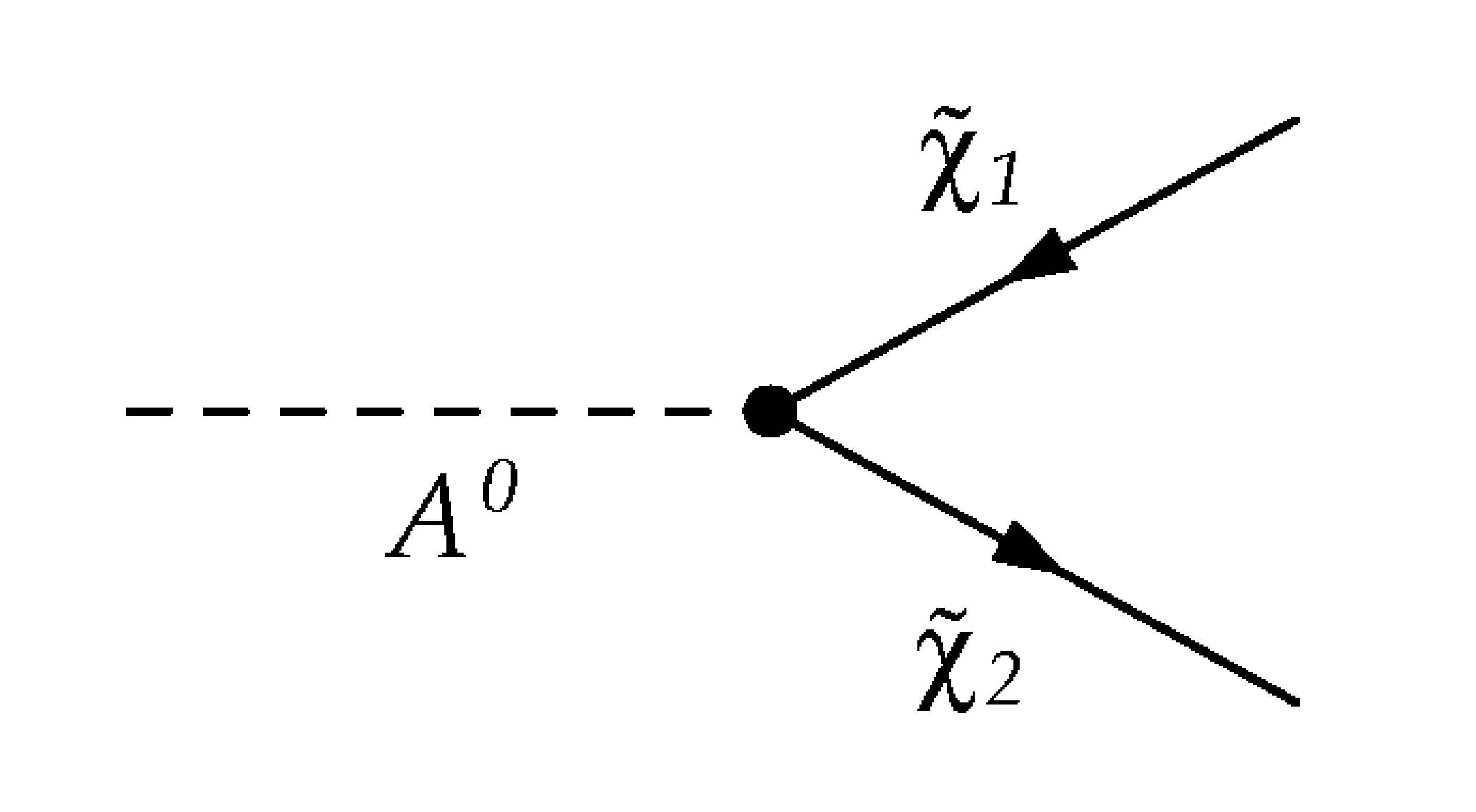}
\caption{Tree-level diagrams for the decay of neutral Higgs bosons $h$,
$H$ and $A$ to charginos $\tilde\chi^+_{1}$ and 
$\tilde\chi^-_{2}$.\label{fig:hXXtreeDiagrams}}
\end{center}
\end{figure}

The tree-level three-point vertex function for the interaction of 
charginos with neutral Higgs bosons $h_k=\{h,H,A\}$ is
\begin{equation}
 G_{\tilde{\chi}^+_i\tilde{\chi}^-_j h_k}^{\mathrm{Born}}\equiv
\omega_R C^R_{ijh_k}  +\omega_L (-1)^{\delta_{k3}}
C^{L}_{ijh_k}.\label{eq:3-point}
\end{equation}
A minus sign appears between the $\omega_R$ and $\omega_L$ terms for the
CP-odd Higgs states, i.e.\ $\delta_{k3}=1$ for $k=3$ and zero otherwise. 
The couplings, $C^{R/L}_{ijh_k}$, are given by
\begin{eqnarray}
\label{eqn:chichihiggscoupling}
 C^R_{ijh_k}=\,C^{L^{\dagger}}_{ijh_k}=\frac{e}{\sqrt{2} s_W} c_{ijh_k} ,
\end{eqnarray}
where 
\begin{align}
c_{ijh_k} =\,&i(a_k U_{j2}V_{i1}+b_k U_{j1}V_{i2}),\nonumber\\
a_k=\,&\{s_{\alpha},-c_{\alpha},\,s_{\beta_n}\},\nonumber\\
b_k=\,&\{-c_{\alpha},-s_{\alpha},c_{\beta_n}\}.\label{eqn:chichihiggscoupling2} 
\end{align}
Here 
$s_{\alpha} \equiv \sin\alpha$,
$c_{\alpha} \equiv \cos\alpha$, etc.,
and the matrices $U$, $V$ have been defined in Sec.~\ref{sec:3}.
The diagrams for these decays at tree-level are shown in 
Fig.~\ref{fig:hXXtreeDiagrams} for the example of the final state
$\tilde\chi_1^+\tilde\chi_2^-$.

The tree-level decay width for the two-body decay 
$h_k \rightarrow \tilde{\chi}^+_i\tilde{\chi}^-_j $ can therefore be written as
\begin{align}
\nonumber\mathit{\Gamma}^{\mathrm{tree}}(h_k \rightarrow \tilde{\chi}^+_i\tilde{\chi}^-_j )=\,&\frac{1}{16 \pi
m_{h_k}^3}\,
\left((|C^R_{ijh_k}|^2+|C^L_{ijh_k}|^2)\kappa(m_{h_k}^2,m_{\tilde{\chi}^\pm_i}^2,m_{\tilde{\chi}^\pm_j}^2)\,\right.\big(m_{h_k}^2\\
&-m_{\tilde{\chi}^\pm_i}^2-m_{\tilde{\chi}^\pm_j}^2\big)\left.-4C^R_{ijh_k}C^L_{ijh_k}
(-1)^{\delta_{k3}} m_{\tilde{\chi}^\pm_i} m_{\tilde{\chi}^\pm_j}\right) ,
\label{eqntreeh}
\end{align}
where
\begin{equation}
 \kappa (x,y,z)=((x-y-z)^2-4 y z)^{1/2}.
\end{equation}
As explained in Sec.~\ref{sec:2}, we ensure the correct on-shell properties 
of the mixed neutral Higgs bosons by the use of finite wave function
normalisation factors $\bf\hat Z_{ij}$, which contain universal
propagator-type contributions up to the two-loop level.
With the aim to investigate the effect of the genuine vertex
contributions for this process, we will compare our full one-loop result
to an improved Born result which incorporates the (process-independent) 
normalisation factors $\bf\hat Z_{ij}$. Accordingly, we define the
improved Born result by summing over the tree-level amplitudes for the 
three neutral Higgs bosons $h_k$ weighted by the appropriate 
$\bf\hat Z$ factor,
\begin{equation}
\hat{G}_{\tilde{\chi}^+_i\tilde{\chi}^-_jh_a}^{\rm Imp.
Born}=\sum_k\hat{\bf Z}_{ak}G^{\rm
Born}_{\tilde{\chi}^+_i\tilde{\chi}^-_jh_k}(M^2_{h_a}).
\label{eq:hXXimpBorn}
\end{equation}
As mentioned above, by definition the $\bf \hat Z$ factors do not include 
contributions due to mixing with the neutral Goldstone boson or the
$Z$ boson, and therefore the relevant one-loop contributions of this
type must be explicitly included in the calculation.

\begin{figure}[htb]
\begin{center}
\includegraphics[scale=0.8]{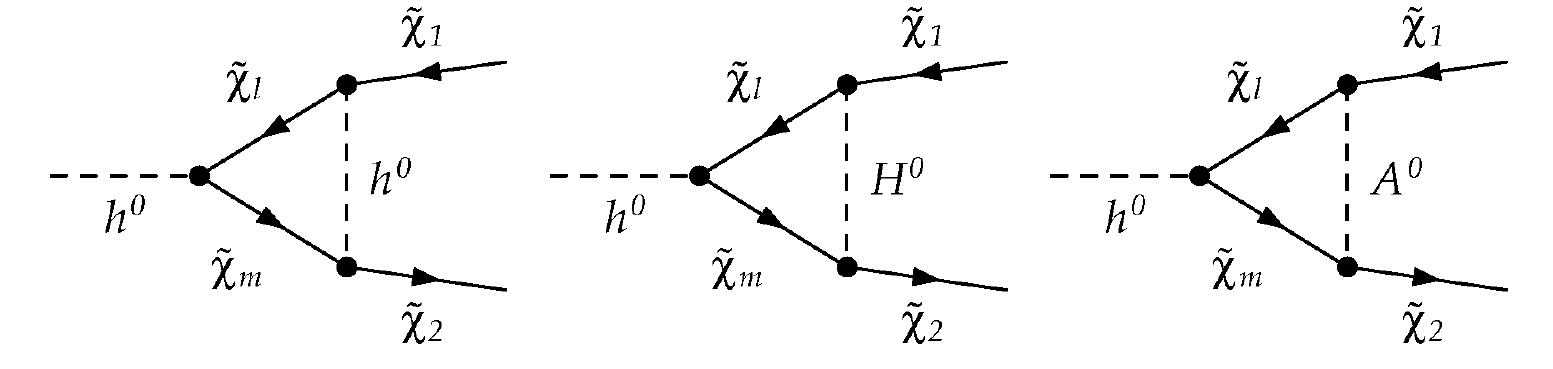}\\
\includegraphics[scale=0.8]{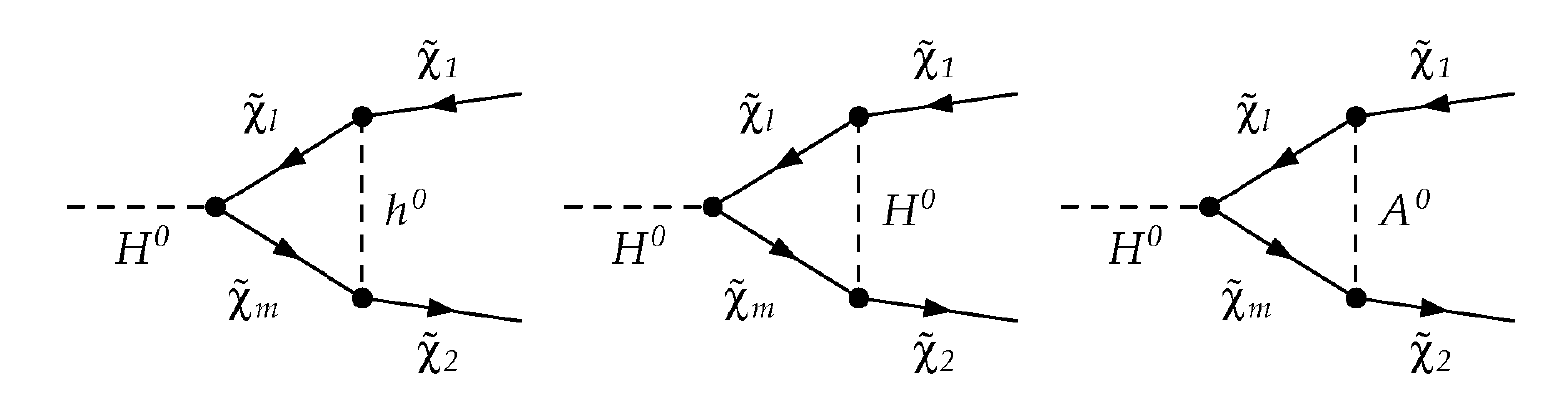}\\
\includegraphics[scale=0.8]{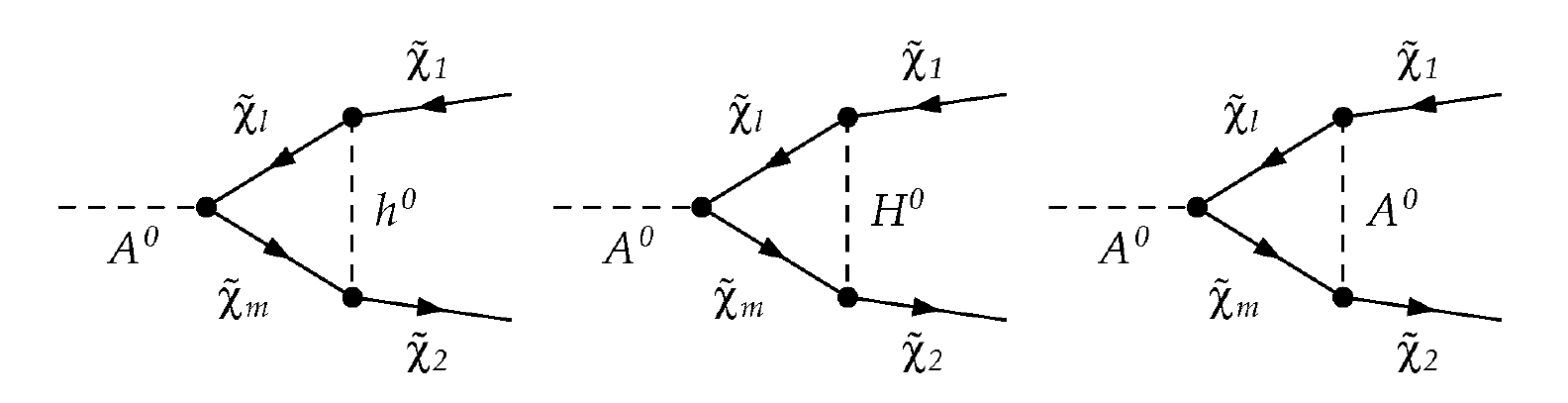}
\caption{A selection of one-loop vertex diagrams for the decay of
neutral Higgs bosons $h$, $H$ and $A$ to charginos $\tilde\chi^+_{1}$
and $\tilde\chi^-_{2}$.\label{fig:hXXvertDiagrams}}
\end{center}
\end{figure}

\begin{figure}[htb]
\begin{center}
\includegraphics[scale=0.83]{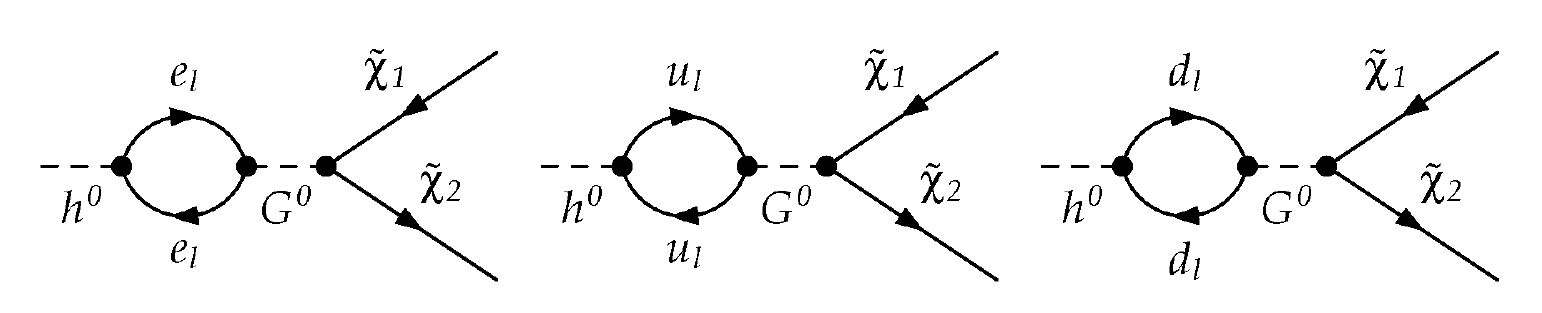}\\
\includegraphics[scale=0.83]{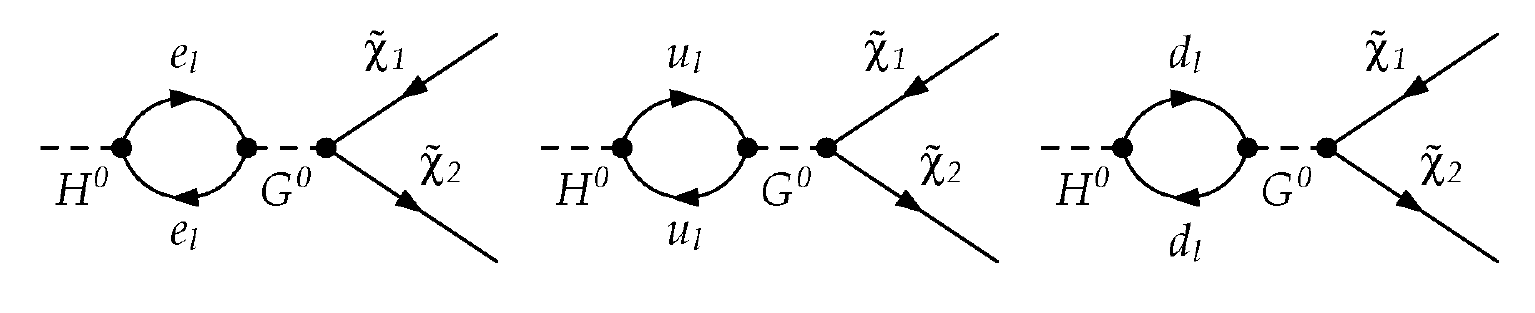}\\
\includegraphics[scale=0.83]{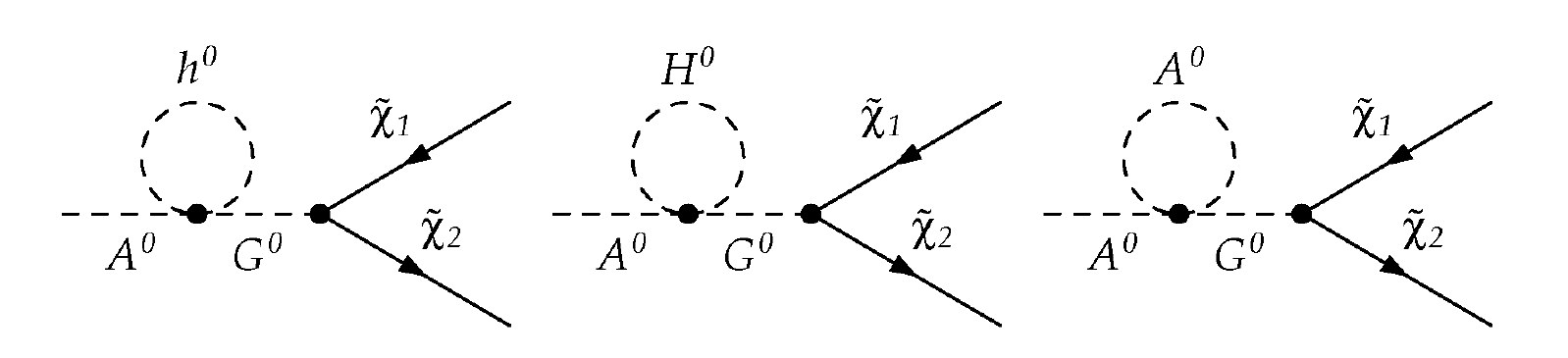}
\caption{A selection of one-loop self-energy diagrams for the decay of
neutral Higgs bosons $h$, $H$ and $A$ to charginos $\tilde\chi^+_{1}$
and $\tilde\chi^-_{2}$.\label{fig:hXXGZDiagrams}}
\end{center}
\end{figure}

\begin{figure}[htb]
\begin{center}
\includegraphics[scale=0.8]{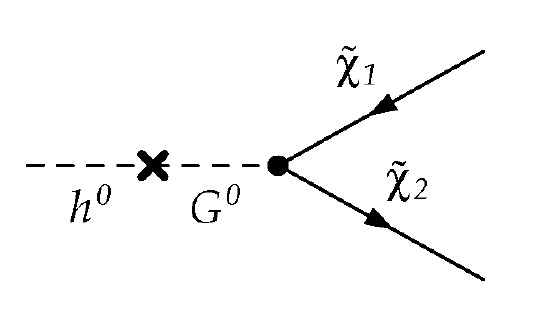}\includegraphics[scale=0.8]{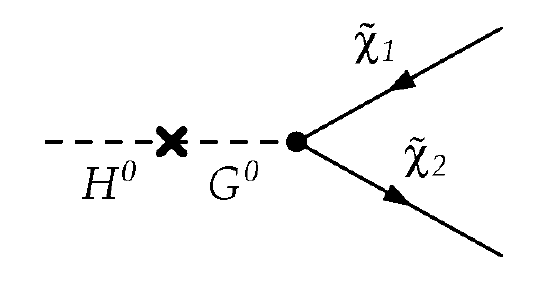}\\
\includegraphics[scale=0.8]{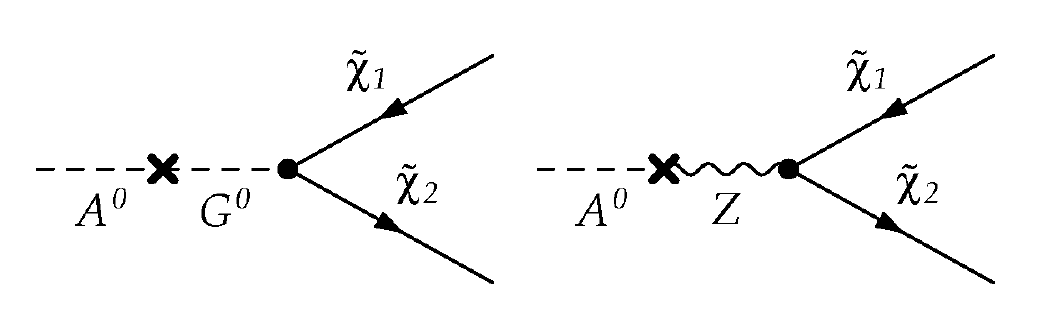}\\
\includegraphics[scale=0.16]{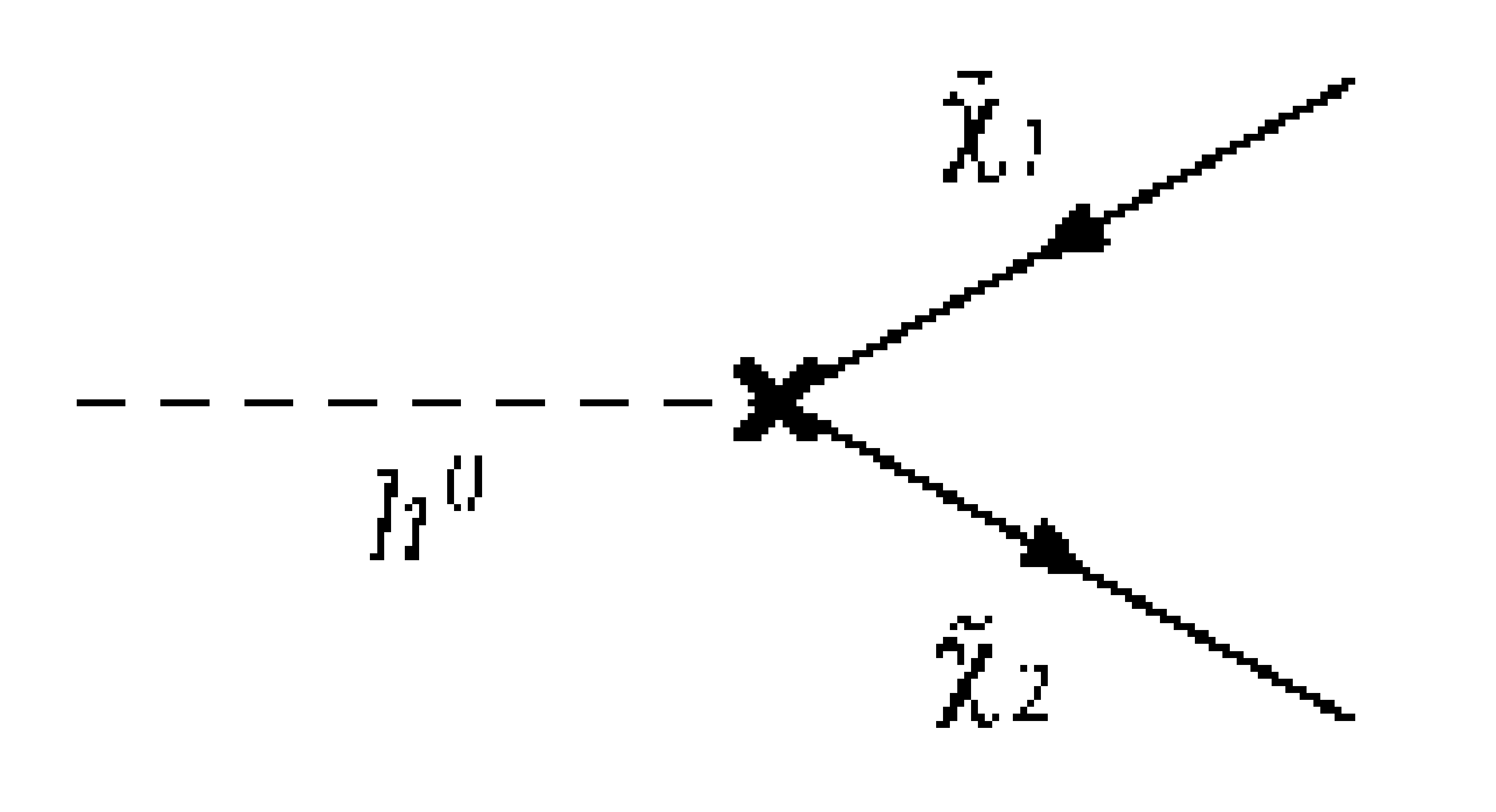}\includegraphics[scale=0.8]{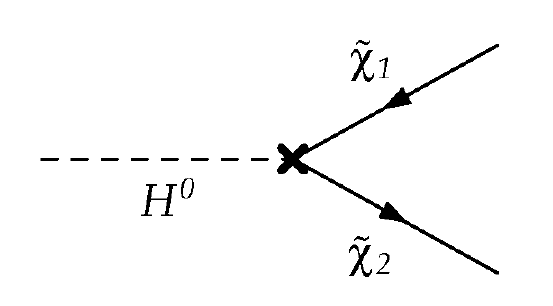}\includegraphics[scale=0.8]{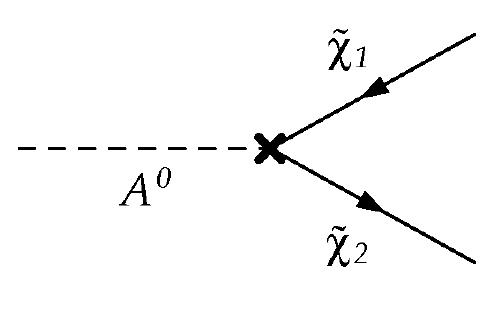}\caption{
Counter-term diagrams for the decay of neutral Higgs bosons $h$, $H$ and
$A$ to charginos $\tilde\chi^+_{1}$ and 
$\tilde\chi^-_{2}$.\label{fig:hXXcounterDiagrams}}
\end{center}
\end{figure}

The one-loop corrections therefore involve vertex diagrams, examples of
which are shown in Fig.~\ref{fig:hXXvertDiagrams}, and the self-energy
diagrams involving the $Z$ and the Goldstone boson, examples of which
are shown in Fig.~\ref{fig:hXXGZDiagrams}. These are calculated
following the procedure outlined earlier. In order to obtain UV finite
results at the one-loop level, the three-point vertex function defined at
tree-level in Eq.~(\ref{eq:3-point}) must be renormalised, i.e.\ we need to calculate the diagrams shown in the bottom row of Fig.~\ref{fig:hXXcounterDiagrams} for the vertex diagrams, and we also need to renormalise the self-energy corrections, i.e. we need to calculate the diagrams shown in the upper rows of Fig.~\ref{fig:hXXcounterDiagrams}.

The counterterm for the three-point vertex function defined in Eq.~(\ref{eq:3-point}) is of the form
\begin{equation}
\label{eqn:vertexcounterterm}
 \delta G_{\tilde{\chi}^+_i\tilde{\chi}^-_j h_k}\equiv\omega_R
\delta C^R_{ijh_k}  +\omega_L (-1)^{\delta_{k3}}
\delta C^{L}_{ijh_k} ,
\end{equation}
and the coupling counterterms are given by
\begin{align}
\delta C^{R/L}_{ijh_k} =\,&\,C^{R/L}_{ijh_k}(\delta Z_e-\frac{\delta s_W}{s_W})+\frac{1}{2}\sum^2_{l=1}(\delta Z^{R/L}_{li} C^{R/L}_{ljh_k}+
\delta \bar{Z}^{L/R}_{jl} C^{R/L}_{ilh_k})\nonumber\\
&+\frac{1}{2}(\delta Z_{h_k h} C^{R/L}_{ijh}+ \delta Z_{h_k H}C^{R/L}_{ijH}+\delta Z_{h_k A}C^{R/L}_{ijA}+\delta Z_{h_k G}C^{R/L}_{ijG}),
\label{eqn:cchcounterterm}
\end{align} 
where, in analogy to Eqs.~(\ref{eqn:chichihiggscoupling}) and (\ref{eqn:chichihiggscoupling2}),
\begin{align}
\nonumber C^R_{ijG}=\,&\,C^{L^{\dagger}}_{ijG}=\frac{e}{\sqrt{2} s_W} c_{ijG},\\
\label{eqn:chichiGcoupling}c_{ijG} =\,&i(-c_{\beta_n} U_{j2}V_{i1}+s_{\beta_n} U_{j1}V_{i2}).
\end{align}
Here $\delta Z_e$ and $\delta s_W$ are defined in Eqs.~(\ref{eq:RCZe})
and (\ref{eq:RCswcw}) respectively, and for the chargino field
renormalisation constants, given in Eqs.~(\ref{eqn:chargdiagfieldren})
and (\ref{eqn:chargoffdiagfieldren}), we have dropped the ``$\pm$''
as our tree-level diagrams do not contain any neutralinos.
Note that the parameter
renormalisation which enters these renormalisation constants is performed in the NCC
scheme where the relevant counterterms are defined in Eqs.~(\ref{eq:NCCa}) to~(\ref{eq:NCCc}).
The Higgs field renormalisation constants $\delta Z_{hh}$, 
$\delta Z_{hH}$, etc.\ appearing in Eq.~(\ref{eqn:cchcounterterm}) are
linear combinations of the field renormalisation constants 
$\delta Z_{\mathcal{H}_1}^{\overline{\mathrm{DR}}}$,
$\delta Z_{\mathcal{H}_2}^{\overline{\mathrm{DR}}}$
given in Eqs.~(\ref{eqn:dZH1}) and (\ref{eqn:dZH2}), as specified in 
Ref.~\cite{Frank:2006yh}.

In order to account for the mixing of the neutral Higgs bosons, 
the result for the amplitude is obtained by summing over the
contributions of the three neutral Higgs bosons, multiplied with the
corresponding wave function normalisation factors
$\bf\hat Z_{ij}$, and by furthermore adding the mixing contributions of
the Higgs boson with the Goldstone boson and the Z~boson, 
\begin{align}
\hat{G}_{\tilde{\chi}^+_i\tilde{\chi}^-_jh_a}^{\rm
Imp.\,1-Loop}=\,&\sum_k\hat{\bf Z}_{ak}\bigg(\hat{G}^{\rm
1PI}_{\tilde{\chi}^+_i\tilde{\chi}^-_jh_k}(M^2_{h_a})+\hat{G}^{\rm
G,Z,se}_{h_k\tilde{\chi}^+_i\tilde{\chi}^-_j}(m^2_{h_k})\bigg).
\label{eq:impamp}
\end{align}
Here $\hat{G}^{\rm 1PI}_{\tilde{\chi}^+_i\tilde{\chi}^-_jh_k}$ 
contains both the tree-level and the complete one-loop contributions in
the MSSM, and
$M^2_{h_a},M^2_{h_b},M^2_{h_c}$ denote the loop-corrected masses, while 
$m^2_{h_k}$ refers to the tree-level masses, see 
Refs.~\cite{Williams:2007dc,Williams:2011bu} for further details.

\begin{figure}[htb!]
\begin{center}
\includegraphics[scale=0.8]{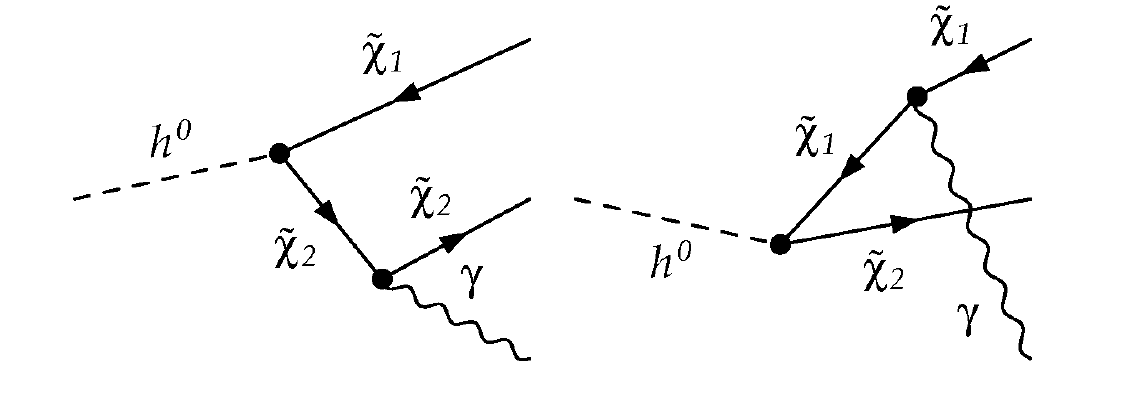}\\
\includegraphics[scale=0.8]{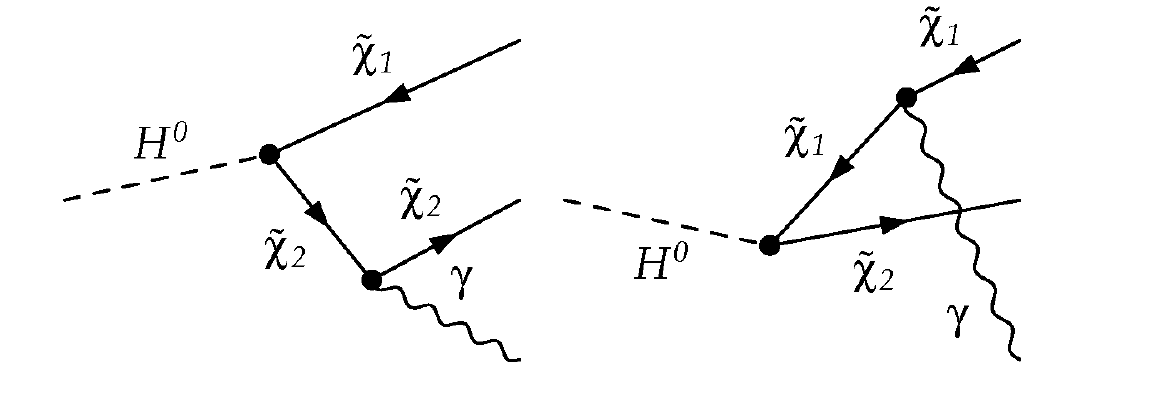}\\
\includegraphics[scale=0.8]{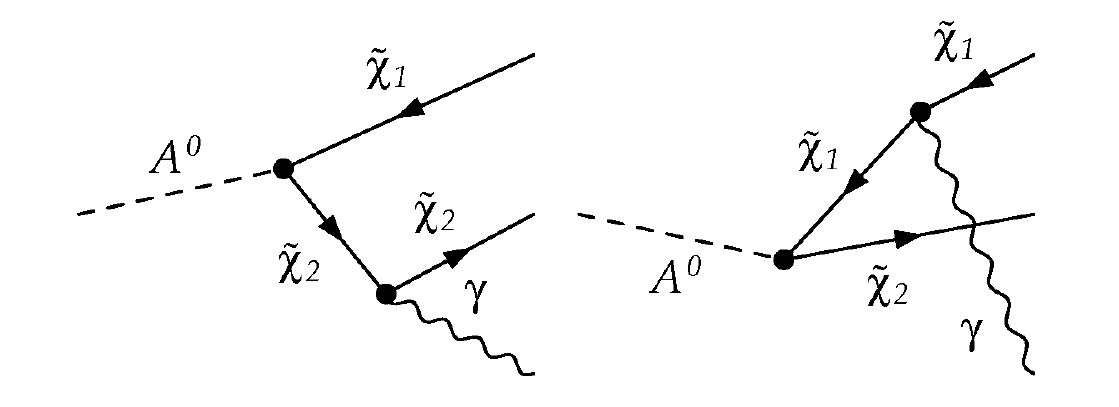}
\caption{Diagrams for the decay of neutral Higgs bosons $h$, $H$ and $A$
to charginos $\tilde\chi^+_{1}$ and $\tilde\chi^-_{2}$ with real photon emission.\label{fig:hXXgDiagrams}}
\end{center}
\end{figure}

Since the virtual contributions to the decay width in
Eq.~(\ref{eq:impamp}) may involve virtual photons giving rise to IR 
divergences, we add the corresponding bremsstrahlung contribution from
diagrams with real photon emission, as shown in Fig.~\ref{fig:hXXgDiagrams}. 
In this way we obtain the complete 1-loop result for the decay width,
\begin{align}
\mathit{\Gamma}^{\rm Full}=\mathit{\Gamma}^{\rm \rm Imp.\,1-Loop}(h_a
\to \tilde{\chi}^+_i\tilde{\chi}^-_j )+\mathit{\Gamma}^{\rm Imp.\,
Born}(h_a\to\tilde{\chi}^+_i\tilde{\chi}^-_j\gamma). \label{eq:hXXfull}
\end{align}
We have compared our result for the general case of complex parameters
to the existing result that was restricted to the case of real
parameters, given in Ref.~\cite{Eberl:2004ic}, and obtainable via the package HFOLD~\cite{Frisch:2010gw}. There the renormalisation
prescriptions used for the chargino and neutralino mixing matrices, for 
$\tan\beta$ (where the renormalisation condition given in
Ref.~\cite{Dabelstein:1994hb} has been applied) and for the charge
renormalisation differ from the ones used in the present work. The numerical
evaluation in Ref.~\cite{Eberl:2004ic} was carried out in the SPS~1a benchmark
scenario~\cite{Allanach:2002nj}. We find agreement between our result
and the result of Ref.~\cite{Eberl:2004ic} within the expected accuracy.

\begin{figure}[htb!]
\hspace{-.8cm}
\begin{tabular}{cc}
\includegraphics[scale=.55]{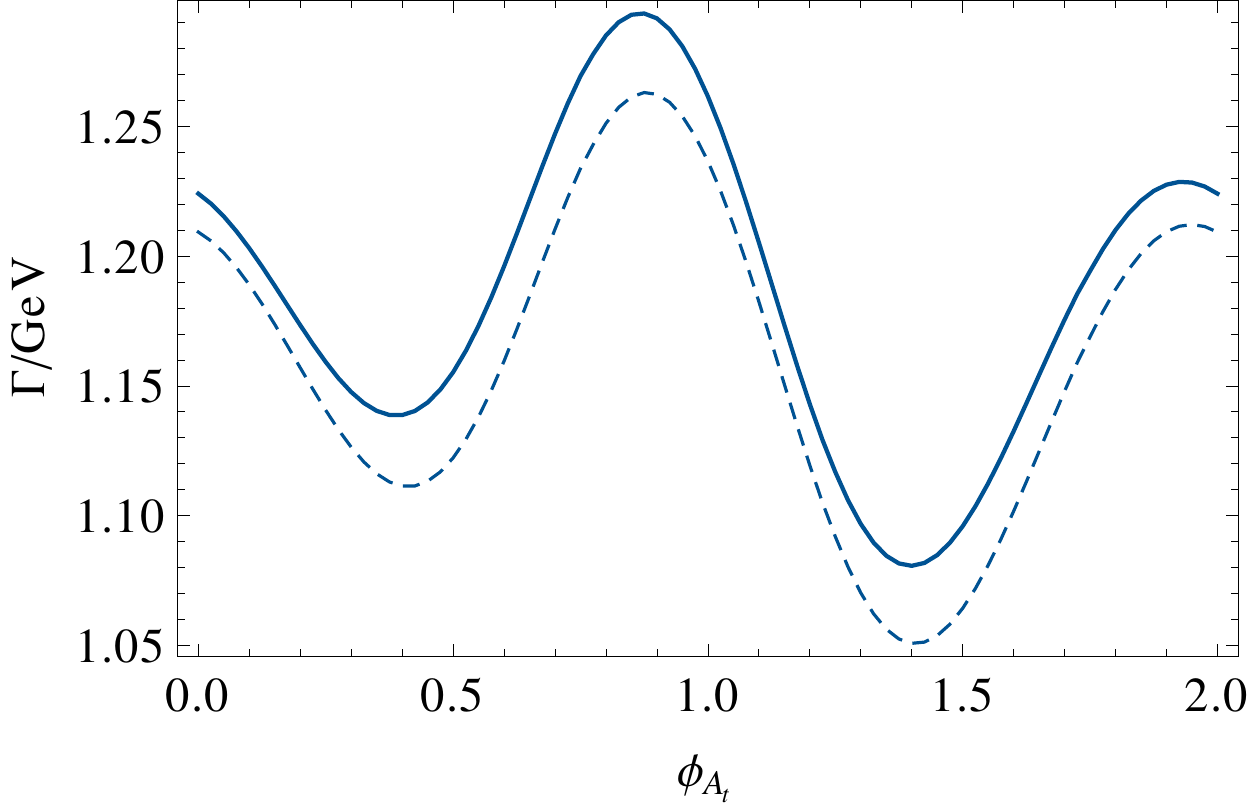} \includegraphics[scale=.55]{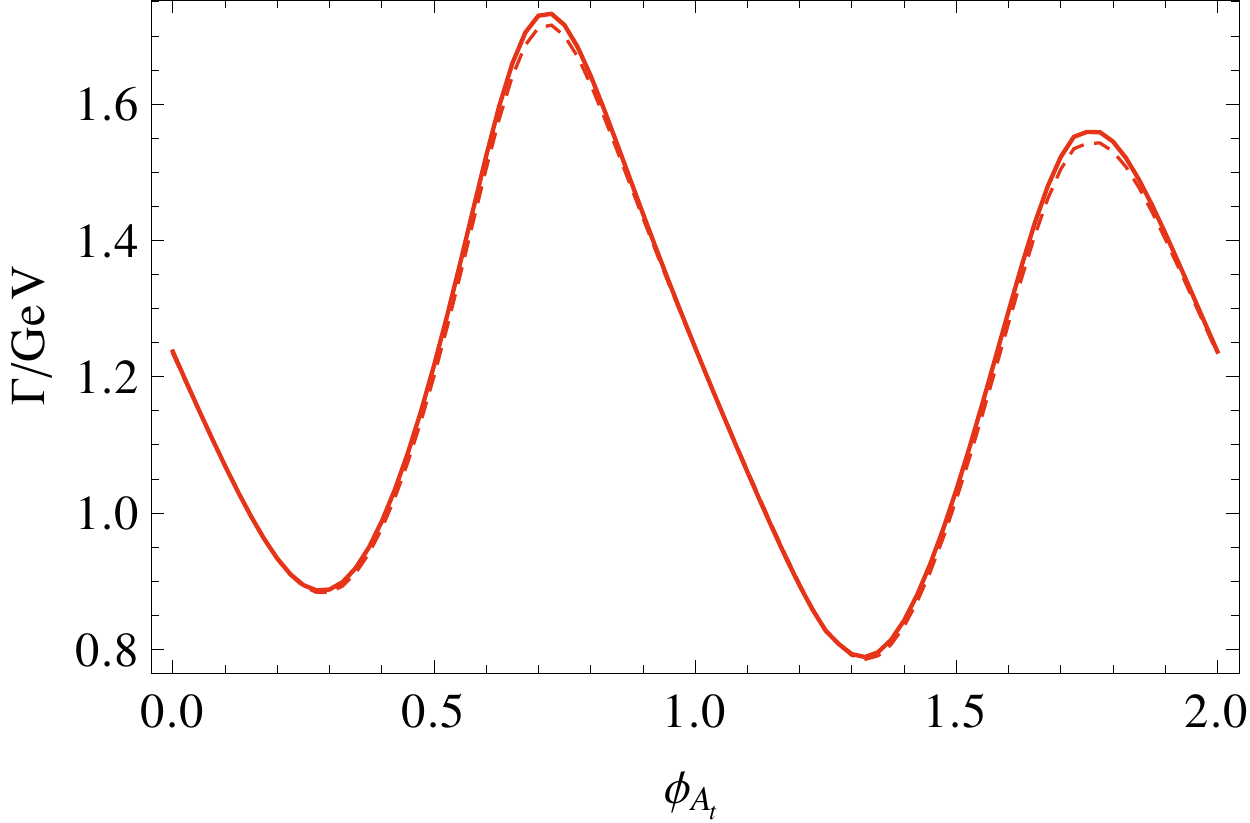}\\
\includegraphics[scale=.55]{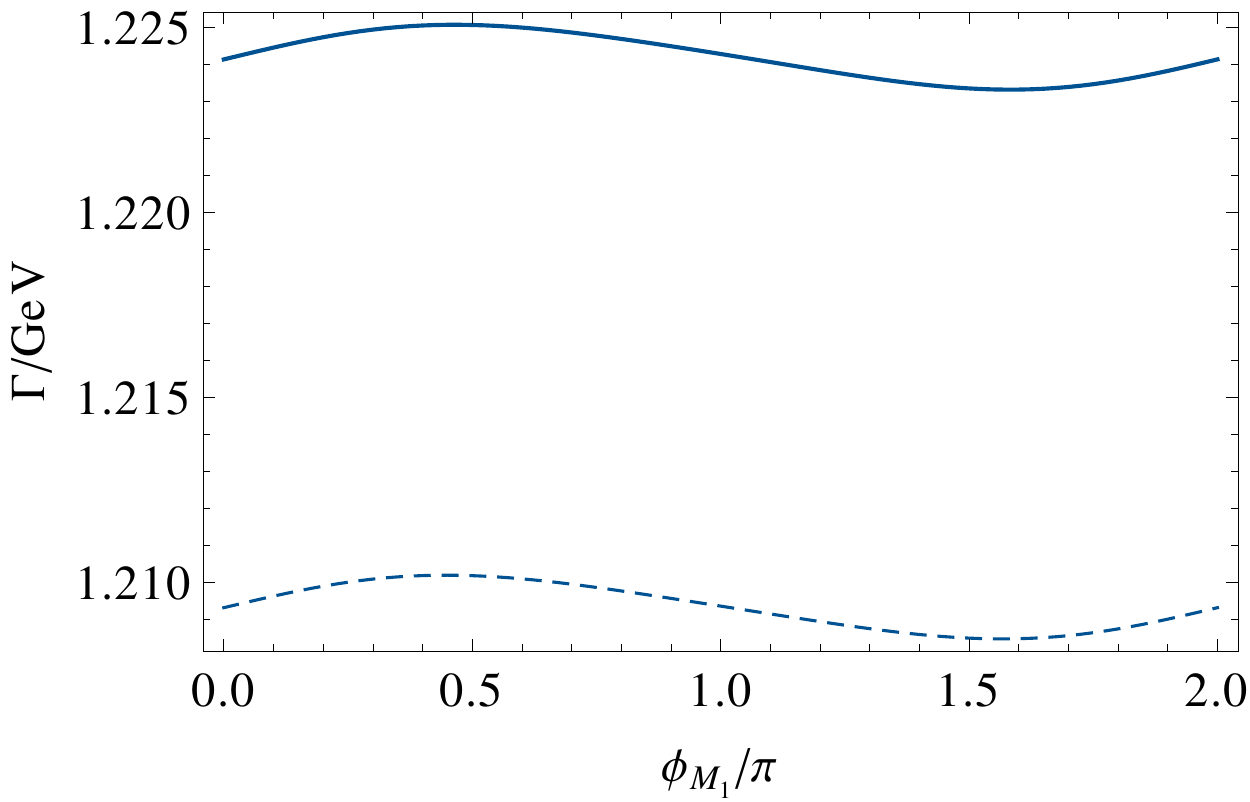} \includegraphics[scale=.55]{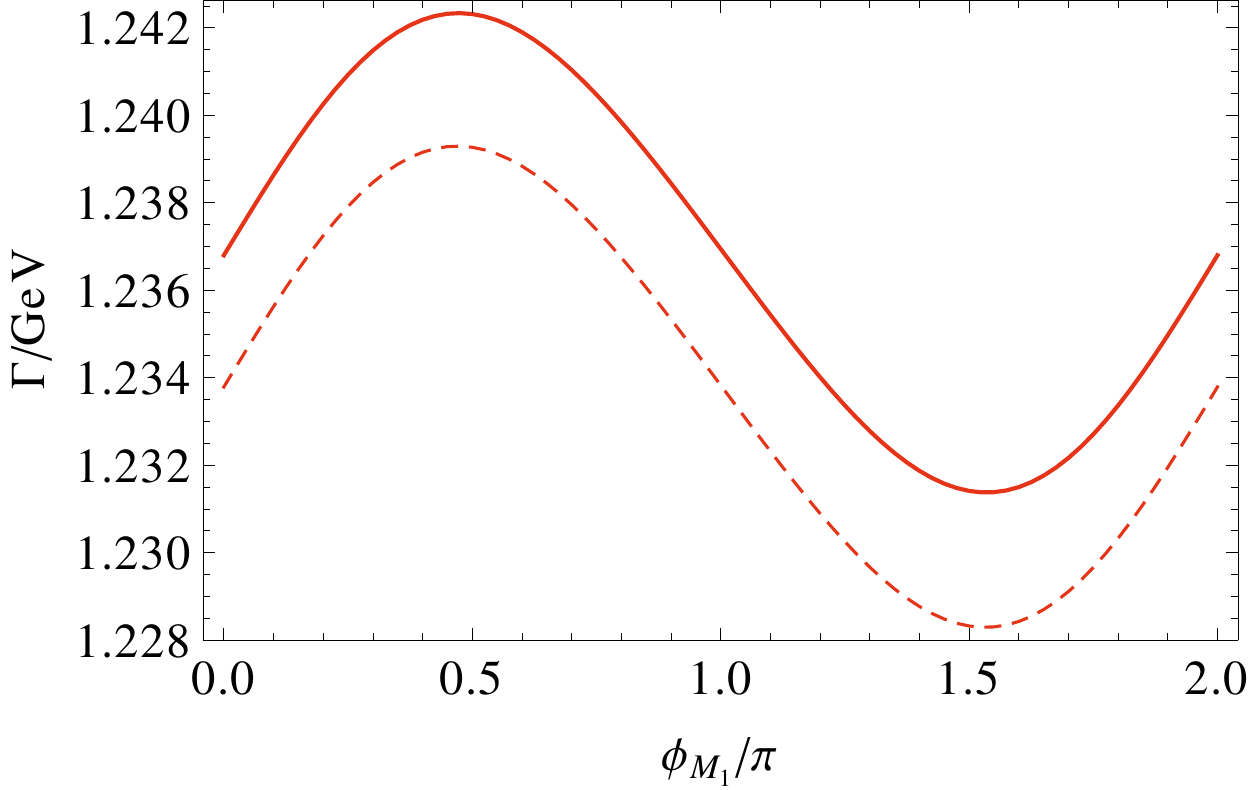}\\
\includegraphics[scale=.55]{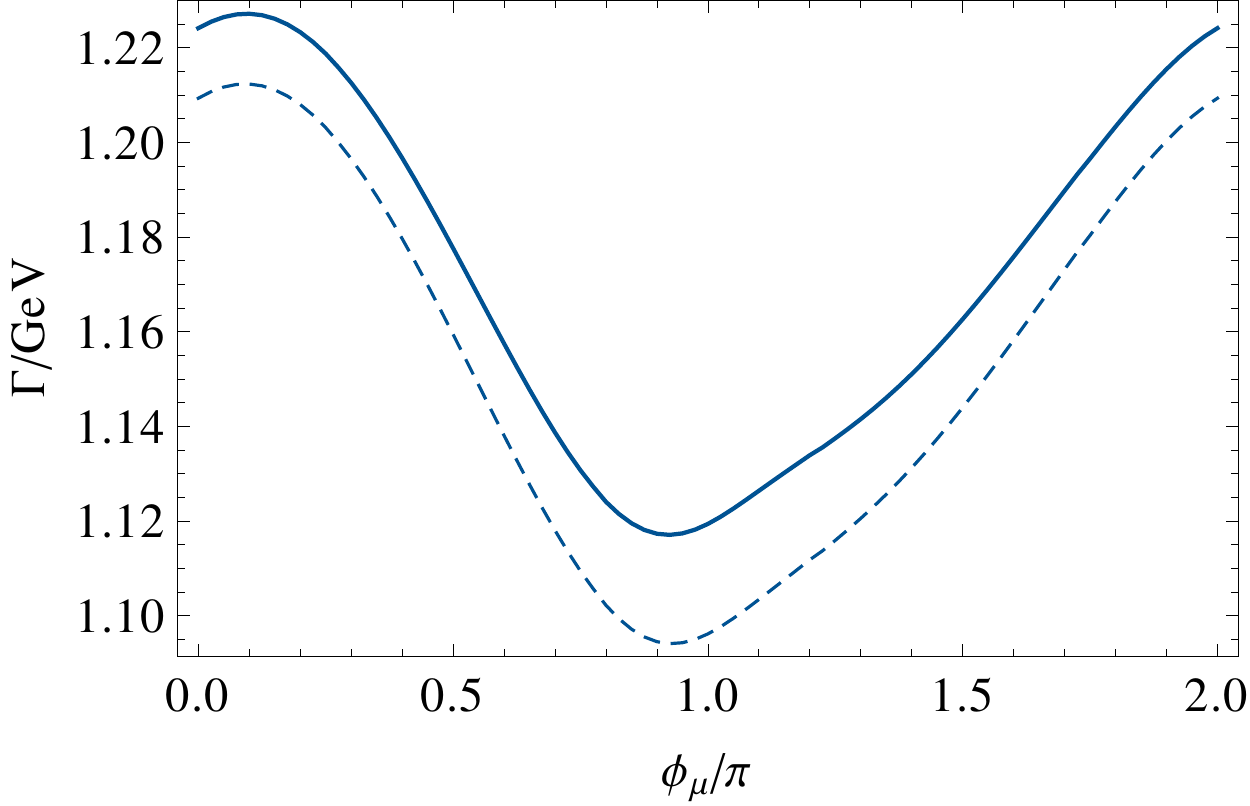} \includegraphics[scale=.55]{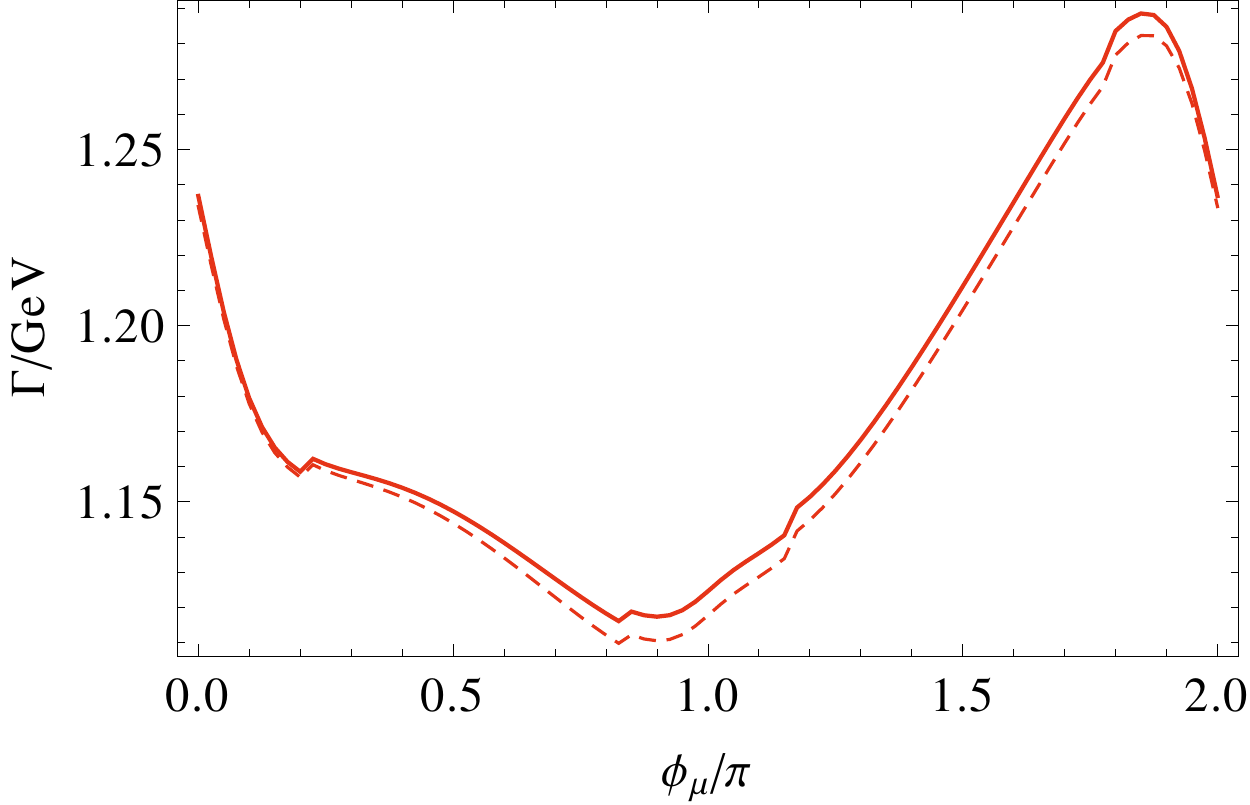}
\end{tabular}
\caption{Decay width in GeV for $h_{2}\to
\tilde{\chi}_{1}^+\tilde{\chi}_{2}^-$ as a function of the phases
$\phi_{A_t}$ (upper), $\phi_{M_1}$ (middle) and $\phi_\mu$ (lower). The
solid (dashed) lines show the 1-loop corrected (improved Born) results
for $M_{\tilde q_{3}} = 600$ (blue, left) and 800 (red, right) GeV.\label{fig:1}}
\end{figure}

For the numerical analysis of our results, we first consider the effects
of the phases $\phi_{A_t}$, $M_1$ and $\mu$ for the parameters as
defined in Tab.~\ref{tab:Params} (as mentioned above, we use the
convention where the phase of the parameter $M_2$ is rotated away).%
\footnote{We also investigated the effects of the phases $\phi_{A_\tau}$,
$\phi_{A_b}$, $\phi_{M_3}$ for the scenario described in Tab.~\ref{tab:Params} and
found the maximum relative deviation from the decay width when the phases are zero to be far below the percent level, i.e.\ a maximum 
of 0.03\%, 0.04\% and 0.03\%, respectively.}
In Fig.~\ref{fig:1} we show our results for 
$\Gamma(h_{2}\to \tilde{\chi}_{1}^+\tilde{\chi}_{2}^-)$ as a function of
the phases $\phi_{A_t}$, $M_1$ and $\mu$. The full result corresponding
to Eq.~(\ref{eq:hXXfull}) is compared with the improved Born result based
on the amplitude given in Eq.~(\ref{eq:hXXimpBorn}). As mentioned above,
the improved Born result incorporates higher-order contributions not
only from the calculation of the mass of the decaying Higgs boson, but
we have also dressed the lowest-order result with the universal wave 
function normalisation factors $\bf\hat Z_{ij}$. As one can see in 
Fig.~\ref{fig:1}, the dominant contribution to the dependence of the decay 
width on the three phases $\phi_{A_t}$, $\phi_{M_1}$ and $\phi_\mu$ is 
already present in the improved Born result.
Despite the fact that
$\phi_{A_t}$ only enters at loop level, its numerical impact on the
decay width is found to be very significant, which is a consequence of
the large (Yukawa enhanced) stop loop corrections in the Higgs sector.
Compared to the decay width at $\phi_{A_t}=0$, the full MSSM one-loop
corrected decay width is modified by up to 11\% and 40\% upon varying 
$\phi_{A_t}$ for $M_{\tilde q_{3}} = 600 \gev$ and 800~GeV, respectively.
Note that here the larger impact of the phase at $M_{\tilde q_{3}}=800\gev$ 
is due to a threshold effect, as
the mass of the decaying Higgs lies very close to the mass of the stops.
The deviations from the improved Born result amount up to 
3\% and 1\% for $M_{\tilde q_{3}} = 600 \gev$ and 800~GeV, respectively.

The effects induced by the phase $M_1$, on the other hand, are
relatively small. The shift compared to the decay width with
$\phi_{M_1}=0$ and the deviations from the improved Born result are both
at the percent level for $M_{\tilde q_{3}} = 600 \gev$, and at the 
sub-percent level for $M_{\tilde q_{3}} = 800 \gev$.
As in the case of $\phi_{A_t}$, the phase of $M_1$ only arises in the expressions
for the decay width at loop level, however the lower sensitivity of the 
decay width to this phase as compared to $\phi_{A_t}$ is expected as the 
pertinent loops are not Yukawa enhanced.

Variation of the phase of $\mu$ can in principle give rise to larger
effects on the decay width of up to 8\% and 10\%, and deviations of 
2\% and 1\% from the improved Born result, for 
$M_{\tilde q_{3}} = 600 \gev$ and $M_{\tilde q_{3}} = 800 \gev$, respectively, 
as it appears in the tree level couplings.
However, taking into account the tight constraints on this phase from 
the EDM bounds discussed earlier, the impact of the phase variation on
the decay width is reduced to the sub-percent level.
The kinks seen in the bottom right-hand plot of
Fig.~\ref{fig:1} showing the dependence of the decay width on $\phi_\mu$
for $M_{\tilde q_{3}} = 800 \gev$ arise due to the crossing of the masses 
of $h_2$ and $h_3$ at these points. 

\begin{figure}[htb!]
\hspace{-.8cm}
\begin{tabular}{cc}
\includegraphics[scale=.55]{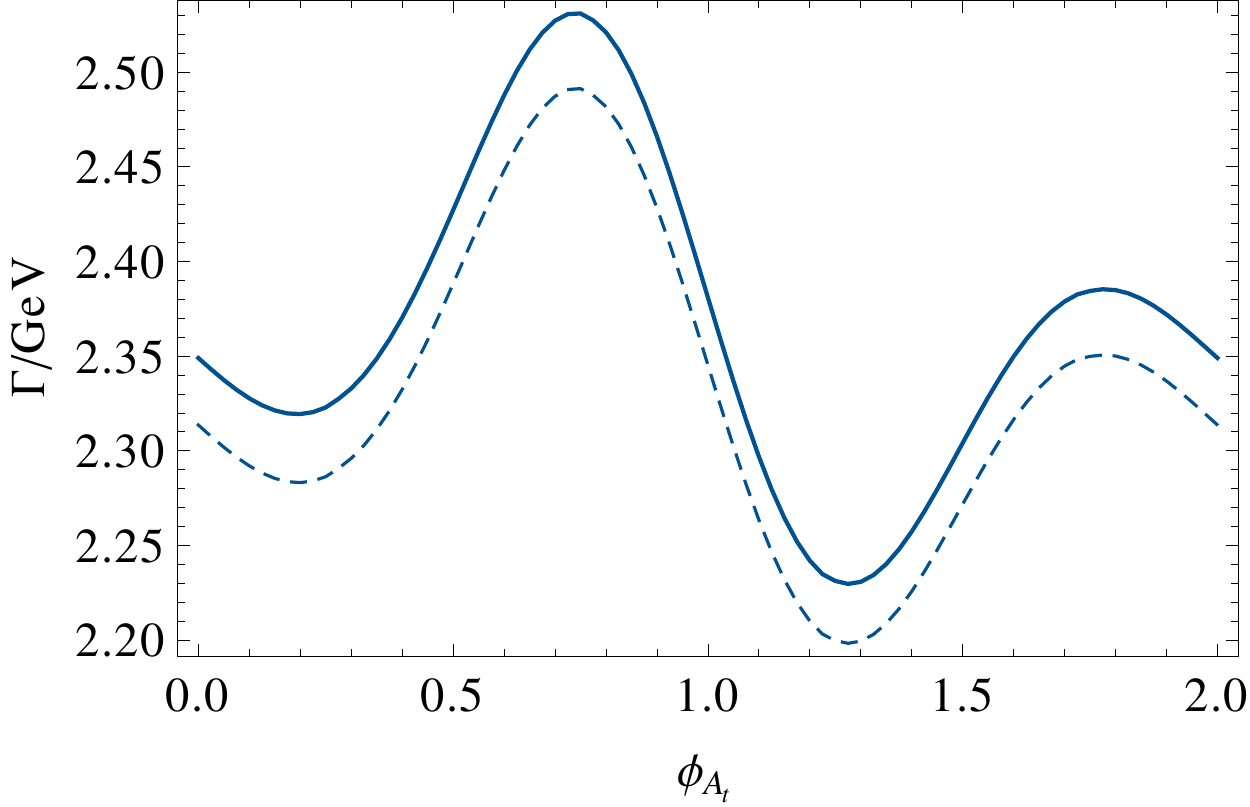} \includegraphics[scale=.55]{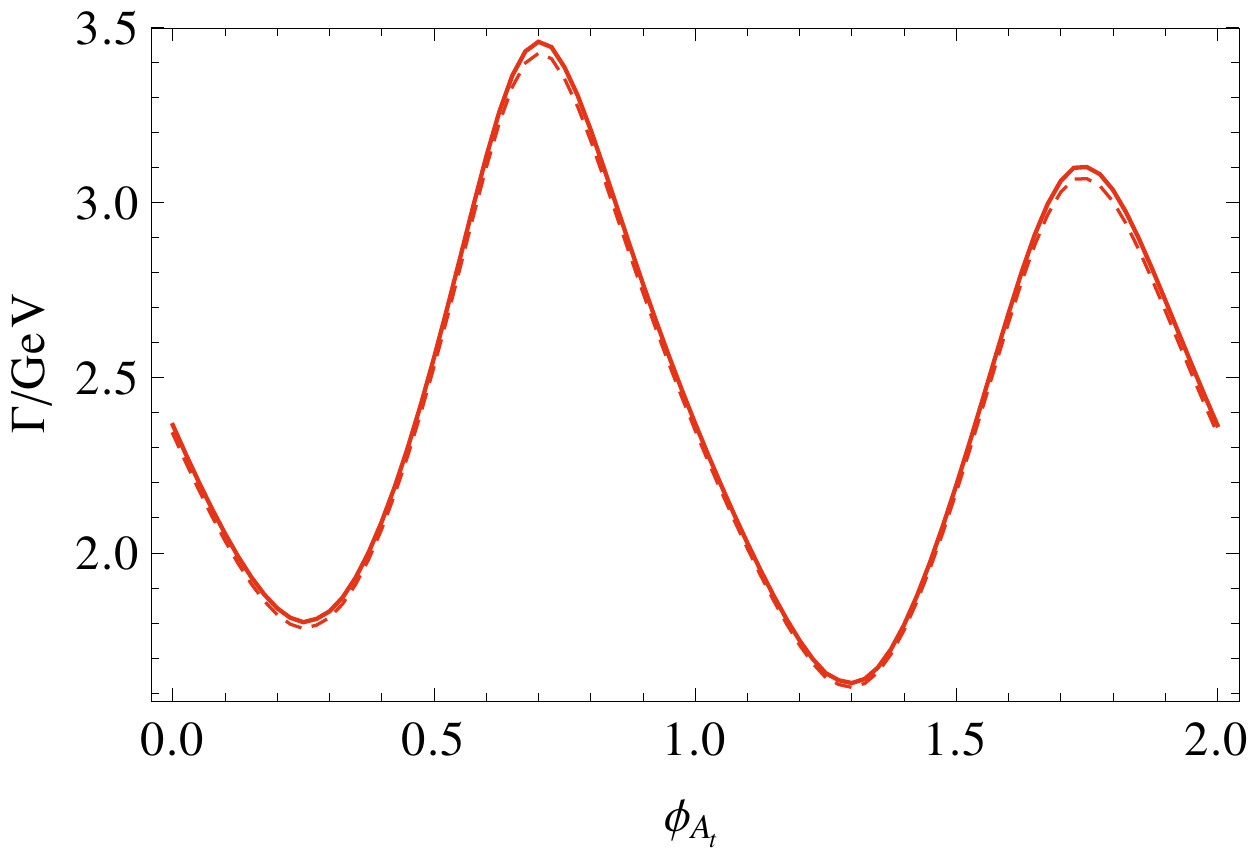}
\end{tabular}
\caption{Sum of the decay widths in GeV for $h_{2}\to
\tilde{\chi}_{1}^+\tilde{\chi}_{2}^-$ and $h_{3}\to
\tilde{\chi}_{1}^+\tilde{\chi}_{2}^-$ as a function of the phase
$\phi_{A_t}$. The solid (dashed) lines show the 1-loop corrected
(improved Born) results for $M_{\tilde q_{3}} = 600$ (blue, left) and 800 (red, right) GeV.\label{fig:3}}
\end{figure}

Accordingly, the phase having the most important impact on the decay
width $\Gamma(h_{2}\to \tilde{\chi}_{1}^+\tilde{\chi}_{2}^-)$ is
$\phi_{A_t}$. 
Due to the prospective difficulty in resolving the decays of $h_2$ and
$h_3$ experimentally, we have further investigated the dependence 
of the sum of the two decay widths, 
$\Gamma(h_{2}\to \tilde{\chi}_{1}^+\tilde{\chi}_{2}^-) + 
\Gamma(h_{3}\to \tilde{\chi}_{1}^+\tilde{\chi}_{2}^-)$,
on $\phi_{A_t}$. As shown in Fig.~\ref{fig:3}, the marked dependence on
$\phi_{A_t}$ is also present for the sum of the two decay widths, giving
rise to shifts of up to 9\% and 48\% for $M_{\tilde q_{3}} = 600 \gev$
and 800~GeV, respectively.

\begin{figure}[htb!]
\begin{center}
\includegraphics[scale=.55]{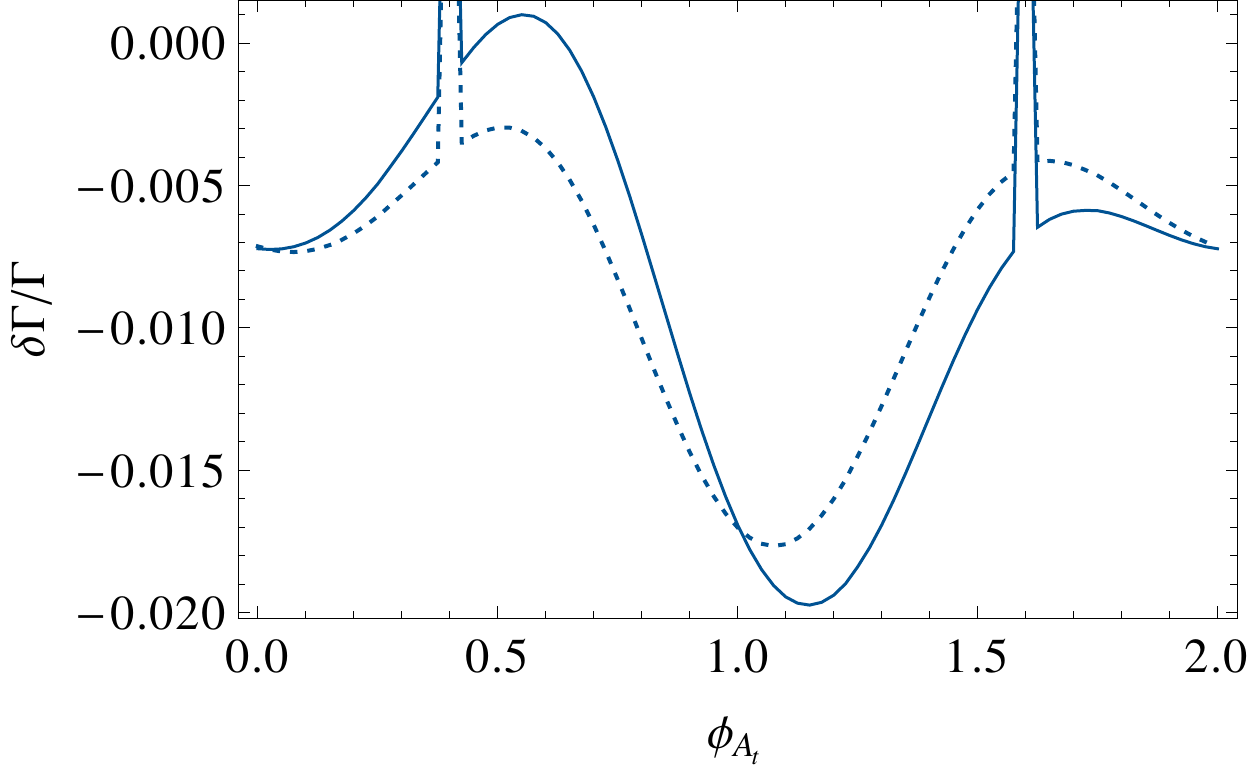}\includegraphics[scale=.55]{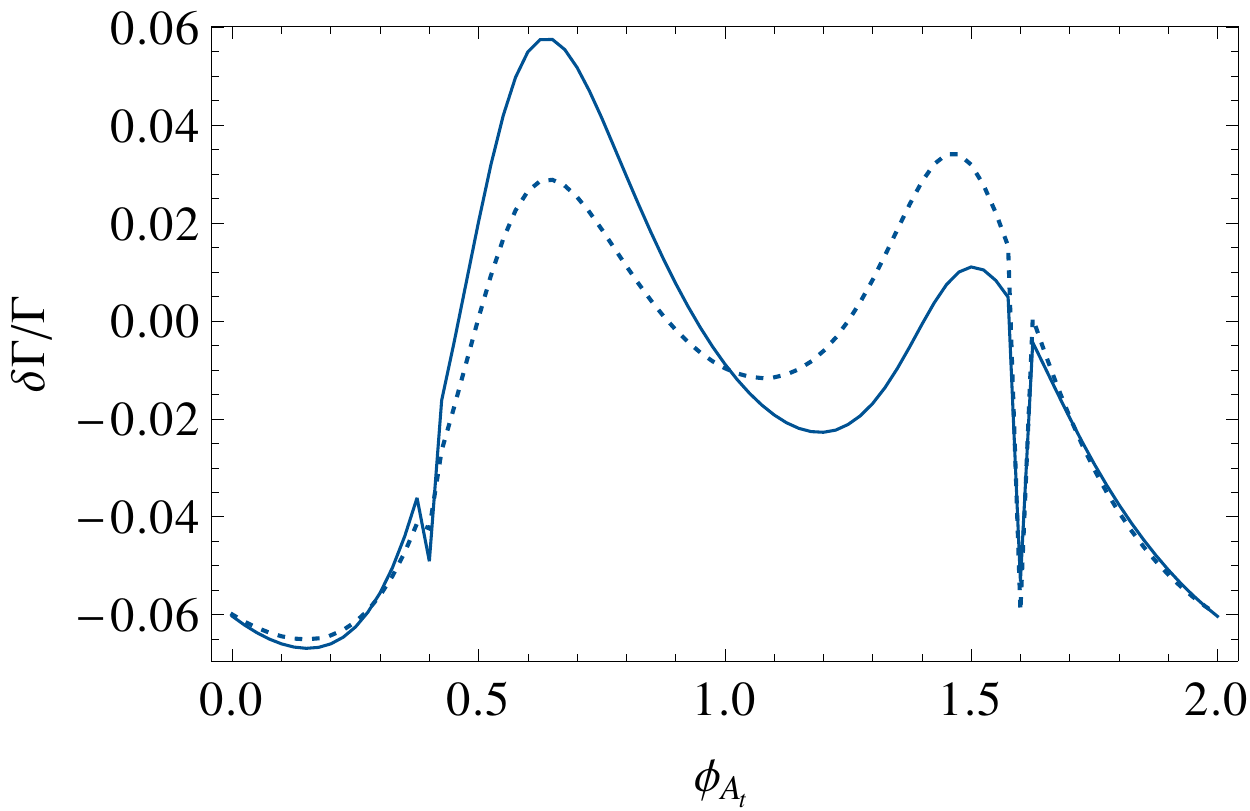}
\includegraphics[scale=.55]{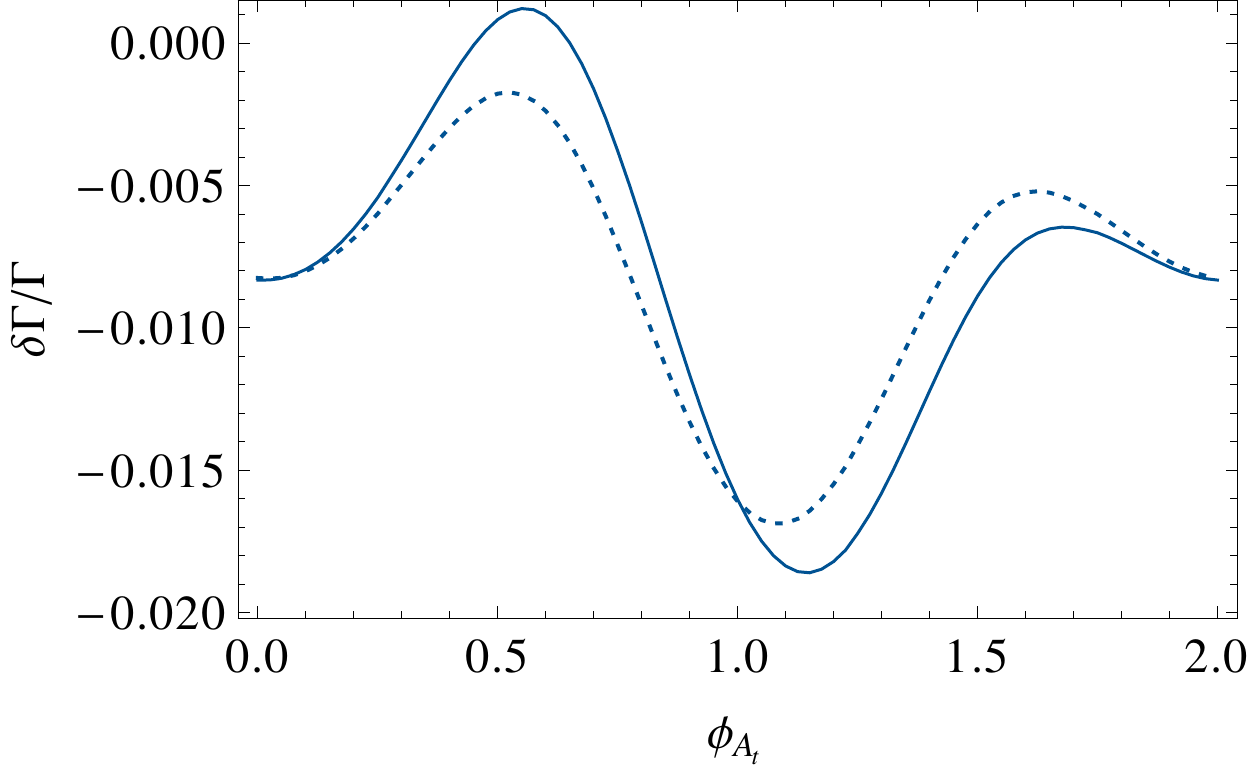}\includegraphics[scale=.55]{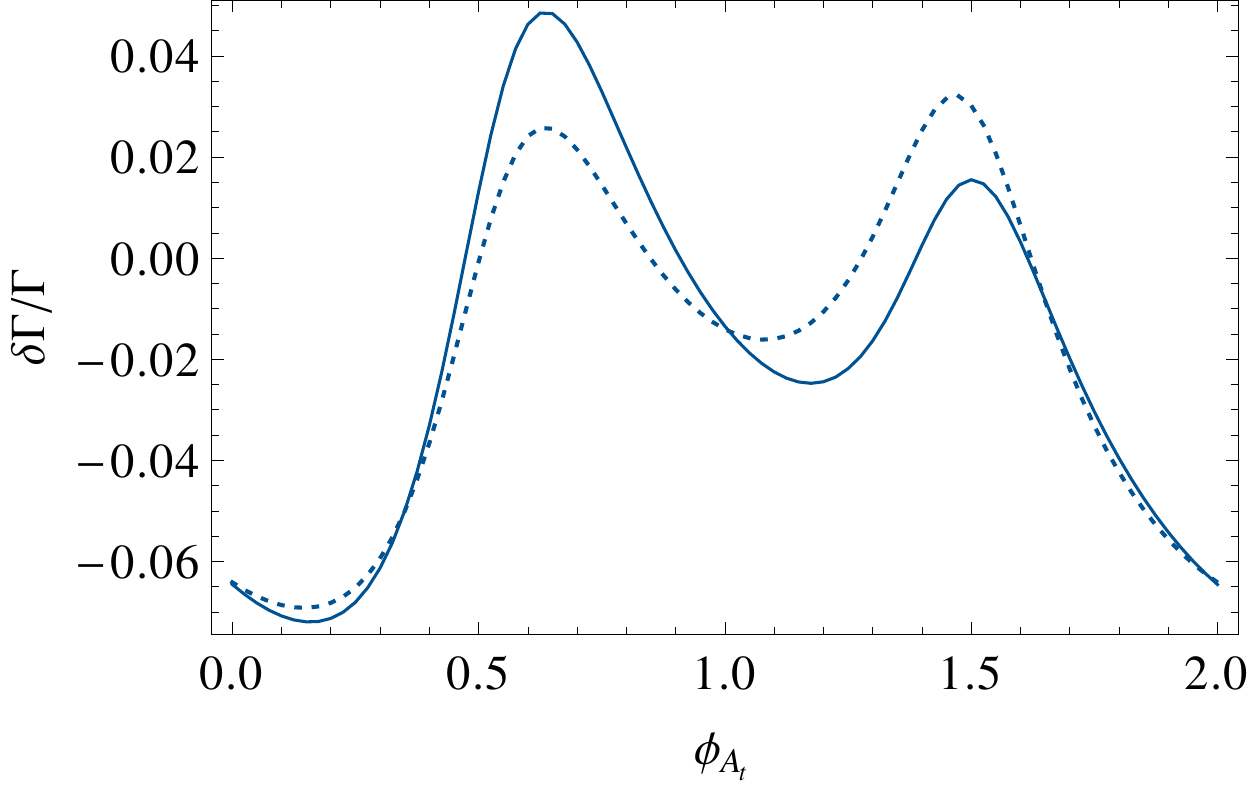}
\caption{Ratio of the 1-loop corrected decay width to the tree-level
decay width, as defined in Eq.~(\ref{eq:deltaGamma}), for $h_{2}\to
\tilde{\chi}_{1,L}^+\tilde{\chi}_{2,R}^-$ as a function of the phase of
$A_t$, $\phi_{A_t}$, showing the effect of absorptive parts of the
self-energies in the chargino field renormalisation constants. The solid
(dotted) line indicates the result with (without) taking the absorptive
part into account. Parameters are as in Tab.~\ref{tab:Params}, except
$M_{\tilde q_{3}} = 500\gev$ for the upper plots and 
$M_{\tilde q_{3}} = 520\gev$ for the lower plots.\label{fig:2}}
\end{center}
\end{figure}

We now turn to the impact of the renormalisation procedure on our final
result, i.e.\ we investigate the numerical relevance of the consistent
treatment of the absorptive parts, which, as discussed in the previous
section, affects in particular the field renormalisation prescription. 
In Fig.~\ref{fig:2} we plot 
\begin{equation}
\frac{\delta
\mathit{\Gamma}}{\mathit{\Gamma}}=\frac{\mathit{\Gamma}^{\rm Full}-
\mathit{\Gamma}^{\rm Imp.\,Born}}{\mathit{\Gamma}^{\rm Imp.\,Born}},
\label{eq:deltaGamma}
\end{equation}
where we compare the result including the absorptive parts with an
approximation where the absorptive parts are neglected.
For unpolarised charginos in the final state, the proper treatment of the
absorptive parts affects the decay width by, at most, 0.4\%.
In Fig.~\ref{fig:2} we display 
the decay width for polarised charginos in the final state, 
$\Gamma(h_{2}\to \tilde{\chi}_{1,L}^+\tilde{\chi}_{2,R}^-)$.
We find that the numerical impact of the proper treatment of the
absorptive parts can amount up to a 3\% effect in the decay width. On
the other hand, as expected, the effect is seen to vanish for the case
of real parameters, i.e.\ $\phi_{A_t} = 0, \pi$.
The spikes seen in these plots arise due to the fact that at these
values of $\phi_{A_t}$ the masses of the $h_2$ and $h_3$ bosons cross. 
The spikes 
are seen to vanish for example on changing $M_{\tilde q_{3}}$ to 520 GeV, 
as the Higgs masses no longer cross for any value of $\phi_{A_t}$,
shown in the lower row of Fig.~\ref{fig:2}.
While there may be a chance to determine the polarisation of charginos 
through the angular distribution of their decays products, a detailed 
study of the prospects at the LHC 
is yet to be undertaken.

\subsection{Chargino production at a future Linear Collider}

As a second example, we now investigate chargino production at a 
Linear Collider, 
$\sigma(\mathit{e}^+\mathit{e}^-\to \tilde{\chi}_i^+\tilde{\chi}_j^-)$.
High-precision measurements of this process in the clean experimental
environment of an $e^+e^-$ Linear Collider could be crucial for
uncovering the fundamental parameters of this sector and for determining
the nature of the underlying physics. A treatment addressing the most
general case of complex parameters is mandatory in this context.

\begin{figure}[htb]
\begin{center}
\includegraphics[scale=0.83]{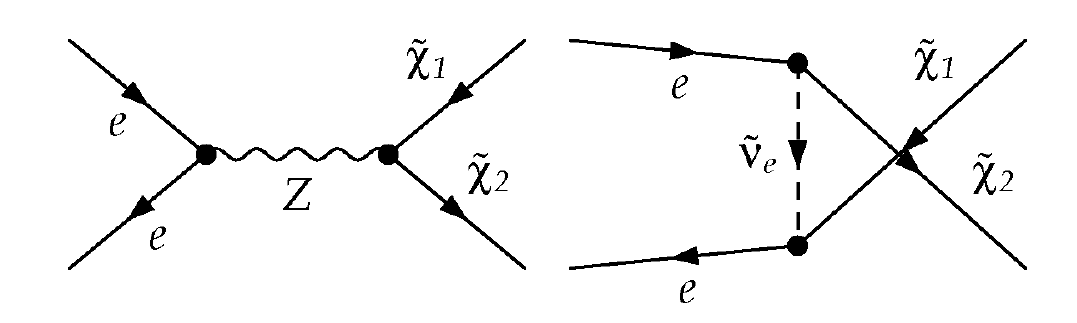}
\caption{Tree-level diagrams for the production of charginos
$\tilde\chi^+_{1}$ and $\tilde\chi^-_{2}$ at the LC.\label{fig:eeXXtree}}
\end{center}
\end{figure}

At leading order, in the limit of massless electrons, the process
$\sigma(\mathit{e}^+\mathit{e}^-\to \tilde{\chi}_1^+\tilde{\chi}_2^-)$
is described by the two diagrams shown in Fig.~\ref{fig:eeXXtree} (there
is one additional diagram for the $\tilde{\chi}_1^+\tilde{\chi}_1^-$ and
$\tilde{\chi}_2^+\tilde{\chi}_2^-$ final states).
The transition matrix element can be written as~\cite{Choi:2000ta},
\begin{equation}
 \mathcal M_{\alpha\beta}(e^+e^-\to\tilde\chi^+_i\tilde\chi^-_j)=i \frac{e^2}{s}Q_{\alpha\beta}\left[\bar v(e^+)\gamma_\mu\omega_\alpha u(e^-)\right]\left[\bar u(\tilde\chi^-_j)\gamma^\mu\omega_\beta v(\tilde\chi^+_i)\right],\label{eq:transAmp}
\end{equation}
in terms of the helicity amplitudes $Q_{\alpha\beta}$, where $\alpha$ refers to the chirality of the $e^+e^-$ current, $\beta$ to that of the 
$\tilde{\chi}_i^+\tilde{\chi}_j^-$ current, and summation over $\alpha$
and $\beta$ is implied. 
\begin{align}
\nonumber Q_{LL}=\,&\,\delta_{ij}+D_Z G_L C^L_{\tilde\chi^+_i\tilde\chi_j^-Z}\\
\nonumber Q_{RL}=\,&\,\delta_{ij}+D_Z G_R C^L_{\tilde\chi^+_i\tilde\chi_j^-Z}\\
\nonumber Q_{LR}=\,&\,\delta_{ij}+D_Z G_L \left(C^{R}_{\tilde\chi^+_i\tilde\chi_j^-Z}\right)^*+D_{\tilde\nu}\frac{1}{2 s_W^2}\left(C^{R}_{\tilde\nu_e e^+\tilde\chi_i^-}\right)^*C^R_{\tilde\nu_e e^+\tilde\chi_j^-},\\
Q_{RR}=\,&\,\delta_{ij}+D_ZG_R \left(C^{R}_{\tilde\nu_e e^+\tilde\chi_i^-}\right)^*.
\end{align}
The $Z\tilde\chi^+_{i}\tilde\chi^-_{j}$ and
$e\tilde\nu_e\tilde\chi^+_{i}$ couplings are given by
\begin{align}
\nonumber C^L_{\tilde\chi^+_i\tilde\chi_j^-Z}=\,&s_W^2\delta_{ij}-U^*_{j1}U_{i1}-\frac{1}{2}U^*_{j2}U_{i2},\\
\nonumber C^R_{\tilde\chi^+_i\tilde\chi_j^-Z}=\,&C^L_{\tilde\chi^+_i\tilde\chi_j^-Z}(U\to V^*),\\
C^R_{\tilde\nu_e e^+\tilde\chi_i^-}=\,&-V_{i1},
\end{align}
and $G_L$, $G_R$, $D_Z$ and $D_{\tilde\nu}$ are defined via
\begin{align}
\nonumber G_L=\,& \frac{s_W^2-\tfrac{1}{2}}{s_W^2 c_W^2},& G_R=\,&\frac{1}{c_W^2}\, , \\
D_Z=\,&\frac{s}{s-M_Z^2},& D_{\tilde\nu}=\,&\frac{s}{t-m_{\tilde \nu}^2}. 
\end{align}
Here $D_Z$ and $D_{\tilde\nu}$ refer to the propagators of the $Z$ boson
and sneutrino, respectively, in terms of the Mandelstam variables 
$s$ and $t$, and we can neglect the non-zero $Z$ width for the considered energies.
The tree-level cross section in the unpolarised case is then obtained by summing over the squared matrix elements,
\begin{equation}
 \sigma^{\rm
tree}=\frac{\kappa^{1/2}(s,m_{\tilde\chi^+_i},m_{\tilde\chi^-_j})}{64
\pi^2 s^2}\int d\Omega\sum_{\alpha,\beta}|{\mathcal M_{\alpha\beta}}|^2 .
\end{equation}

\begin{figure}[htb]
\begin{center}
\includegraphics[scale=0.8]{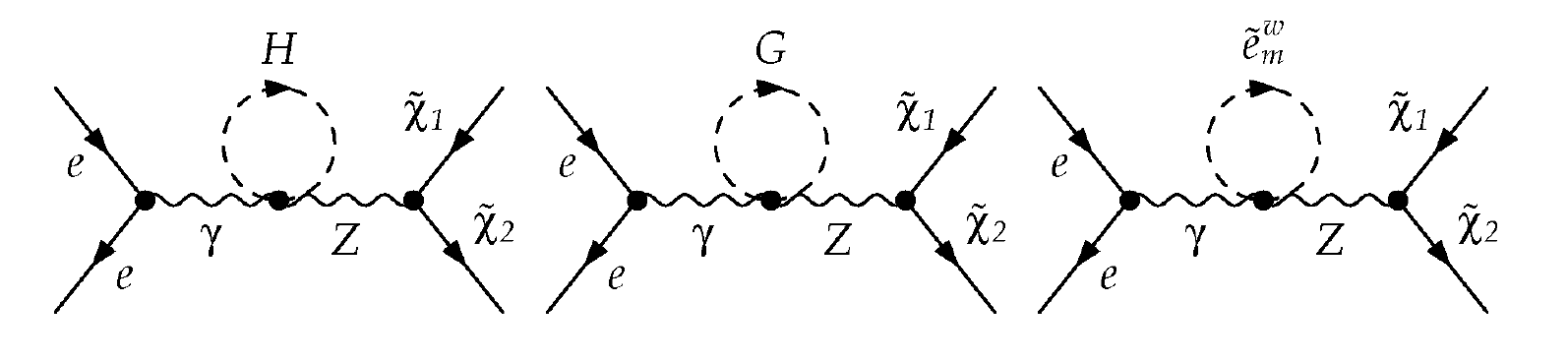}\\
\includegraphics[scale=0.8]{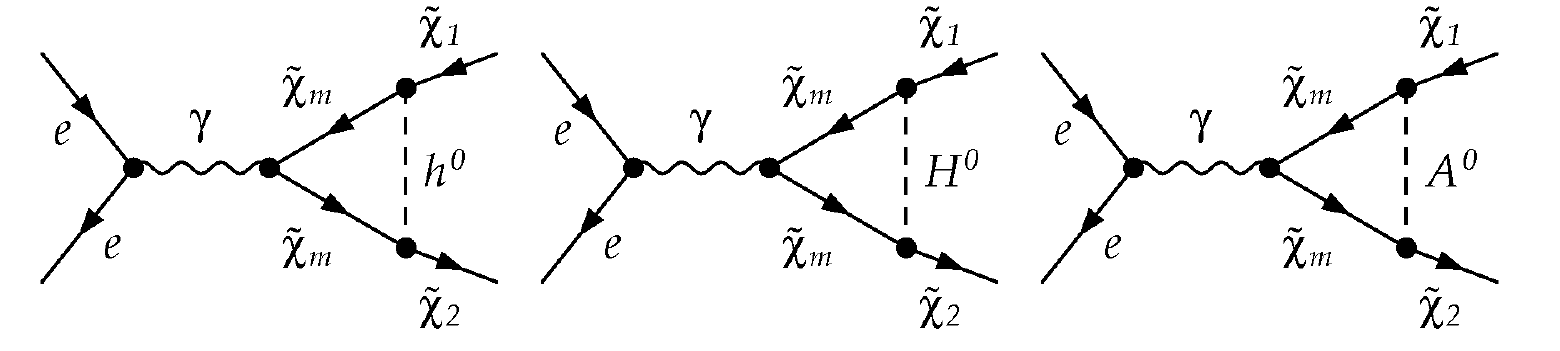}\\
\includegraphics[scale=0.8]{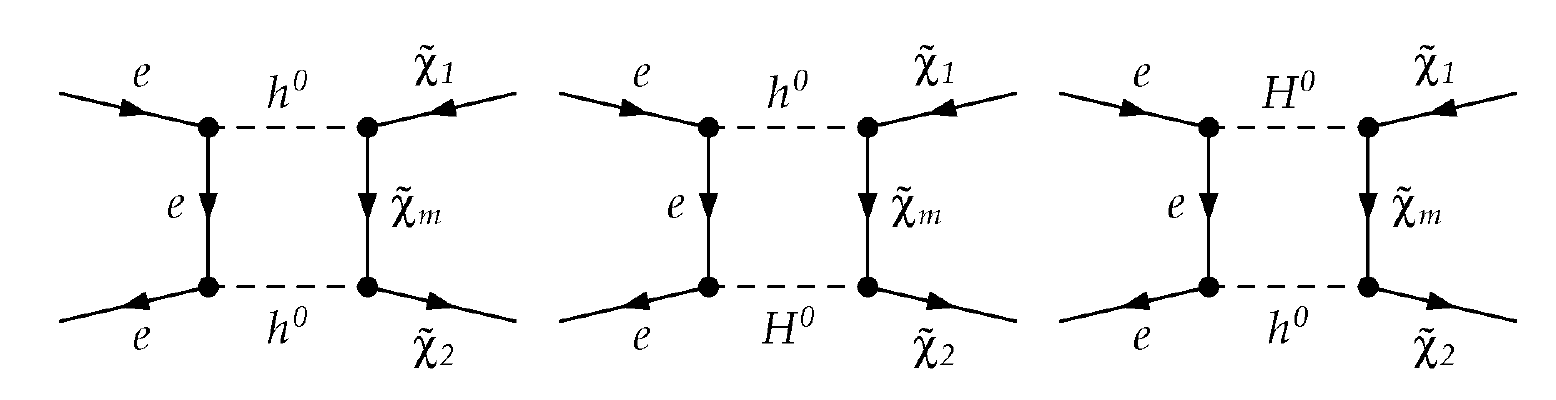}
\caption{Example one-loop self-energy (upper), vertex (middle) and box 
(lower) diagrams for the production of charginos $\tilde\chi^+_{1}$ and
$\tilde\chi^-_{2}$ at the LC.\label{fig:eeXXloopDiagrams}}
\end{center}
\end{figure}

The one-loop corrections involve self-energy, vertex and box diagrams, 
examples of which are shown in Fig.~\ref{fig:eeXXloopDiagrams}. The
diagrams are calculated following the procedure outlined earlier.

\begin{figure}[htb]
\begin{center}
\includegraphics[scale=0.8]{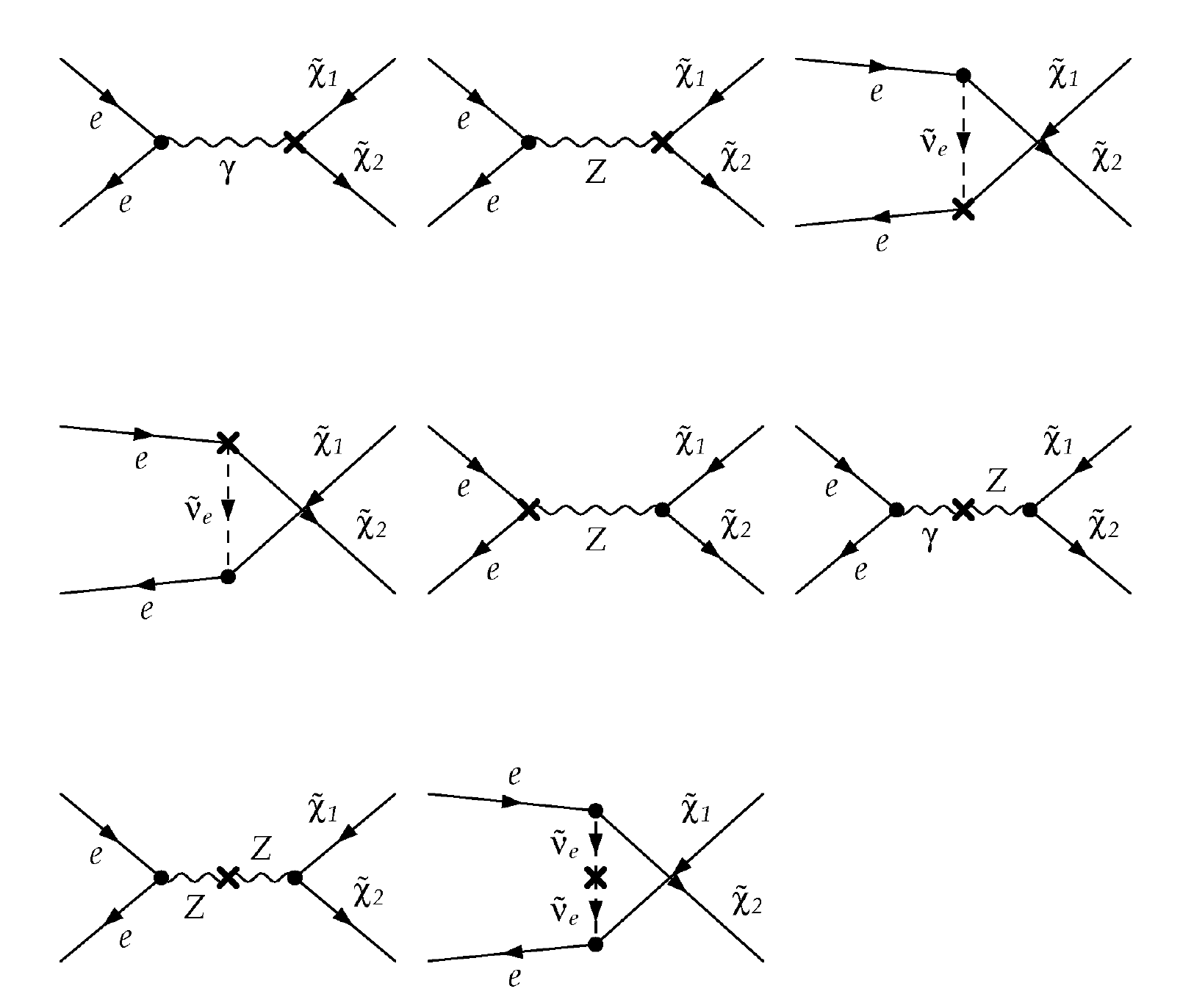}
\caption{Counterterm diagrams for the production of charginos $\tilde\chi^+_{1}$ and $\tilde\chi^-_{2}$ at the LC.\label{fig:eeXXcounterDiagrams}}
\end{center}
\end{figure}

In order to obtain finite results at one-loop, we need to renormalise
the couplings defined at tree-level in Eq.~(\ref{eq:transAmp}), i.e.\ we need to calculate the diagrams shown in Fig.\ref{fig:eeXXcounterDiagrams}.
This involves renormalising the $\gamma\tilde\chi^+_{i}\tilde\chi^-_{j}$, $Z\tilde\chi^+_{i}\tilde\chi^-_{j}$ and $e\tilde\nu_e\tilde\chi^+_{i}$ vertices as follows,
\begin{align}
\nonumber\delta\mathrm{G}^L_{\tilde\chi^+_i\tilde\chi_j^-\gamma}=\,& \frac{i e}{2}\left(\delta_{ij} \left(2 \delta Z_e+\delta Z_{\gamma\gamma}\right)-\frac{\delta Z_{Z\gamma}}{c_W s_W} C^L_{\tilde\chi^+_i\tilde\chi_j^-Z}+\delta Z^L_{ij}+\delta \bar Z^L_{ij}\right),\\
\nonumber\delta\mathrm{G}^L_{\tilde\chi^+_i\tilde\chi_j^-Z} =\,& \frac{-i e}{c_W s_W}\bigg(\delta C^L_{\tilde\chi^+_i\tilde\chi_j^-Z}+C^L_{\tilde\chi^+_i\tilde\chi_j^-Z}\left(\delta Z_e-\frac{\delta c_W}{c_W}-\frac{\delta s_W}{s_W}+\frac{\delta  Z_{ZZ}}{2}\right)\\
&-\delta_{ij}\frac{c_W s_W}{2}\delta Z_{\gamma
Z}+\frac{1}{2}\sum_{n=1,2}\left(\delta
Z^L_{nj}C^L_{\tilde\chi^+_i\tilde\chi_n^-Z}+C^L_{\tilde\chi^+_n\tilde\chi_j^-Z}\delta
\bar Z^L_{in}\right)\bigg).
\end{align}
where the analogous right-handed parts are obtained by the replacement
$L\to R$. Furthermore,
\begin{align}
\nonumber \delta\mathrm{G}^R_{\tilde\nu_e e^+\tilde\chi_i^-}=\,&\frac{i e\delta_{ij}}{s_W}\bigg(C^R_{\tilde\nu_e e^+\tilde\chi_i^-}\bigg(\delta Z_e-\frac{\delta s_W}{s_W}
+\frac{1}{2}\left(\delta Z_{\tilde \nu_e}+\delta Z^{e^*}_{L}\right)\\
\nonumber &+\frac{1}{2}\left(\delta Z^R_{1i}V^*_{12}+\delta Z^R_{2i} V^*_{22}\right)\bigg)+\delta C^R_{\tilde\nu_e e^+\tilde\chi_i^-}
\bigg),\\
\delta\mathrm{G}_{\overline{\tilde\nu}_i \tilde\nu_j}=\,&i
\delta_{ij}\left(\frac{1}{2}(\delta Z_{\tilde\nu_i}+\delta
Z^*_{\tilde\nu_i})p^2-\delta
m_{\tilde\nu_i}^2-\frac{m_{\tilde\nu_i}^2}{2}(\delta
Z_{\tilde\nu_i}+\delta Z^*_{\tilde\nu_i})\right),
\end{align}
for $\tilde\nu_i=\tilde\nu_e,\tilde\nu_\mu,\tilde\nu_\tau$, and the
counterterm contributions of the coupling factors are given by
\begin{align}
\nonumber\delta C^L_{\tilde\chi^+_i\tilde\chi_j^-Z}=\,&\delta C^R_{\tilde\chi^+_i\tilde\chi_j^-Z}=2s_W\delta s_W\delta_{ij},\\
\delta C^R_{\tilde\nu_e e^+\tilde\chi_i^-}=\,&0.
\end{align}
Note that again for brevity we drop the $\pm$ for the chargino field renormalisation constants.
Using this prescription to renormalise the vertices we obtain UV-finite results.

As the incoming and outgoing particles are charged, in order to obtain
an infra-red finite result one must furthermore include soft photon
radiation, which introduces the dependence on a cut-off.
Using the phase-space slicing method the full phase space for the real
photonic corrections can be divided into a soft, a hard collinear and a 
hard non-collinear (IR finite) region,
\begin{equation}
\sigma^{\rm brems} = \sigma^{\rm soft}(\Delta E) + 
  \sigma^{\rm hard}_{\rm coll}(\Delta E, \Delta\theta) + 
  \sigma^{\rm hard}_{\rm non-coll}(\Delta E, \Delta\theta) .
\end{equation}
Here the singular soft and hard collinear regions are defined
by $E<\Delta E$ and $\theta<\Delta \theta$, respectively. Accordingly,
the full cross section at next-to-leading order, including also the virtual
contributions, is given by
\begin{equation}
\sigma^{\rm full}=\sigma^{\rm tree}+\sigma^{\rm virt+soft}(\Delta E) +
  \sigma^{\rm hard}_{\rm coll}(\Delta E, \Delta\theta) + 
  \sigma^{\rm hard}_{\rm non-coll}(\Delta E, \Delta\theta) .
\end{equation}
Since in our analysis we are particularly interested in the relative
size of the weak SUSY corrections, it is useful to consider a 
``reduced genuine SUSY cross-section'', as defined by the SPA 
convention~\cite{AguilarSaavedra:2005pw}, where the numerically large 
logarithmic terms of the QED-type corrections depending on $\Delta E$
and the terms proportional to $L_e \equiv \log s/m_e^2$ are subtracted
in a consistent and gauge-independent way. Accordingly, our
numerical analysis below is done for the quantity (see
Refs.~\cite{AguilarSaavedra:2005pw,Oller:2005xg})
\begin{align}
\sigma^{\rm weak}=\,&\,\sigma^{\rm tree} +
\sigma^{\rm virt+soft}(\Delta E)+\sigma^{\Delta E},\\
\sigma^{\Delta E}=\,&-\frac{\alpha}{\pi}\sigma^{\rm
tree}\bigg(\frac{3}{2} L_e+\log \frac{4 (\Delta
E)^2}{s}(L_e-1+\Delta_\gamma)\bigg),
\end{align}
where $\Delta_\gamma$ is given by the coefficient of the terms in the
soft photon correction involving $\Delta E$ that arise from final state
radiation (ISR) and from the interference between
initial and final state radiation (IFI). To be more explicit, we express
the soft photon contribution as a sum of initial state radiation (ISR),
FSR and IFI,
\begin{equation}
 \sigma^{\rm soft}\,=\,-\sigma^{\rm tree}\frac{\alpha}{4
\pi^2}\left(\delta_{\rm soft}^{\rm ISR}+\delta_{\rm soft}^{\rm
FSR}+\delta_{\rm soft}^{\rm IFI}\right) .
\end{equation}
Defining $\delta_{\rm ISR}$ etc., in terms of the soft photon integrals
$I_{ij}$~\cite{Denner:1991kt},
\begin{align}
\nonumber\delta_{\rm soft}^{\rm
ISR}\,=\,&\,I_{p_1p_1}+I_{p_2p_2}-2I_{p_1p_2}\,,\\
\nonumber\delta_{\rm soft}^{\rm
FSR}=\,& I_{k_1k_1}+I_{k_2k_2}-2I_{k_1k_2}\,,\\
\delta_{\rm soft}^{\rm
IFI}\,=\,&\,-2\left(I_{p_1k_1}+I_{p_2k_2}-I_{p_1k_2}-I_{p_2 k_1}\right)\,,
\end{align}
the quantity $\Delta_\gamma$ can be obtained
by taking the coefficient of 
$4\pi\log\frac{4(\Delta E)^2}{s}$ in $\delta_{\rm soft}^{\rm FSR}+\delta_{\rm
soft}^{\rm IFI}$.

In the soft limit, the photonic contributions can be factorised into analytically integrable expressions proportional to the tree-level cross-section for $\sigma(\mathit{e}^+\mathit{e}^-\to \tilde{\chi}_i^+\tilde{\chi}_j^-)$.
In our calculation carried out with \texttt{FormCalc} the contributions
from soft photon radiation have been incorporated 
using the soft photon factor as given explicitly in 
Ref.~\cite{Denner:1991kt}.

Restricting our general result for complex MSSM parameters to the
special case of vanishing phases, we have compared with the results 
given in Refs.~\cite{Oller:2005xg,Fritzsche:2005}, which were evaluated 
in the $\rm SPS1a^\prime$ benchmark scenario. The renormalisation
prescription in Ref.~\cite{Oller:2005xg} differs from the one used in
the present work in the renormalisation of the chargino and neutralino
mixing matrices as well as of the electric charge and $\tan\beta$.
Furthermore, a different choice has been made in
Ref.~\cite{Oller:2005xg} for the masses chosen as input in the selectron /
sneutrino sector (as a consequence, the sneutrino mass in
Ref.~\cite{Oller:2005xg} receives a shift at the one-loop level, while
we have chosen an on-shell condition for the sneutrino mass). On the
other hand, in the special case of real parameters our renormalisation
prescription is the same as the one used in Ref.~\cite{Fritzsche:2005},
with the exception of the renormalisation of $\tan\beta$. We find
numerical agreement within the expected accuracy with the results given in 
Refs.~\cite{Oller:2005xg,Fritzsche:2005}.

\begin{figure}
\hspace{-.8cm}
\begin{tabular}{cc}
\includegraphics[scale=.56]{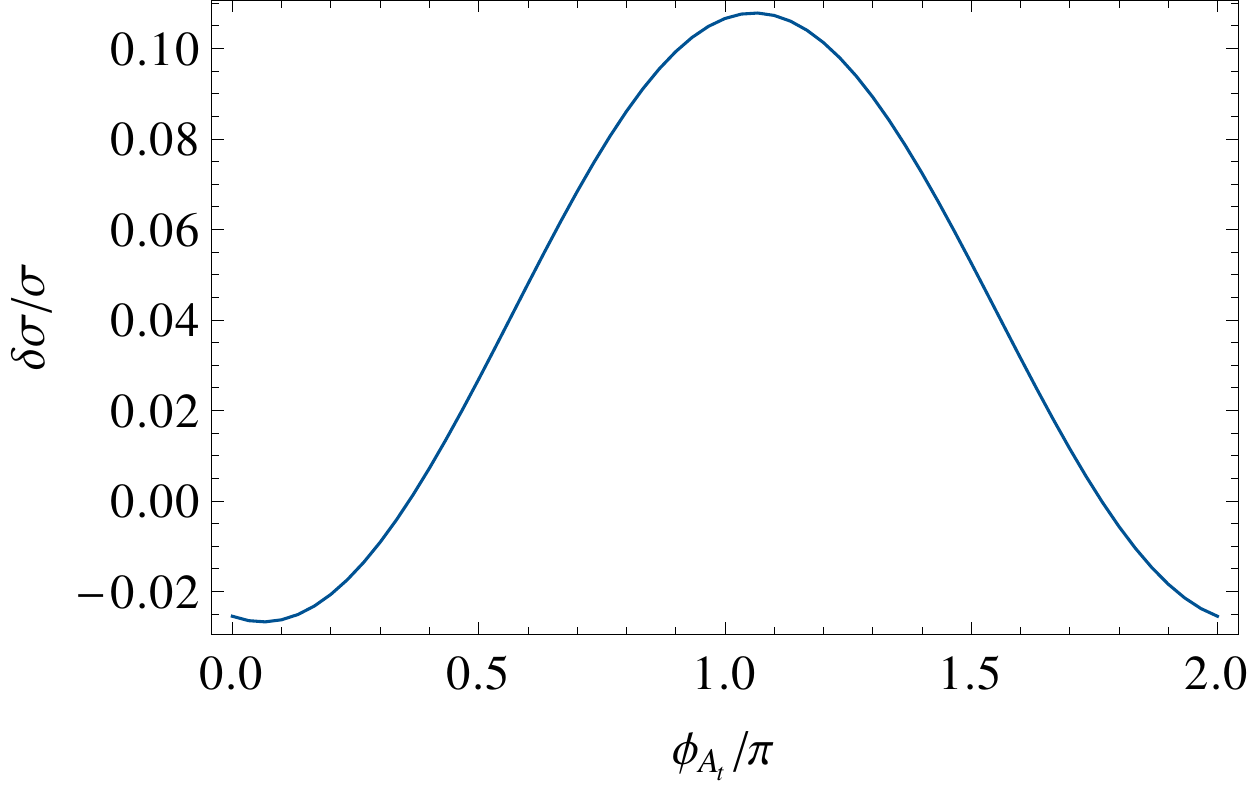}
\includegraphics[scale=.56]{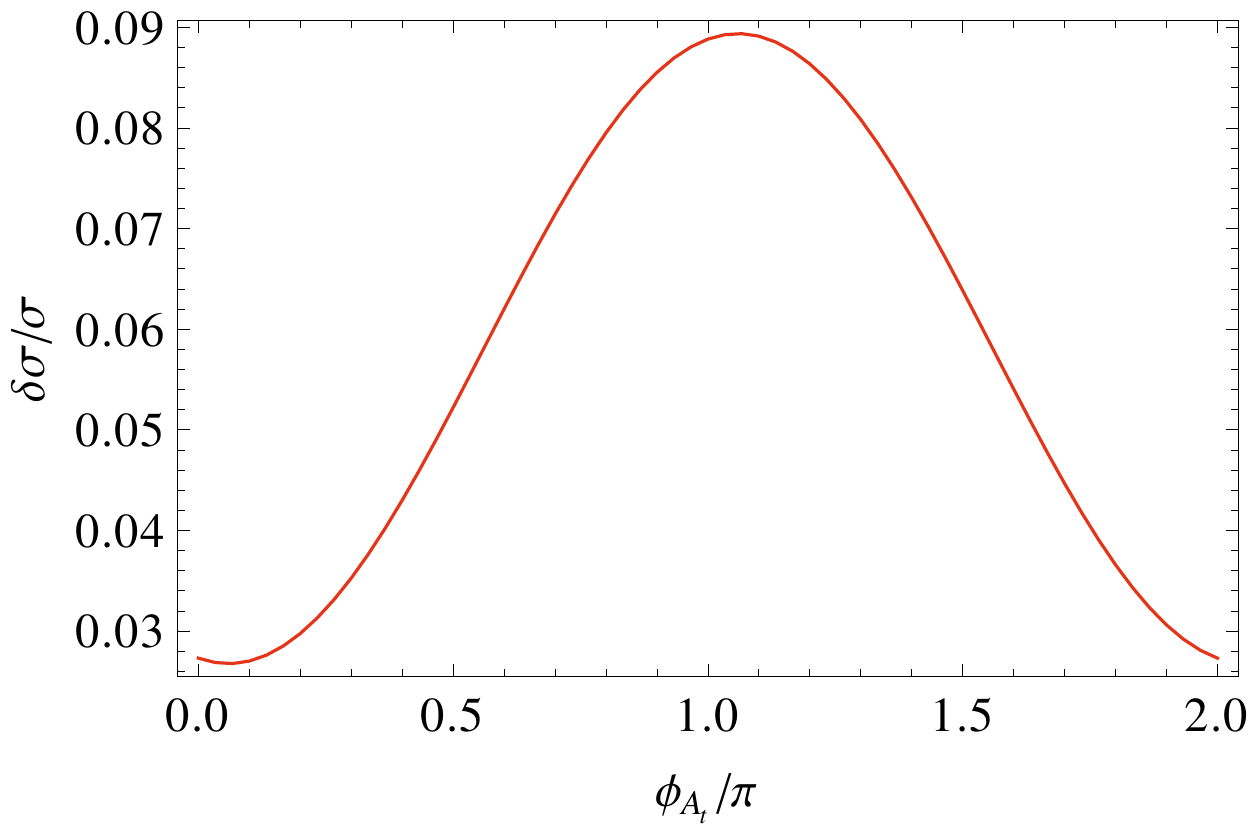}\\
\includegraphics[scale=.56]{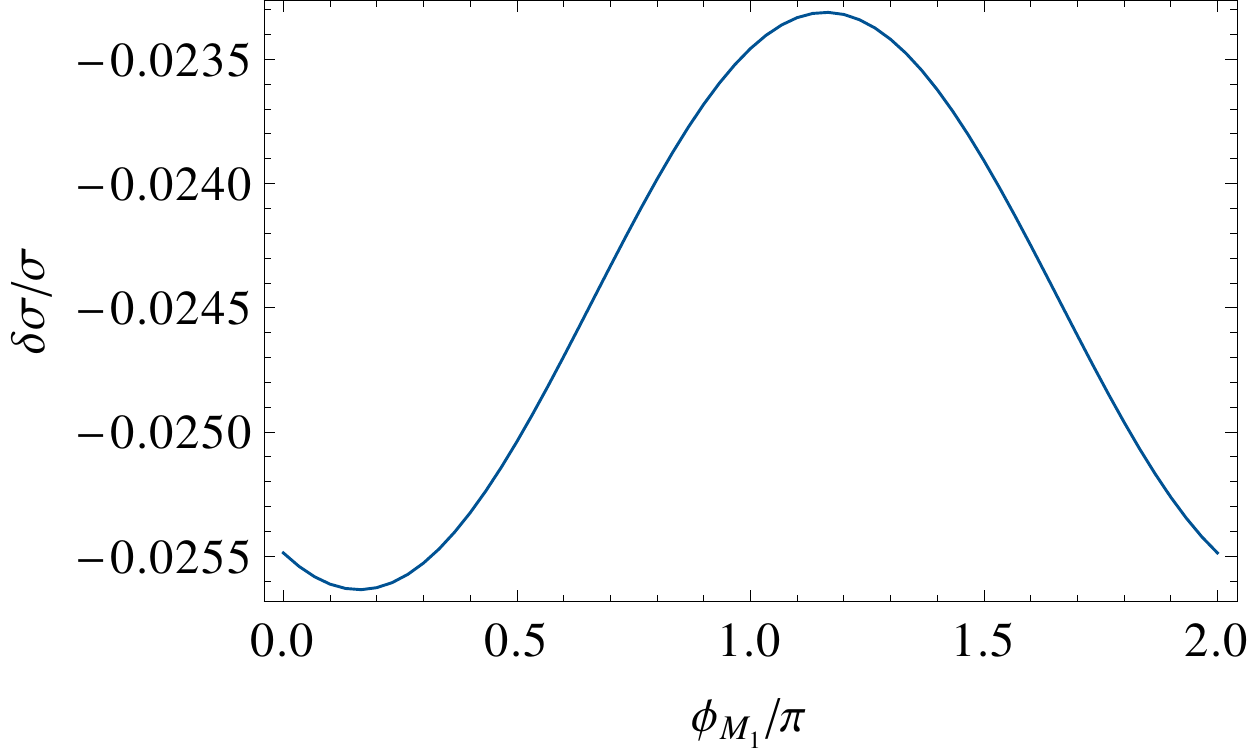} \includegraphics[scale=.56]{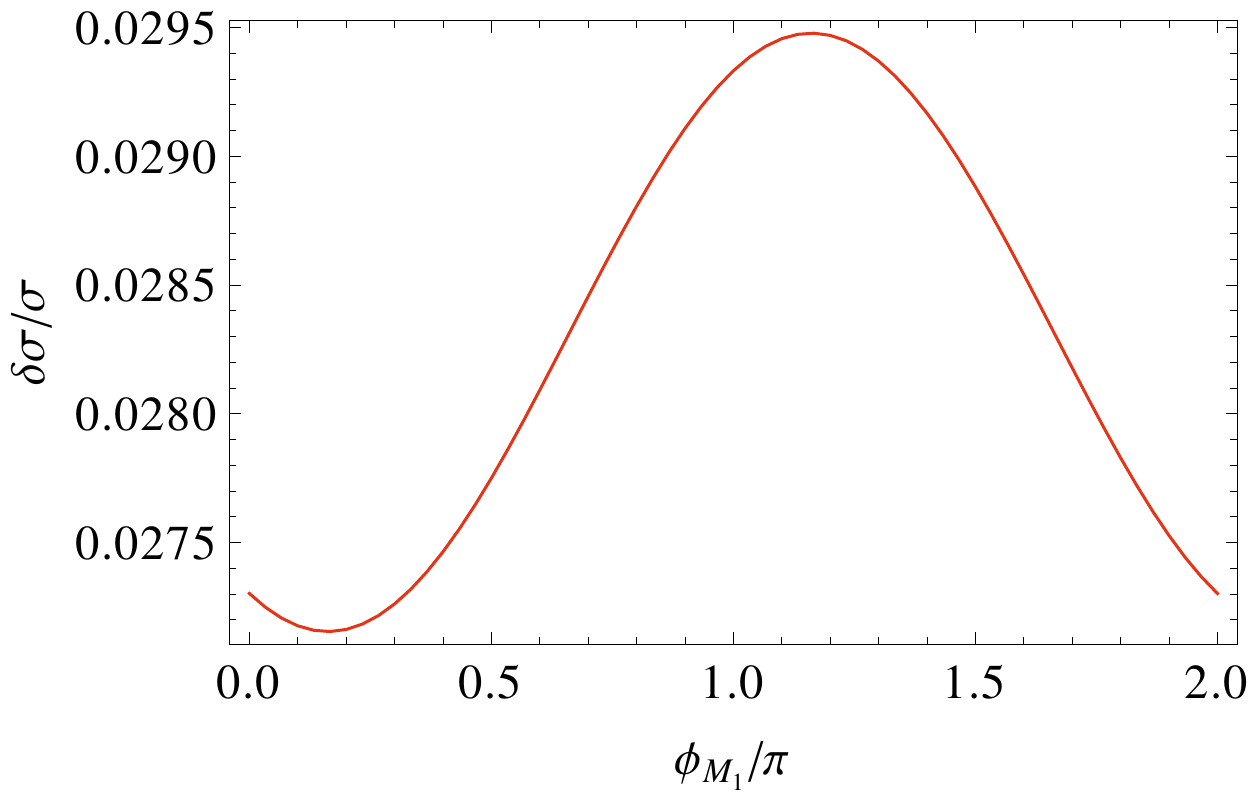}\\
\includegraphics[scale=.56]{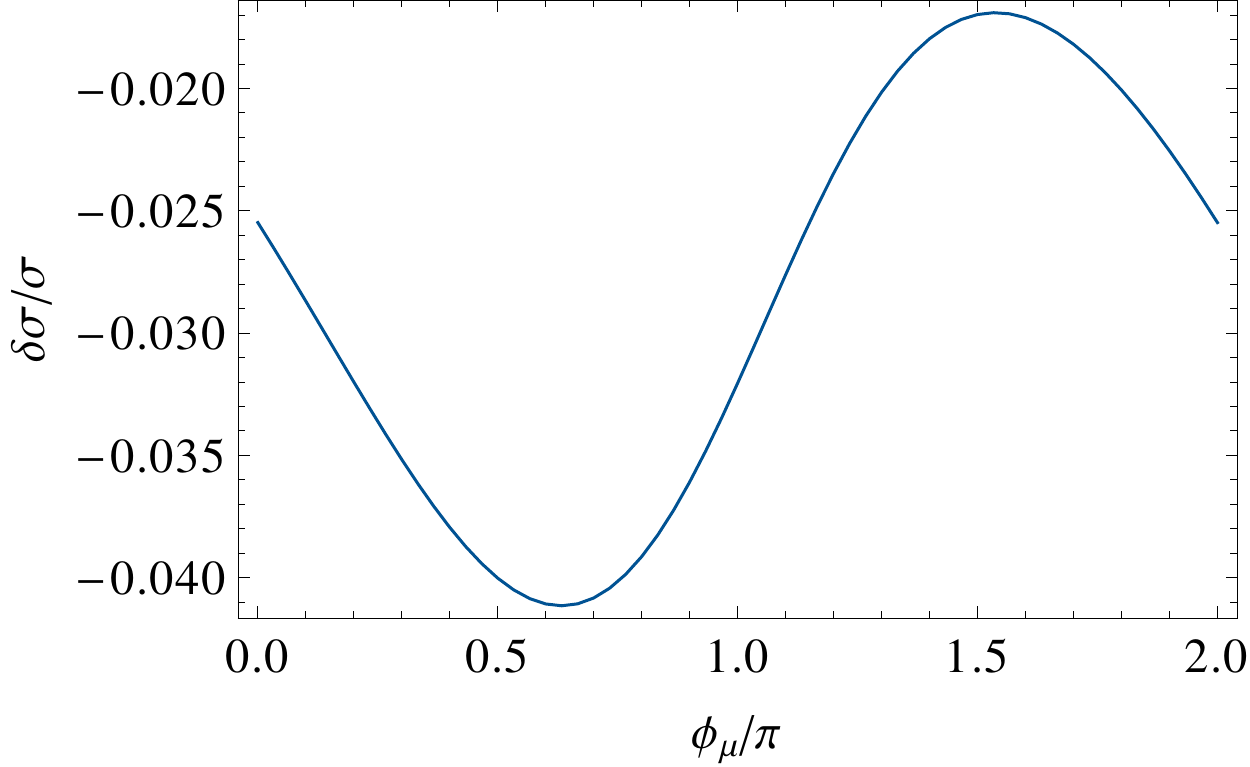} \includegraphics[scale=.56]{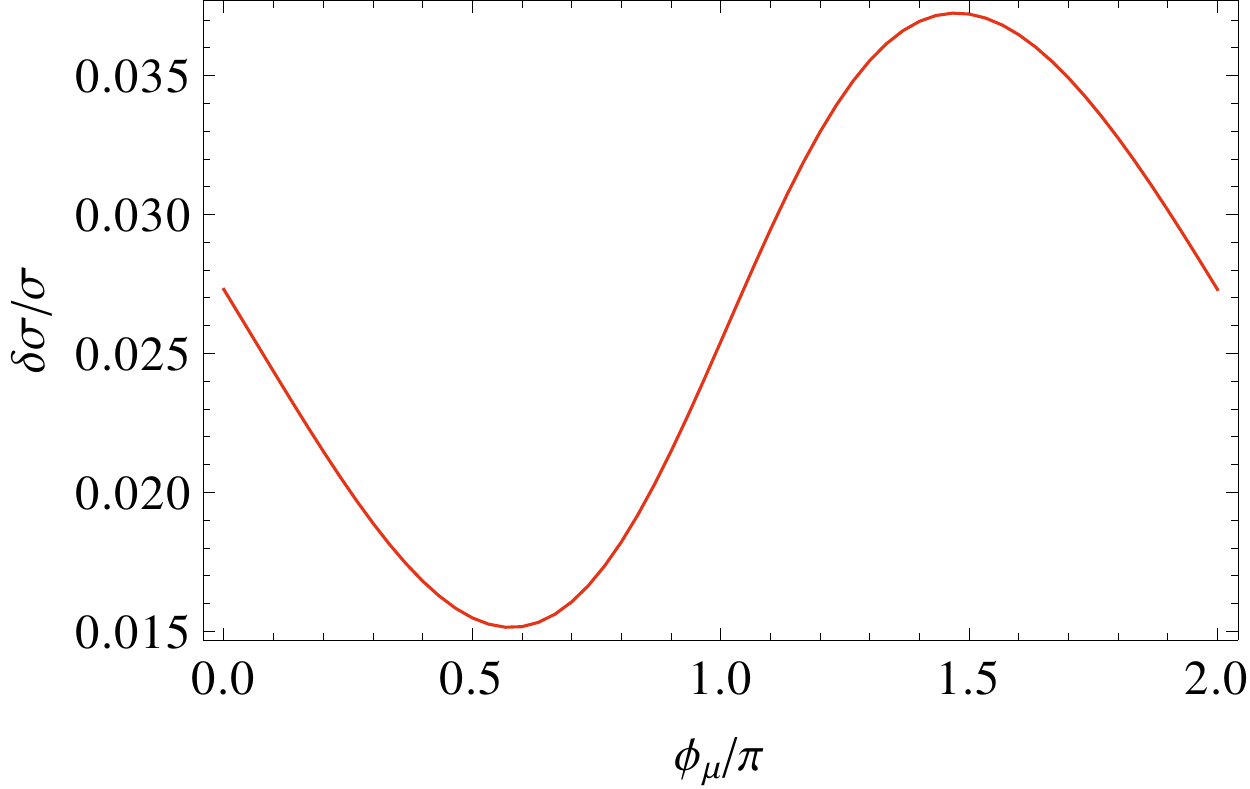}
\end{tabular}
\caption{$\delta\sigma/\sigma$ for $e^+e^-\to
\tilde{\chi}_1^+\tilde{\chi}_2^-$ as a function of the phases
$\phi_{A_t}$ (upper), $\phi_{M_1}$ (middle) and $\phi_\mu$ (lower row) for 
$M_{\tilde q_{3}} = 600$ (left) and 800 (right) GeV.\label{fig:4}}
\end{figure}

We now turn to the investigation of our results for the case of complex
MSSM parameters. In particular, we study the relative size of the one-loop 
corrections,
\begin{equation}
\frac{\delta\sigma}{\sigma}=\frac{(\sigma^{\rm weak}-\sigma^{\rm tree})}{\sigma^{\rm tree}}
\label{eq:deltaSigma}
\end{equation}
as a function of $\phi_{A_t}$, $\phi_\mu$ and $\phi_{M_1}$, for a
$\sqrt{s}= 800\gev$ LC.
In Fig.~\ref{fig:4}, the dependence on each of the phases is seen to be
qualitatively the same for $M_{\tilde q_{3}} = 600\gev$ and 800~GeV.
In the case of $\phi_{A_t}$, the dependence is sizeable, due to the
Yukawa enhancement for the stop loops, leading to effects of up to
$\sim 12\%$ for $M_{\tilde q_{3}}=600 \gev$ and up to $\sim6\%$ for 
$M_{\tilde q_{3}} =800 \gev$.
For the EDM-allowed regions of $\phi_{M_1}$, which enters only at
loop-level, and $\mu$, which is highly constrained, the numerical impact
of the phase variations is rather small, at most $\sim0.2\%$ for $\phi_{M_1}$.
Thus, particularly for low values of $M_{\tilde q_{3}}$, the
phenomenologically most relevant effect arises from varying the 
phase $\phi_{A_t}$.

\begin{figure}[h]
\begin{center}
\includegraphics[scale=.56]{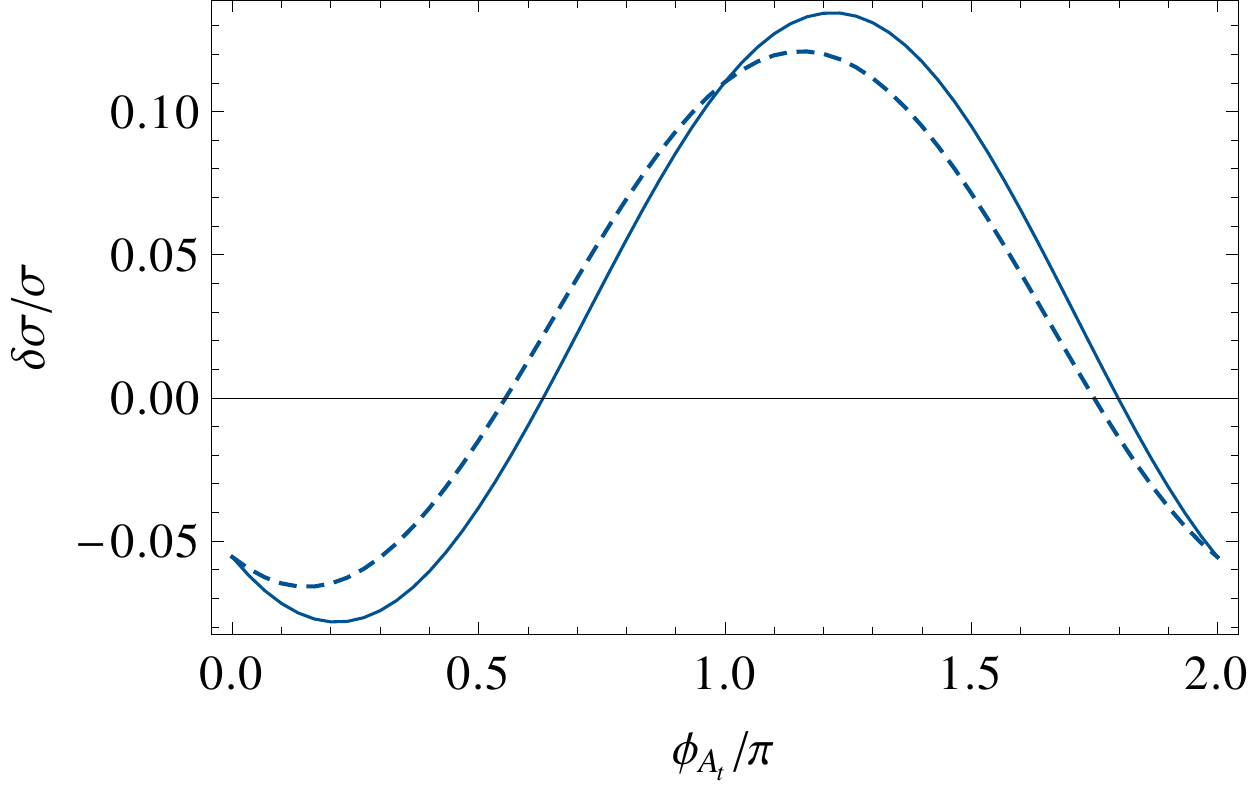}\includegraphics[scale=.56]{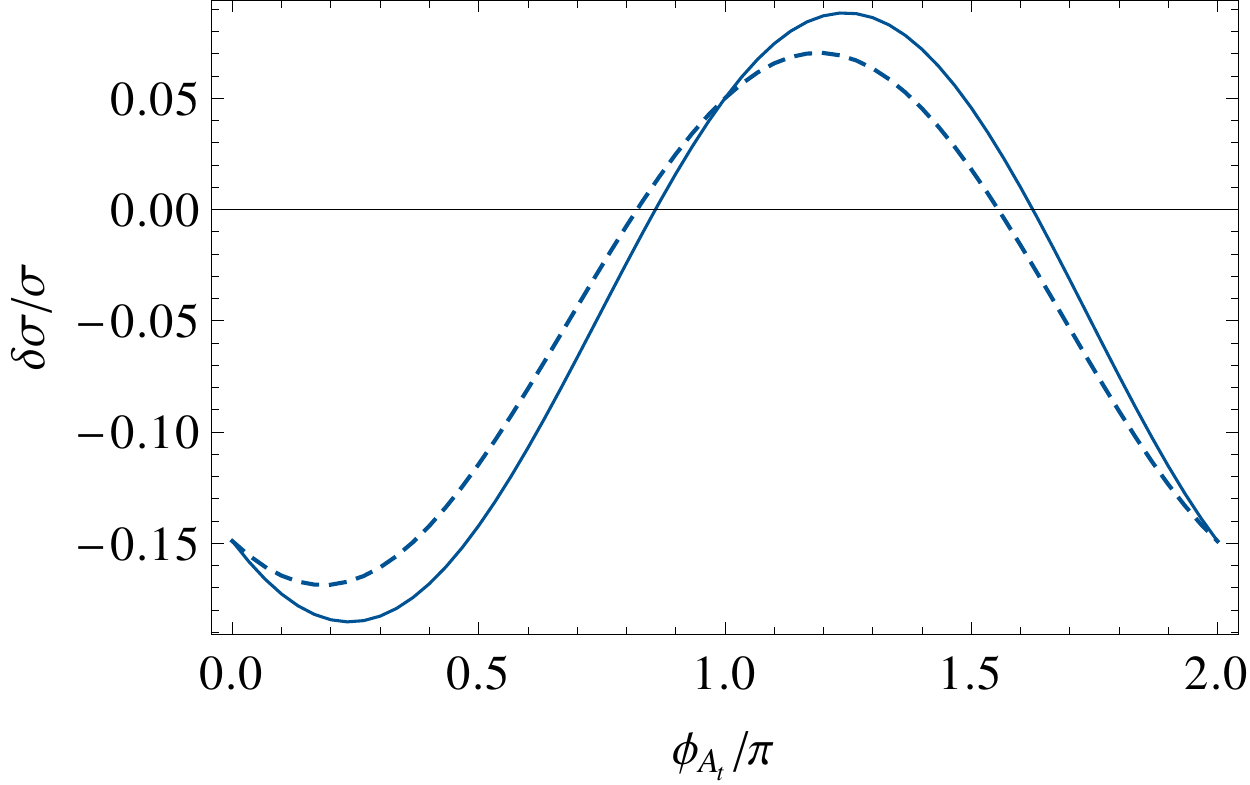}
\caption{$\delta\sigma/\sigma$ for $e^+e^-\to
\tilde{\chi}_{1}^+\tilde{\chi}_{2}^-$ (left) and $e^+e^-\to
\tilde{\chi}_{1,L}^+\tilde{\chi}_{2,R}^-$ (right) as a function of the phase
$\phi_{A_t}$, for $M_{\tilde q_{3}} = 500 \gev$, including (solid) and 
ignoring (dashed) the absorptive part of the loop integrals in the 
field renormalisation.\label{fig:5}}
\end{center}
\end{figure}

In Fig.~\ref{fig:5} we plot $\delta\sigma/\sigma$ as a function of
$\phi_{A_t}$, comparing the results of including and ignoring the 
absorptive parts of the loop integrals in the field renormalisation.
The left plot shows the unpolarised cross section for 
$e^+e^-\to \tilde{\chi}_{1}^+\tilde{\chi}_{2}^-$, while the right
plot shows the result for specific polarisation states of the produced
charginos, $e^+e^-\to \tilde{\chi}_{1,L}^+\tilde{\chi}_{2,R}^-$.
The impact of properly accounting for the absorptive parts of the loop
integrals in the field renormalisation is clearly visible in
Fig.~\ref{fig:5}. The resulting difference amounts to 
up to $\sim 2\%$, which could be phenomenologically relevant at 
linear collider precisions. Thus, a consistent inclusion of the
absorptive parts of loop integrals is not only desirable from a
conceptual point of view, but can also give rise to phenomenologically
relevant effects in the MSSM with complex parameters.
Another feature that can be seen in Fig.~\ref{fig:5} is the fact that 
for $\phi_{A_t}=\pi$ (in the present case where all other phases are set to
zero) the results of including and ignoring the absorptive parts
coincide. This is as expected, confirming that in the case of real
parameters the absorptive parts can be neglected, see Sec.~\ref{sec:4.1}.

\begin{figure}[tb]
\begin{center}
\includegraphics[scale=.56]{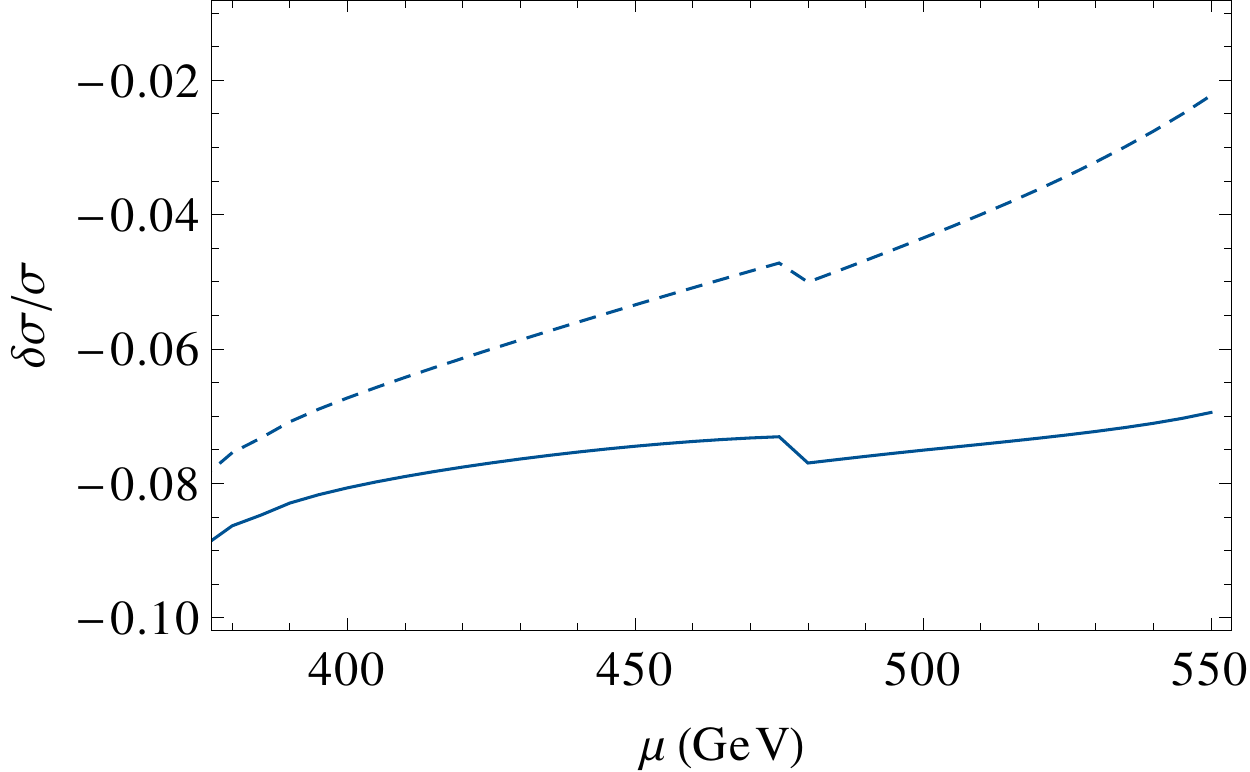}
\caption{$\delta\sigma/\sigma$ for $e^+e^-\to
\tilde{\chi}_{1}^+\tilde{\chi}_{2}^-$ as a function of $\mu$, for
$M_{\tilde q_{3}} = 500 \gev$ and $\phi_{A_t}=\pi/4$, including (solid) 
and ignoring (dashed) the absorptive part of the loop integrals in 
the field renormalisation.\label{fig:6}}
\end{center}
\end{figure}

We find that the numerical impact of properly incorporating the
absorptive parts increases with an increasing hierarchy between the
parameters $M_1$, $M_2$ and $\mu$.
This is illustrated in Fig.~\ref{fig:6}, where the cross section for
$e^+e^-\to \tilde{\chi}_{1}^+\tilde{\chi}_{2}^-$ is
shown as a function of $\mu$. The difference between the results 
including and neglecting the absorptive parts is seen to increase for
increasing $\mu$.
The behaviour of the results around $\mu = 480 \gev$ is caused by a 
threshold effect due to the sneutrino mass lying in this region.

\section{Conclusions}\label{sec:5}

We have derived a renormalisation scheme for the chargino and neutralino
sector of the MSSM that is suitable for the most general case of complex
parameters. We have put particular emphasis on a consistent treatment of 
imaginary parts, which arise on the one hand from the complex parameters
of the model and on the other hand from absorptive parts of loop
integrals. We have demonstrated that products of imaginary parts can 
contribute to predictions for physical observables in the MSSM already
at the one-loop level and therefore need to be taken into account in
order to obtain complete one-loop results.

Concerning the parameter renormalisation in the chargino and neutralino
sector, we have shown that the phases of the parameters in the chargino
and neutralino sector do not need to be renormalised at the one-loop
level. We have therefore adopted a renormalisation scheme where only the
absolute values of the parameters $M_1$, $M_2$ and $\mu$ are subject to
the renormalisation procedure. In order to perform an on-shell
renormalisation for those parameters one needs to choose three out of
the six masses in the chargino and neutralino sector that are
renormalised on-shell, while the predictions for the physical masses of
the other three particles receive loop corrections. We have
demonstrated, using the examples of gaugino-like and higgsino like
scenarios with complex parameters, that the appropriate choice for the
mass parameters
used as input for the on-shell conditions depends both on the process
and the region of MSSM parameter space under consideration. In order to
avoid unphysically large contributions to the counterterms and the 
mass predictions one needs to choose for the on-shell renormalisation
one bino-like, one wino-like and one higgsino-like particle.
We have provided full expressions for the renormalisation constants of 
$|M_1|$, $|M_2|$ and $|\mu|$ for the case where $M_1$ and $\mu$ can 
be complex (i.e., we have adopted the convention where the phase of 
$M_2$ has been rotated away) and for all possible combinations of charginos 
and neutralinos being chosen on-shell.

For the field renormalisation, the consistent incorporation of
absorptive parts gives rise to the fact that full on-shell conditions,
which ensure that all mixing contributions of the involved fields vanish
on-shell, can only be satisfied if independent
field renormalisation constants are chosen for incoming and outgoing
fields. If instead one works with a scheme where those renormalisation
constants are related to each other by the usual hermiticity relations, 
non-trivial corrections associated with the external legs of the
considered diagrams (i.e., finite wave function normalisation factors)
need to be incorporated in order to obtain the correct on-shell
properties of the incoming and outgoing particles.

Within the described renormalisation framework we have derived
complete one-loop results for the processes 
$h_a\to\tilde\chi_i^+\tilde\chi_j^-$ (supplemented by Higgs-propagator
corrections up to the two-loop level) and 
$e^+e^-\to\tilde\chi_i^+\tilde\chi_j^-$ in the MSSM with complex
parameters. For both processes we have investigated the dependence of
the results on the phases of the complex parameters. In particular, we have
analysed in this context the numerical relevance of the absorptive 
parts of loop integrals.

Concerning our results for heavy Higgs decays to charginos, 
$h_a\to\tilde\chi_i^+\tilde\chi_j^-$,
which may be of interest for SUSY Higgs searches at the LHC, 
we find that the phase variations have a significant 
numerical impact on the prediction for the decay width. In particular,
varying the phase $\phi_{A_t}$, which is so far almost unconstrained by
the EDMs, can lead to effects of up to 40\% in the decay width.
We find that the impact of the absorptive parts in the field
renormalisation constants is most pronounced for the case of polarised
charginos in the final state, for which the impact of a proper treatment
of the absorptive parts can amount up to a 3\% effect in the decay
width.

For chargino pair-production at an 
$e^+e^-$ Linear Collider, $e^+e^-\to\tilde\chi_i^+\tilde\chi_j^-$, 
we find that the dependence of the cross-section on the phase
$\phi_{A_t}$ is sizable, yielding effects of up to 12\% in our example.
The impact of a proper treatment of the absorptive parts in the field
renormalisation constants turns out to be numerically relevant in view
of the prospective experimental accuracy of measurements at a future 
Linear Collider. We find effects of 2--5\% in our numerical example. 
Our results for the one-loop contributions to chargino pair-production
at a Linear Collider for the general case of complex MSSM parameters may
also be of interest for investigating the accuracy with which the
parameters of the MSSM Lagrangian can be determined from high-precision
measurements at a Linear Collider, since in this context the
incorporation of higher-order effects in the theoretical predictions,
which lead to a non-trivial dependence on a variety of MSSM parameters,
is inevitable.

\section*{Acknowledgements}
The authors gratefully acknowledge support of the DFG through the grant SFB 676, ``Particles, Strings, and the Early Universe'', as well as the Helmholtz Alliance, ``Physics at the
 Terascale''. AKMB would also like to thank Karina Williams and Krzysztof Rolbiecki for many helpful discussions.

\end{document}